\documentclass{aa}  

\usepackage{amsmath, amssymb, amsfonts, amscd}
\usepackage{natbib}
\usepackage{graphicx}
\usepackage{txfonts}
\usepackage[final]{hyperref}
\usepackage{mathrsfs}
\usepackage{color}
\usepackage{mathabx}
\usepackage{supertabular}
\usepackage{empheq}
\usepackage{multirow}
\usepackage{numprint}
\usepackage{siunitx}
\hypersetup{
   colorlinks=true,
   citecolor=blue,
   linkcolor=blue,
   urlcolor=black
   }

\def\define{\equiv}				
\def\sign{{\rm sign}}				
\def\scale{\, \propto \,}			
\def\scalprod{\cdot}				
\def\rond{\circ}
\def\function{f}
\def\fsmooth{h}

\newcommand{\ddroit}{{\rm d}}				
\newcommand{\vect}[1]{\boldsymbol{#1}}		
\newcommand{\dd}[2]{\partial_{#2} #1}		
\newcommand{\DD}[2]{\dfrac{\ddroit #1}{\ddroit #2}}			
\newcommand{\DDn}[3]{\dfrac{\ddroit^{#3} #1}{\ddroit #2^{#3}}} 
\newcommand{\Dpart}[2]{\dfrac{{\rm D} #1}{{\rm D} #2}} 
\newcommand{\abs}[1]{\left| #1 \right|}		

\def\logdix{\log_{\rm 10}}

\def\pressure{p}

\def\nab{\nabla}					
\def\grad{\nab}					
\def\gradp{\grad_{\pressure}}
\def\gradp{\grad_{\pressure}}

\newcommand{\expo}[1]{{\rm e}^{#1}}				
\newcommand{\integ}[4]{\int_{#3}^{#4} #1 {\rm d} #2 }	
\newcommand{\normal}[1]{\tilde{#1}}					

\def\Ggrav{G}					

\def\rr{r}						
\def\time{t}					
\def\zz{z}
\def\XX{X}

\def\luminosity{L}				
\def\ipla{{\rm p}}				
\def\istar{\star}					
\def\smaxis{a}					
\def\norb{n_\istar}				
\def\spinrate{\Omega}			
\def\Lstar{\luminosity_{\istar}}

\def\Mbody{M}					
\def\Rbody{R}					
\def\Mpla{\Mbody_\ipla}			
\def\Mstar{\Mbody_\istar}			
\def\Rpla{\Rbody_\ipla}			

\def\iearth{\Earth}				
\def\Mearth{\Mbody_{\iearth}}		
\def\Rearth{\Rbody_{\iearth}}		

\def\isun{\sun}					
\def\Msun{\Mbody_{\isun}}			
\def\Lsun{\luminosity_{\isun}}		

\def\chartime{t}				
\def\ggravi{g}					

\def\irot{\ipla}

\def\period{P}					
\def\Porb{\period_{\istar}}			
\def\irot{{\rm rot}}

\def\itide{{\rm T}}
\def\rtidlock{\rr_{\itide}}
\def\Qtide{Q}
\def\tage{\chartime_{{\rm pf}}}

\def\Protor{\period_{\irot ; 0}}
\def\ihz{{\rm HZ}}
\def\rhz{\rr_{\ihz}}
\def\Thz{\temperature_{\ihz}}
 
\def\vel{V}						
\def\isurf{{\rm s}}				
\def\Vvect{\vect{\vel}}			
\def\Vr{\vel_\rr}					

\def\ieq{{\rm eq}}

\def\genvar{\bullet}
\def\iatm{{\rm a}}
\def\iinc{\rm i}

\def\sigmaSB{\sigma_{\rm SB}}

\newcommand{\mean}[1]{\overline{#1}}	
\def\up{\uparrow}					
\def\down{\downarrow}				
\def\wiflux{F}				
\def\Fnet{\wiflux_{-}}					
\def\Fstar{\wiflux_{\star}}				
\def\Fearth{\wiflux_{\Earth}}
\def\optdepth{\tau}					
	
\def\Fup{\wiflux_{\up}}					
\def\Fdown{\wiflux_{\down}}				

\def\Fupsw{\wiflux_{\up \sw}}
\def\Fuplw{\wiflux_{\up \lw}}

\def\Fdownsw{\wiflux_{\down \sw}}
\def\Fdownlw{\wiflux_{\down \lw}}
\def\Fsurf{\wiflux_{\isurf}}				
\def\Fatm{\wiflux_{\iatm}}				

\def\Finc{\wiflux_{\iinc}}				

\def\opacity{\kappa}					
\def\psurf{\pressure_{\isurf}}

\def\lw{{\rm L}}						
\def\sw{{\rm S}}						
\def\klw{\opacity_{\lw}}				
\def\ksw{\opacity_{\sw}}				

\def\betalw{\beta_{\lw 0}}				
\def\betasw{\beta_{\sw 0}}				
\def\Fnetlw{\wiflux_{- \lw}}				
\def\Fnetsw{\wiflux_{- \sw}}			
\def\Ftotlw{\wiflux_{+ \lw}}				
\def\Ftotsw{\wiflux_{+ \sw}}			

\def\Asurfsw{A_{\isurf}}				

\def\Aintlw{\mathscr{A}_{\lw}}
\def\Bintlw{\mathscr{B}_{\lw}}
\def\Aintsw{\mathscr{A}_{\sw}}
\def\Bintsw{\mathscr{B}_{\sw}}

\def\ulw{\betalw}					
\def\usw{\betasw}					

\def\ftrans{\mathcal{T}}				
\def\ftranslw{\ftrans_{\lw}}				
\def\ftranssw{\ftrans_{\sw}}			
\def\gammbeta{\zeta}				
\def\gamswplus{\gammbeta_{\sw}^{+}}	
\def\gamswminus{\gammbeta_{\sw}^{-}}	
\def\gamswpm{\gammbeta_{\sw}^{\pm}}
\def\gamlwpm{\gammbeta_{\lw}^{\pm}}
\def\gamlwplus{\gammbeta_{\lw}^{+}}	
\def\gamlwminus{\gammbeta_{\lw}^{-}}	

\def\optdepthgen{\optdepth_{\genvar}}	
\def\optdepthlw{\optdepth_{\lw}}		
\def\optdepthsw{\optdepth_{\sw}}		

\def\temperature{T}					
\def\temp{\temperature}
\def\Abudget{\mathcal{C}}				
\def\Kbudget{\mathcal{K}}				
\def\Agrbudget{\mathcal{A}}			
\def\Alw{\Abudget_{\lw}}				
\def\Asw{\Abudget_{\sw}}				
\def\Klw{\Kbudget_{\lw}}				
\def\Ksw{\Kbudget_{\sw}}				
\def\Agrlw{\Agrbudget_{\lw}}			
\def\Agrsw{\Agrbudget_{\sw}}			

\def\iconv{{\rm sen}}
\def\iadv{{\rm adv}}
\def\irad{{\rm rad}}
\def\Vadv{\vel_{\iadv}}
\def\Vconv{\vel_{\iconv}}
\def\Dynbudget{\wiflux}			
\def\Dconv{\Dynbudget_{\iconv}}
\def\Dadv{\Dynbudget_{\iadv}}
\def\meanDyn{\mean{\Dynbudget}}
\def\meanDconv{\meanDyn_{\iconv}}
\def\meanDadv{\meanDyn_{\iadv}}
\def\unitsurf{S}						
\def\density{\rho}					
\def\iday{{\rm d}}					
\def\inight{{\rm n}}					
\def\Cd{C_{\rm D}}					
\def\Tatm{\temperature_{\iatm}}			
\def\Tsurf{\temperature_{\isurf}}			
\def\Tday{\temperature_{\isurf ; \iday}}			
\def\Tnight{\temperature_{\isurf ; \inight}}		
\def\Taday{\temperature_{\iatm ; \iday}}
\def\Tanight{\temperature_{\iatm ; \inight}}
\def\Trad{\temperature_{\irad}}
\def\Tskin{\temperature_{\rm skin}}

\def\Cp{C_{\pressure}}				
\def\rhoatm{\density_{\iatm}}			
\def\rhoaday{\density_{\iatm ; \iday}}
\def\absV{\abs{\Vvect}}				
\def\meantemp{\mean{\temperature}}	
\def\Bhemis{B}						
\def\Bday{\Bhemis_{\isurf ; \iday}}	 	
\def\Bnight{\Bhemis_{\isurf ; \inight}}		
\def\Baday{\Bhemis_{\iatm ; \iday}}	 	
\def\Banight{\Bhemis_{\iatm ; \inight}}	
\def\Batm{\Bhemis_{\iatm}}			

\newcommand{\perturb}[1]{#1^{\prime}}
\def\zenithphi{\phi}					

\def\deltaVr{\perturb{\Vr}}
\def\deltatemp{\perturb{\temp}}
\def\meanV{\mean{\vel}}

\newcommand{\fricquant}[1]{#1_{\ast}}
\def\Vfric{\fricquant{\vel}}
\def\tempfric{\fricquant{\temp}}

\def\karman{\mathcal{K}}
\def\mixingL{\ell}
\def\roughheight{\zz_0}
\def\mixingLmax{\mixingL_{0}}
\def\zsurf{\zz_{\rm h}}

\def\Rspec{R_{\rm s}}
\def\Rgp{R_{\rm GP}}

\def\Lnorm{L}
\def\Lconv{\Lnorm_{\iconv}}
\def\Ladv{\Lnorm_{\iadv}}
\def\Fnorm{\normal{\wiflux}_{\istar}}

\def\Tnorm{\normal{\temp}_{\iday}}
\def\Tanorm{\normal{\temp}_{\iatm}}
\def\Tanorminf{\normal{\temp}_{\iatm ; {\rm inf}}}
\def\Dconvnorm{\normal{\Dynbudget}_{\iconv}}
\def\Dadvnorm{\normal{\Dynbudget}_{\iadv}}
\def\DTanorm{\Delta \Tanorm}
\def\fTanorm{\function}
\def\fTanormlog{\hat{\fTanorm}}

\def\massemol{\mathcal{M}}
\def\icond{{\rm cond}}
\def\carbondiox{{\rm CO_2}}
\def\nitrogen{{\rm N_2}}
\def\water{{\rm H_2O}}
\def\monoxcarb{{\rm CO}}
\def\Tcondcarbdiox{\temp_{\icond , \carbondiox}}
\def\Tcondcarbdioxc{\temp_{\icond , \carbondiox}^{\rm 1mbar}}
\def\Mmolatm{\massemol_{\iatm}}

\def\pressvelocity{\omega}	
\def\iup{{\rm up}}			
\def\isub{{\rm sub}}			
\def\area{A}				
\def\Aup{\area_{\iup}}		
\def\Asub{\area_{\isub}}		

\def\pvup{\pressvelocity_{\iup}}			
\def\pvsub{\pressvelocity_{\isub}}		

\def\Ddiff{\mathcal{D}}				

\def\tautoplw{\optdepth_{\rm top ; \lw}}
\def\nexptau{n}						
\def\betagas{\beta}					
\def\tadv{\chartime_{\rm adv}}			
\def\tupd{\chartime_{\iup}}				
\def\tsub{\chartime_{\isub}}			
\def\trad{\chartime_{\rm rad}}			
\def\tdrag{\chartime_{\rm drag}}

\def\taugrlw{\optdepth_{\isurf ; \lw}}		
\def\taugrsw{\optdepth_{\isurf ; \sw}}		
\def\taugrlwconv{\taugrlw^{\iconv}}
\def\taugrlwrad{\taugrlw^{\irad}}
\def\ieq{{\rm eq}}
\def\Teq{\temp_{\ieq}}				
\def\Feq{\wiflux_{\ieq}} 				
\def\thermalpower{Q}
\def\Qin{\thermalpower_{\rm in}}
\def\Qconv{\Qin}
\def\Qadv{\Qin}

\def\effconv{\varepsilon_{\iconv}}
\def\effadv{\varepsilon_{\iadv}}
\def\effcirc{\varepsilon_{\rm circ}}
\def\work{W}

\def\efftherm{\eta}
\def\effthermconv{\efftherm_{\iconv}}
\def\effthermadv{\efftherm_{\iadv}}
\def\efftot{\varepsilon_{\rm eff}}

\def\Rossbyrad{L_{\rm Ro}}
\def\Rossbyradnorm{\normal{L}_{\rm Ro}}
\def\BVfreq{N}
\def\cwave{c_{\rm wave}}
\def\Hpress{H}
\def\pcrit{\pressure_{\rm crit}}
\def\pcritrad{\pcrit^{\irad}}
\def\pcritconv{\pcrit^{\iconv}}

\newcommand{\eq}[1]{Eq.~(\ref{#1})}
\newcommand{\eqs}[2]{Eqs.~(\ref{#1}) and~(\ref{#2})}
\newcommand{\eqsto}[2]{Eqs.~(\ref{#1}-\ref{#2})}
\newcommand{\eqsthree}[3]{Eqs.~(\ref{#1}), (\ref{#2}), and~(\ref{#3})}
\newcommand{\eqsfour}[4]{Eqs.~(\ref{#1}), (\ref{#2}), (\ref{#3}) and~(\ref{#4})}
\newcommand{\append}[1]{Appendix~\ref{#1}}
\newcommand{\units}[1]{~${\rm #1}$}
\newcommand{\fig}[1]{Fig.~\ref{#1}}
\newcommand{\figs}[2]{Figs.~\ref{#1} and~\ref{#2}}
\newcommand{\sect}[1]{Sect.~\ref{#1}}

\newcommand{\tab}[1]{Table~\ref{#1}}

\definecolor{emerald}{rgb}{0.3,0.85,0.2}
\definecolor{smcolor}{rgb}{0.7,0.3,0.0}
\definecolor{lightseegreen}{rgb}{0.12,0.698,0.67}
\definecolor{darkcyan}{rgb}{0,0.55,0.55}

\newcommand{\kc}[1]{#1}
\newcommand{\rec}[1]{#1}
\newcommand{\recc}[1]{#1}

\npstyleenglish

\begin{document} 
  \title{Atmospheric stability and collapse on tidally locked rocky planets }
  
  \subtitle{}
  
  \author{P. Auclair-Desrotour\inst{1,2} \and K. Heng\inst{1,3}
          }

  \institute{University of Bern, Center for Space and Habitability, Gesellschaftsstrasse 6, CH-3012, Bern, Switzerland 
  	\and ASD/IMCCE, CNRS-UMR 8028, Observatoire de Paris, PSL, UPMC, 77 Avenue Denfert-Rochereau, 75014 Paris, France 
	\and University of Warwick, Department of Physics, Astronomy \& Astrophysics Group, Coventry CV4 7AL, U. K. \\
  \email{pierre.auclair-desrotour@csh.unibe.ch} 
  }
 

  \date{Received ...; accepted ...}

  \abstract
   {Over large timescales, a terrestrial planet may be driven towards spin-orbit synchronous rotation by tidal forces. In this particular configuration, the planet exhibits permanent dayside and nightside, which may induce strong day-night temperature gradients. The nightside temperature depends on the efficiency of the day-night heat redistribution and determines the stability of the atmosphere against collapse.} 
   {To better constrain the atmospheric stability, climate, and surface conditions of rocky planets located in the habitable zone of their host star, it is thus crucial to understand the complex mechanism of heat redistribution. }
   {Building on early works and assuming dry thermodynamics, we developed a hierarchy of analytic models taking into account the coupling between radiative transfer, dayside convection, and large-scale atmospheric circulation in the case of slowly rotating planets. There are two types of these models: a zero-dimensional two-layer approach and a two-column radiative-convective-subsiding-upwelling (RCSU) model. They yield analytical solutions and scaling laws characterising the dependence of the collapse pressure on physical features, which are compared to the results obtained \recc{by early works} using 3D global climate models (GCMs).  }
   {The analytical theory captures (i) the dependence of temperatures on atmospheric opacities and scattering in the shortwave and in the longwave, (ii) the behaviour of the collapse pressure observed in GCM simulations at low stellar fluxes that are due to the non-linear dependence of the atmospheric opacity on the longwave optical depth at the planet's surface, (iii) the increase of stability generated by dayside sensible heating, and (iv) the decrease of stability induced by the increase of the planet size.   }
   {}
   {}



\maketitle


\section{Introduction}

Both observations and theoretical studies achieved over the last decade suggest that rocky planets can be found around stars of different masses and they represent a large fraction of the population of exoplanets \citep[e.g.][]{Cumming2008,Mordasini2009a,Howard2013,Ronco2017}. Many of these planets are located in the habitable zone of their host stars \citep[e.g.][]{Kopparapu2013}, \kc{which} is basically the \rec{region where the incident stellar flux and greenhouse effect enable} them to sustain surface liquid water. This is particularly the case for planets detected around brown \kc{dwarfs} and very-low-mass stars \citep[e.g.][]{PL2007,Raymond2007,Kopparapu2017}. 

\kc{Three out of} seven \kc{of the} Earth-sized rocky planets hosted by the TRAPPIST-1 ultra-cool dwarf star \kc{(\rec{planets e, f, and g}) that were discovered in 2017 are within} the habitable zone \citep[][]{Gillon2017,Grimm2018}. Recently, two temperate Earth-mass planet candidates were detected around the M-dwarf Teegarden's Star by the \textit{CARMENES} spectrograph using radial-velocity measurements \citep[][]{Zechmeister2019}, while an Earth-sized planet orbiting in the habitable zone of the TOI-700 red dwarf was discovered by the \textit{TESS} observatory \citep[][]{Gilbert2020,Rodriguez2020}. This planet \rec{(TOI-700~d)} is the first of its kind detected by \textit{TESS} and it \recc{might} \rec{harbor} temperate surface conditions \citep[][]{Suissa2020}. \rec{The} number of discovered temperate rocky planets will keep growing in the future with the upcoming transit searches of the \textit{TESS} \citep[][]{Barclay2018} and \textit{PLATO} \citep[][]{Ragazzoni2016proc} observatories, which opens a new era of atmospheric characterisation and \kc{motivates a better understanding of} the mechanisms that \kc{govern} planetary climates and surface conditions. 

Most planets located in the habitable zone of sub-Solar-mass stars, and particularly those orbiting very-low-mass \kc{dwarfs} such as TRAPPIST-1, are expected to be circularised and tidally locked in spin-orbit synchronous rotation \citep[e.g.][]{Kasting1993} or in a spin-orbit resonance \citep[e.g.][]{Correia2014} over short timescales. As rocky planets locked into synchronous rotation exhibit permanent dayside and nightside, their climate is affected by a strong gradient of thermal forcing along the axis connecting the sub-stellar and anti-stellar points. 

Since it is continually heated by the stellar incident flux, the dayside is hotter than the nightside, which is cooled by infrared radiation towards space. If the nightside temperature goes below the condensation temperature of greenhouse gases present in the atmosphere, then the nightside becomes a cold trap for these gases, which \kc{condense} and form an ice sheet at \kc{the} planet surface. This triggers atmospheric collapse \citep[e.g.][]{Joshi1997,HK2012,Wordsworth2015}: \kc{the} greenhouse effect \kc{diminishes} and the atmosphere cools down, \kc{thereby} accelerating the condensation process in the cold trap and leading the atmospheric composition, climate, and surface conditions to change radically. It is thus crucial to characterise preliminarily the atmospheric stability of rocky planets against collapse to better constrain the surface conditions of the Earth-like exoplanets found in the habitable zone of their host star. 

 \cite{Kasting1993}, who investigated the habitability conditions around main sequence stars, initially raised the question \kc{of} atmospheric stability. This question was first addressed using 3D global climate models (GCMs), \kc{which are} codes \kc{that solve} the coupled angular momentum, mass conservation, radiative transfer, and energy equations \kc{in three dimensions on the surface of a sphere}. In their pioneering works, \cite{Joshi1997} and \cite{Joshi2003} examine the case of rocky planets orbiting M-Dwarfs and hosting $\carbondiox$/$\water$ atmospheres, while \cite{MS2010} characterise the large-scale circulation patterns of Earth-like planets tidally locked in spin-orbit synchronisation in various regimes. 

A series of studies continued on this path, such as \cite{Heng2011a} and \cite{HV2011}, who characterise the atmospheric dynamics of a hypothetical tidally locked Earth and treat the case of the Gliese~581g super-Earth as a scaled-up version of Earth. By considering the effects of moist thermodynamics, \cite{Leconte2013} study the circulation patterns of close-in extrasolar terrestrial planets in \kc{the} presence of clouds. They thus show evidence of the climate moist bistability, \rec{which} is the fact that water can be either vaporised -- which acts to increase the runaway greenhouse effect -- or captured in permanent cold traps, \rec{leading} the atmosphere to collapse. \cite{LM2016} examine how surface friction affects the atmospheric stability, while \cite{Turbet2018} computed stability diagrams for TRAPPIST-1 planets assuming $\nitrogen$/$\carbondiox$ atmospheric mixtures. Similarly, \cite{Wordsworth2015} and \cite{KA2016} treat the case of $\monoxcarb$, $\carbondiox$, and $\nitrogen$-dominated dry atmospheres, and \cite{KW2019} investigate the effect of shortwave absorption on collapse. 

Meanwhile, with the rise of next-generation space telescopes, such as the \textit{James Webb Space Telescope} \citep[\textit{JWST};][]{Deming2009}, 
 the mechanism of global heat redistribution itself was examined in order to decipher the upcoming thermal phase curves of tidally locked extrasolar rocky planets and hot Jupiters \citep[e.g.][]{SD2009,CA2011,Selsis2011,PBS2013,KA2015,KK2018,Koll2019}. Although their \kc{goals are} slightly different, these works are complementary with those dealing with atmospheric collapse. 
 
 While they are convenient to treat the complex non-linear physics governing planetary climates, GCMs are far too costly to explore a broad range of the parameter space, \kc{with} any single climate simulation requiring days or weeks of CPU time. Furthermore, the asymptotic regimes of the steady states reached after convergence may actually be characterised using a relatively small set of nondimensional control parameters \citep[e.g.][]{KA2015}.
 This motivated the development of simplified analytic theories reducing both the physics complexity and the dimensionality \rec{of the problem}. 
 
 First, 1D analytic models were introduced to characterise the thermal structure of planetary atmospheres. \kc{\cite{Guillot2010} derived temperature-pressure ($T-P$) profiles in the limit of the 'double-gray' approximation and pure absorption. \cite{RC2012} considered $T-P$ profiles in the presence of convection. \cite{Heng2012} generalised the work of \cite{Guillot2010} to include isotropic scattering. \cite{PG2014} used the 'picket fence' model of \cite{Chandrasekhar1935} and \cite{Mihalas1978} to generalise the work of \cite{Guillot2010} by including the simplified non-grey radiative transfer due to spectral lines versus continua. \cite{Heng2014} generalised the work of \cite{Heng2012} to include non-isotropic/anisotropic scattering, and also demonstrated that the governing equations of these $T-P$ profiles and the two-stream solutions have a common origin. \cite{Mohandas2018} generalised the work of \cite{PG2014} to include isotropic scattering.}
 
 
 Second, the analytic theory of global heat redistribution was developed incrementally through a diversity of approaches starting from analytical scalings of the atmospheric stability of super-Earths orbiting M-dwarf stars \citep[e.g.][]{HK2012}. Laying the foundation of the 0-D theory, \cite{Wordsworth2015} proposed a two-layer model that self-consistently includes both radiative transfer and sensible heat exchanges with planetary surface due to convection within the dayside planetary boundary layer (PBL).  In this box model, the atmosphere is treated as a globally isothermal and optically thin layer, and the complex features of the three-dimensional general circulation are ignored. Despite these simplifications, the model approximately matches results obtained from GCM simulations for $\carbondiox$-dominated planets. Particularly, it convincingly captures the behaviour of the collapse (or critical) pressure $\pcrit$ -- that is the minimum surface pressure for atmospheric stability -- at high stellar fluxes.
  
Comparable \kc{progress was} achieved by \cite{KA2016} regarding the 1D theory. Building on the early work by \cite{RC2012}, these authors developed a two-column radiative-convective model that takes the effect of subsidence (i.e. downwelling flows) on the nightside temperature profile into account. This model is thus referred to as a radiative-convective-subsiding (RCS) model in the following. By performing a series of simulations with a GCM, \cite{KA2016} showed that the large-scale atmospheric circulation acts as a global heat engine \citep[a similar work was done by][in the case of hot Jupiters]{KK2018}, and used this statement to derive scalings for the typical wind speeds parametrising heat fluxes. They thus recovered the stability diagrams of $\carbondiox$-dominated atmospheres previously obtained by \cite{Wordsworth2015} from GCM simulations. 
 
In the present work, we aim to consolidate the analytic theory by building mainly on the developments made by \cite{Wordsworth2015} and \cite{KA2016}. Particularly, our goal is to elucidate analytically how the collapse pressure is affected by the non-linear pressure dependence of the atmospheric optical thickness at low stellar fluxes, by dayside convection, and by large-scale advection. Therefore, we adopt the 0-D approach proposed by \cite{Wordsworth2015} and introduce incrementally in the model the effects of longwave and shortwave absorption and scattering, turbulent heat exchanges with the planet surface -- \kc{which} we call 'sensible heating' -- and advective heat transport by a stellar and anti-stellar cell, which is the regime of slowly rotating planets. Following \cite{KA2016}, scalings of wind speeds are derived in the framework of heat engine theory. \kc{With every step}, intermediate results are compared with those obtained in \kc{earlier} studies. The full model -- which encompasses radiative transfer, dayside convection, and large-scale circulation -- is formally written as one single equation, given by \eq{eqsingle_gen}, and controlled by a small set of dimensionless parameters. This model is used to compute stability diagrams, and finally compared to a 1D two-column radiative-convective-subsiding-upwelling (RCSU) model that we develop by introducing dayside upwelling flows in the RCS model of \cite{KA2016}.

In \sect{sec:physical_setup}, we detail the physical setup of the 0-D model and discuss the main assumptions. In \sect{sec:radiative_model}, we develop a two-layer radiative model of global heat redistribution including absorption and scattering in the two-stream, dual-band, and grey gas approximations. We then derive analytically the shortwave and longwave \kc{flux} profiles, radiative transmission functions, and equilibrium temperatures. In \sect{sec:sensible_heating}, the dayside sensible heating is included in the model. In \sect{sec:advection}, we relax Wordsworth's globally isothermal atmosphere assumption and make the day-night temperature gradient depend on the heat transport \kc{due to} stellar and anti-stellar circulation. In \sect{sec:atmospheric_stability}, the model is used to compute stability diagrams for $\carbondiox$-dominated atmospheres derived from the case treated by \cite{Wordsworth2015}. We thus show that the model captures the evolution of the collapse pressure at low stellar fluxes observed in GCM simulations, and the fact that the atmospheric stability increases with sensible heating and decreases as the planet's size increases. In \sect{sec:RCSU_model}, we introduce our two-column RCSU model and discuss the role played by the atmospheric structure and, finally in \sect{sec:conclusions} we summarise the conclusions and future works.


\section{Physical setup and main assumptions}
\label{sec:physical_setup}

\begin{figure*}[htb]
   \centering
   \includegraphics[width=0.8\textwidth,trim = 0.0cm 2cm 0cm 1cm,clip]{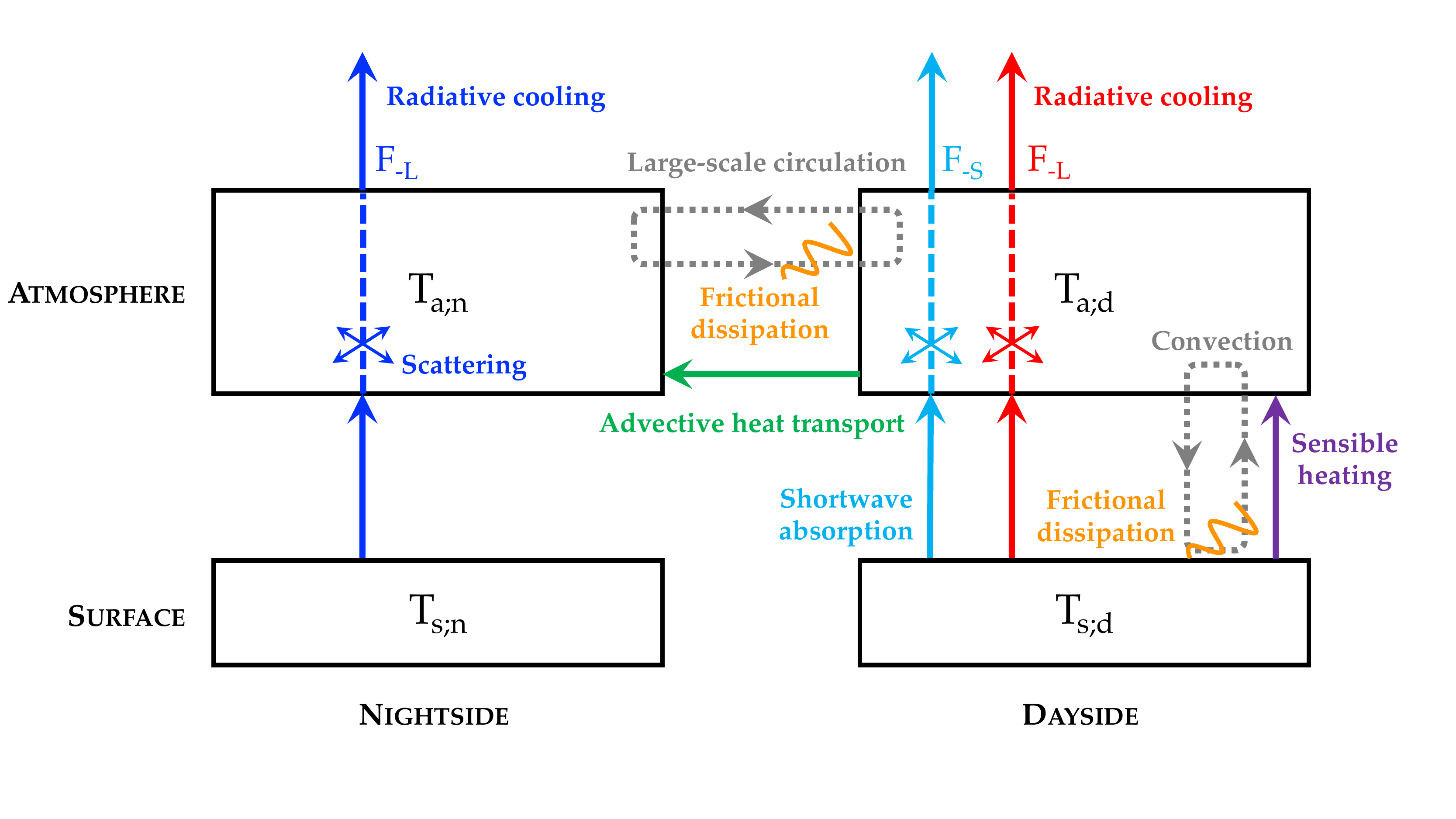}
      \caption{Diagram of the two-layer radiative-convective-advective model detailed in the present study. The diagram shows the processes involved in dayside-nightside heat redistribution that are taken into account in the model. Radiative net fluxes ($\Fnet \define \Fup - \Fdown$) are upwards in the convention \rec{used}, meaning that the shortwave net flux $\Fnetsw $ is negative.}
       \label{fig:model_overview}%
\end{figure*}

We introduce in this section the main features of the 0-D model developed in the study as well as frequently used notations. This provides a global overview of the mechanisms involved in day-night heat redistribution of terrestrial planets that we take into account (\fig{fig:model_overview}). These mechanisms are detailed in the next sections. 

\subsection{Tidal locking in $1{:}1$ spin-orbit resonance}
\label{ssec:tidal_locking}

We consider the simplified case of a tidally-locked rocky planet, of mass $\Mpla$ and radius $\Rpla$, synchronously moving on a coplanar and circular orbit of period $\Porb$. This regime is mainly relevant for planets orbiting small main-sequence stars, such as K and M \kc{dwarfs}, since the habitable zone of the host star is within the tidal lock radius in this case \citep[][]{Kasting1993,Edson2011}. This result may be derived from a simple scale analysis. 

First, the tidal lock radius is usually estimated for dry planets from the formula \citep[][]{Peale1977,Kasting1993,Dobrovolskis2009,Edson2011},

\begin{equation}
\rtidlock \approx 0.024 \left( \frac{\Protor \tage}{\Qtide} \right)^{\frac{1}{6}} \Mstar^{1/3}.
\label{rtidlock}
\end{equation}

\noindent Here, $\Mstar$ is the stellar mass, $\Protor$ the original rotation period of the planet, $\tage$ the time period from formation, and $\Qtide$ the tidal quality factor accounting for tidal dissipation in the planet's interior \citep[the smaller $\Qtide$ and the larger tidally dissipated energy;][]{GS1966}. All quantities are in CGS units (centimetres, grams, seconds) in the formula. 

Second, the typical radius of the habitable zone $\rhz$ can be roughly defined as the star-planet distance at which the black body equilibrium temperature of the planet has a given value $\Thz$. The equilibrium temperature is given as a function of the stellar flux received by the planet $\Fstar$, 

\begin{equation}
\Teq \define \left( \frac{\Fstar}{4 \sigmaSB} \right)^{\frac{1}{4}},
\label{Teq}
\end{equation}

\noindent the parameter $\sigmaSB = 5.670367 \times 10^{-8} $\units{W~m^{-2}~K^{-4}} \citep[][]{codata2014} being the Stefan-Boltzmann constant, \kc{and the symbol $\define$ meaning 'defined by'}. The above definition of $\rhz$ thus leads to $\Fstar = 4 \sigmaSB \Thz^4$. The stellar flux is also expressed as a function of the orbital radius $\smaxis$ and the stellar bolometric luminosity $\Lstar$, 

\begin{equation}
\Fstar = \frac{\Lstar}{4 \pi \smaxis^2}. 
\label{Fstar}
\end{equation}

\noindent By combining the two expressions of $\Fstar$, it follows that 

\begin{equation}
\rhz = \sqrt{\frac{\Lstar}{16 \pi \sigmaSB \Thz^4}} \scale \Lstar^{1/2} . 
\end{equation}

The dependence of the stellar bolometric luminosity of main-sequence stars on the stellar mass has been quantified by empirical formulae, such as \citep[][]{Barnes2008}

\begin{equation}
\Lstar = \Lsun 10^{4.101 \mu^3 + 8.172 \mu^2 + 7.108 \mu},
\label{Lstar_Barnes2008}
\end{equation}

\noindent with $\mu \define \logdix \left( \Mstar/\Msun \right) $, the notations $\Msun$ and $\Lsun$ referring to the Solar mass and luminosity, respectively. A linear regression of $\logdix \left(\sqrt{ \Lstar / \Lsun} \right)$ on the range $-1 \leq \mu \leq 0$ yields

\begin{equation}
\logdix \left( \sqrt{ \frac{\Lstar}{\Lsun}} \right) \approx 1.3196 \mu - 0.2662,
\end{equation}

\noindent and thus $\rhz \scale \Mstar^{1.32}$. By comparing this scaling law to $\rtidlock \scale \Mstar^{1/3}$ (\eq{rtidlock}), we observe that the tidal lock radius ($\rtidlock$) increases slower with the stellar mass than the characteristic radius of the habitable zone ($\rhz$). This means that the probability to be tidally locked in synchronous rotation for a planet orbiting in the habitable zone decays as the stellar mass increases, which is  illustrated for instance by Kasting's plot \citep[][Fig.~16]{Kasting1993}.

In the circular coplanar regime, there is no seasonal variation of the local incident stellar flux because of the absence of obliquity and eccentricity. The stellar forcing at the top of the planet's atmosphere is thus invariant with time. The spin rotation of the planet is defined by the spin angular velocity $\spinrate = \norb$, where $\norb \define 2 \pi / \Porb$ is the orbital frequency. Hence, $\spinrate$ is determined using the third Keplerian law,

\begin{equation}
\spinrate = \sqrt{\frac{\Ggrav \left( \Mstar + \Mpla \right)}{\smaxis^3}},
\end{equation}

\noindent where $\Ggrav$ designates the universal gravitational constant. 

Besides, considering both the expression of the stellar flux received by the planet, given by \eq{Fstar}, and the fact that the stellar bolometric luminosity $\Lstar$ of main-sequence stars may be written as a function of the stellar mass (see \eq{Lstar_Barnes2008}), we remark that the planet's angular velocity is actually fully constrained by the stellar mass and flux through the relationship

\begin{equation}
\spinrate = \sqrt{\Ggrav \left( \Mstar + \Mpla \right)} \left[ \frac{4 \pi \Fstar}{\Lstar \left( \Mstar \right) } \right]^{3/4},
\end{equation}

\noindent and thus cannot be taken as a free parameter. Particularly, in simulations that we perform with general circulation models to benchmark the theory, the planet's rotation rate is specified as a function of $\Mpla$, $\Mstar$, and $\Fstar$ using the preceding expression with the scaling law of $\Lstar$ given by \eq{Lstar_Barnes2008}.

\subsection{\kc{Radiative transfer}}
\label{ssec:radiative_transfers}

  The planet's atmosphere is considered in the standard shallow \kc{atmosphere} framework \citep[][]{Vallis2006}\footnote{The term 'shallow' is used here in the context of radiative \kc{transfer}, where it just refers to the ratio between the vertical and horizontal scales of the fluid layer. Its meaning is more subtle in the context of fluid dynamics since it may designate different approximations in this case \citep[see e.g.][Chapter~3]{Vallis2006}.}, where the thickness of the fluid layer is assumed to be far smaller than horizontal scales, comparable to the planet's radius in order of magnitude. In this case, the horizontal propagation of radiation may be neglected and \rec{radiation only propagates} upwards and downwards. Upwelling fluxes are denoted by $\Fup$ and downwelling fluxes by $\Fdown$. This sets the bases of the two-stream approximation \citep[e.g.][Sect.~3.1]{Heng2017}, which is assumed to derive radiative net fluxes $\Fnet \define \Fup - \Fdown$ in the next section. 

In addition to the two-stream approximation, we consider that the frequency spectra of the stellar and planetary radiative fluxes do not overlap, \rec{which} is the so-called dual-band approximation \citep[][Sect.~4.1]{Heng2017}. The radiative fluxes emitted by the star and the planet are thus split into two decoupled components, which are refereed to as the 'shortwave' and 'longwave' fluxes, and subscripted by $\sw$ and $\lw$, respectively. While this simplification holds for rocky planets hosted by Sun-like stars, where \kc{the} bodies\kc{'} surface temperatures are separated by more than one decade in orders of magnitude, it tends to be less adapted to those hosted by ultra-cool dwarf stars, such as TRAPPIST-1, \kc{although the stellar temperature ($\approx 3000$~K) is still far higher than temperate planetary atmospheres ($\approx 300-500$~K)}. In these cases, the stellar and planetary radiative fluxes \kc{partly} overlap. \kc{Cool stars} are also \kc{those that are more} likely to host tidally-locked exoplanets in their habitable zone than Sun-like stars, as stated in \sect{ssec:tidal_locking}. We should thus bear in mind the limitations of the dual-band approximation when applying the model to such systems. 

Finally, grey gas opacities are assumed, meaning  that atmospheric opacities in the longwave and shortwave are wavelength-independent in the model \citep[][Sect.~4.1]{Heng2017}. In each band, absorption is described by a unique parameter that we call 'effective opacity' to emphasise the fact that it does not really correspond to a mean opacity owing to the existing correlation between radiative spectra and opacity lines \citep[see e.g.][Fig.~10]{Wordsworth2015}. Because of this correlation and of the strong wavelength dependence of opacity lines, this hypothesis appears as one of the strongest regarding radiative transfers. Particularly, as stated by \cite{Leconte2013} from the comparison of GCM simulations using different approaches, the cooling of the planet's surface is considerably underestimated when grey opacities are used, meaning that the nightside temperature is overestimated. 

This is what motivated the strong effort made to include refined treatments of \kc{radiative transfer} \citep[e.g. the k-correlated distribution method;][]{LO1991} in \kc{GCMs} used to study the heat redistribution on tidally-locked exoplanets \citep[e.g.][]{Leconte2013,Wordsworth2015}. In the present work however, we have to admit the grey gas assumption as the \kc{price} to pay for the simplification of the theoretical analysis. For exhaustive discussions of its limitations, as well as those of the shallow-water, two-stream, and dual-band approximations, one may consult the reference books by \cite{Seager2010}, \cite{Pierrehumbert2010}, and \cite{Heng2017}. We note that, in analytic developments, we follow the conventions \rec{and} notations employed by \cite{Heng2017} in Chapters~3 and~4.

As shown by \cite{KA2016} through the development of a two-column 1D model, the self-consistent treatment of the coupling between radiative transfer and the atmospheric structure is a major source of complexity, mainly because of the degeneracies affecting boundary conditions. As our goal is to introduce simplified atmospheric dynamics in the analytical theory, we make the choice to favour the atmospheric circulation over the atmospheric structure, noting that the atmospheric dynamics already induces non-negligible mathematical complications. Hence, following the early work by \cite{Wordsworth2015}, we opt for a zero-dimensional two-layer model where the atmosphere is considered as isothermal across the vertical direction. However, we relax the well-mixed atmosphere approximation made by \cite{Wordsworth2015}, and assume that dayside and nightside atmospheric temperatures, denoted by $\Taday$ and $\Tanight$, are not the same in the general case. Similarly, the planet surface is characterised by its dayside and nightside temperatures, $\Tday$ and $\Tnight$ (\fig{fig:model_overview}).

\subsection{Atmospheric heat transport}
\label{ssec:heat_transport_circulation}

The four temperatures of the system are coupled together by the mechanisms of heat transport taken into account in the model, which are of three types: (i) radiative transfers along the vertical direction (shortwave and longwave absorption, radiation, and scattering); (ii) surface-atmosphere turbulent exchanges in the dayside convective boundary layer; and (iii) day-night horizontal \rec{heat transport}. Each of these components may defined separately. Owing to the above simplifications, radiative transfers are described by simple analytic solutions, which are derived in \sect{sec:radiative_model}. For the sensible heating due to turbulent heat transport, among the different prescriptions existing in literature, we choose to follow that proposed by \cite{KA2016}, which treats the atmosphere as a heat engine where the convective flow acts against friction in the surface boundary layer. This \kc{part} of the theory is detailed in \sect{sec:sensible_heating}. 

The large-scale \rec{day-night heat transport} turns out to be the most complex effect to include since there is no general theory of atmospheric heat flux to our knowledge. This is due to the fact that the mechanisms contributing to the heat transport cannot be decoupled and are spanning over the three dimensions at planetary scales. Nevertheless, as demonstrated by \cite{KA2015}, the involved processes may be disentangled by making use of the Buckingham-Pi theorem \citep[][]{Buckingham1914}, which unravels the nondimensional parameters governing the system. 

Among these parameters, we find the nondimensional Rossby deformation length, 

\begin{equation}
\Rossbyradnorm \define \frac{\Rossbyrad}{\Rpla} = \sqrt{\frac{\cwave}{2 \spinrate \Rpla}},
\label{Ronorm}
\end{equation}

\noindent the length $\Rossbyrad$ being the Rossby radius of deformation \citep[see e.g.][Sect.~3.8.2]{Vallis2006}, and $\cwave$ the characteristic speed of gravity waves, defined hereafter. The nondimensional Rossby deformation length governs the circulation regime of the atmosphere. In the case of slow rotators, $\Rossbyradnorm \gtrsim 1$, meaning that waves can propagate planetwide. Because the effect of rotation is weak, the circulation that develops in this regime is approximately symmetric with respect to the axis connecting the substellar and anti-stellar points, the day-night flow being driven by the balance between advection and pressure-gradient acceleration \citep[e.g.][]{Leconte2013}. In this stellar and anti-stellar circulation, \kc{high}-altitude winds blow from the dayside to the nightside and low-altitude wind from the nightside to the dayside \citep[e.g.][]{MS2010,Leconte2013}. 

\begin{figure*}[htb]
   \centering 
\includegraphics[height=0.32\textwidth,trim = 0.0cm 0.4cm 0.0cm 0.4cm,clip]{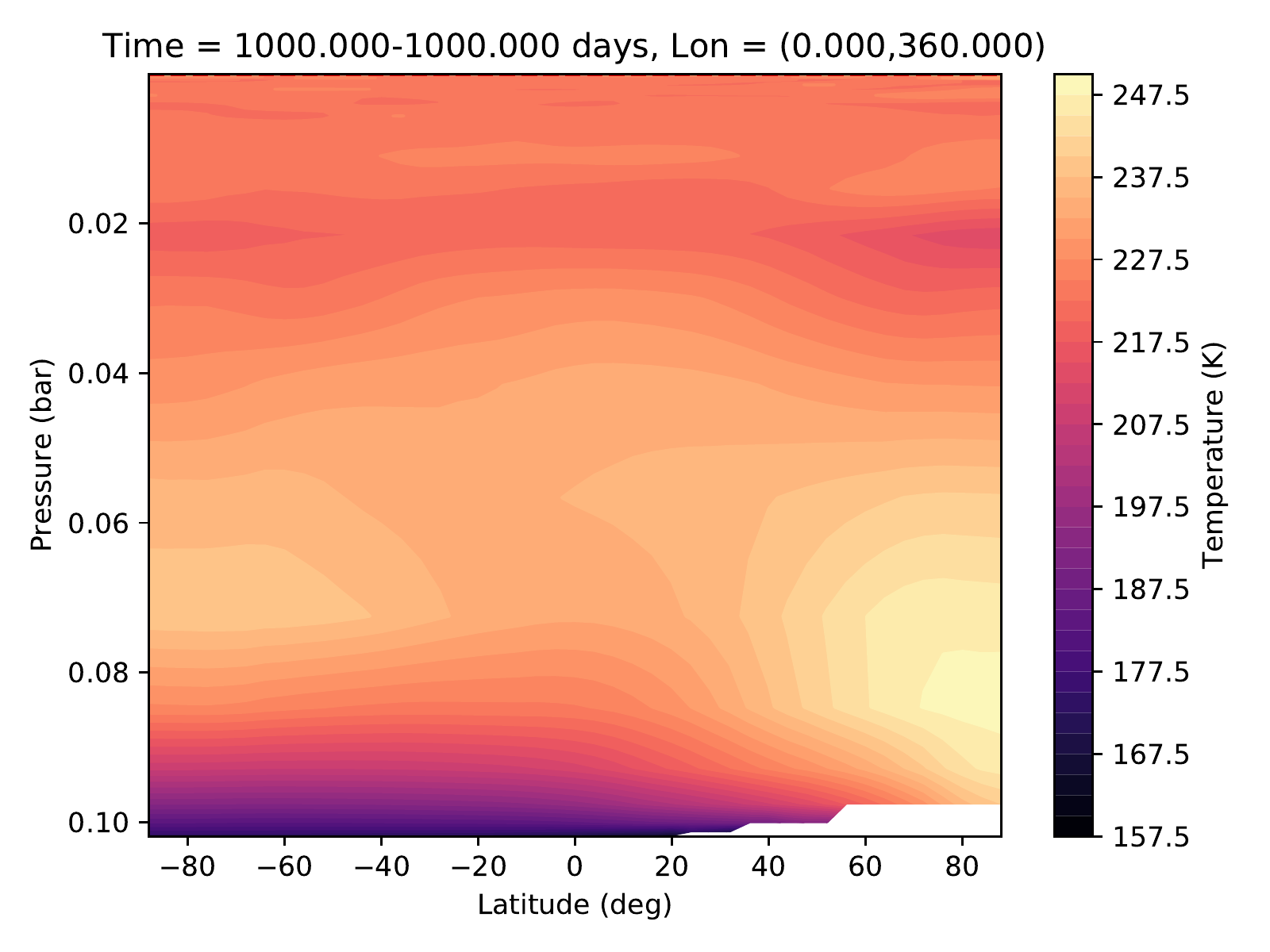}
\includegraphics[height=0.32\textwidth,trim = 0.0cm 0.4cm 0.0cm 0.4cm,clip]{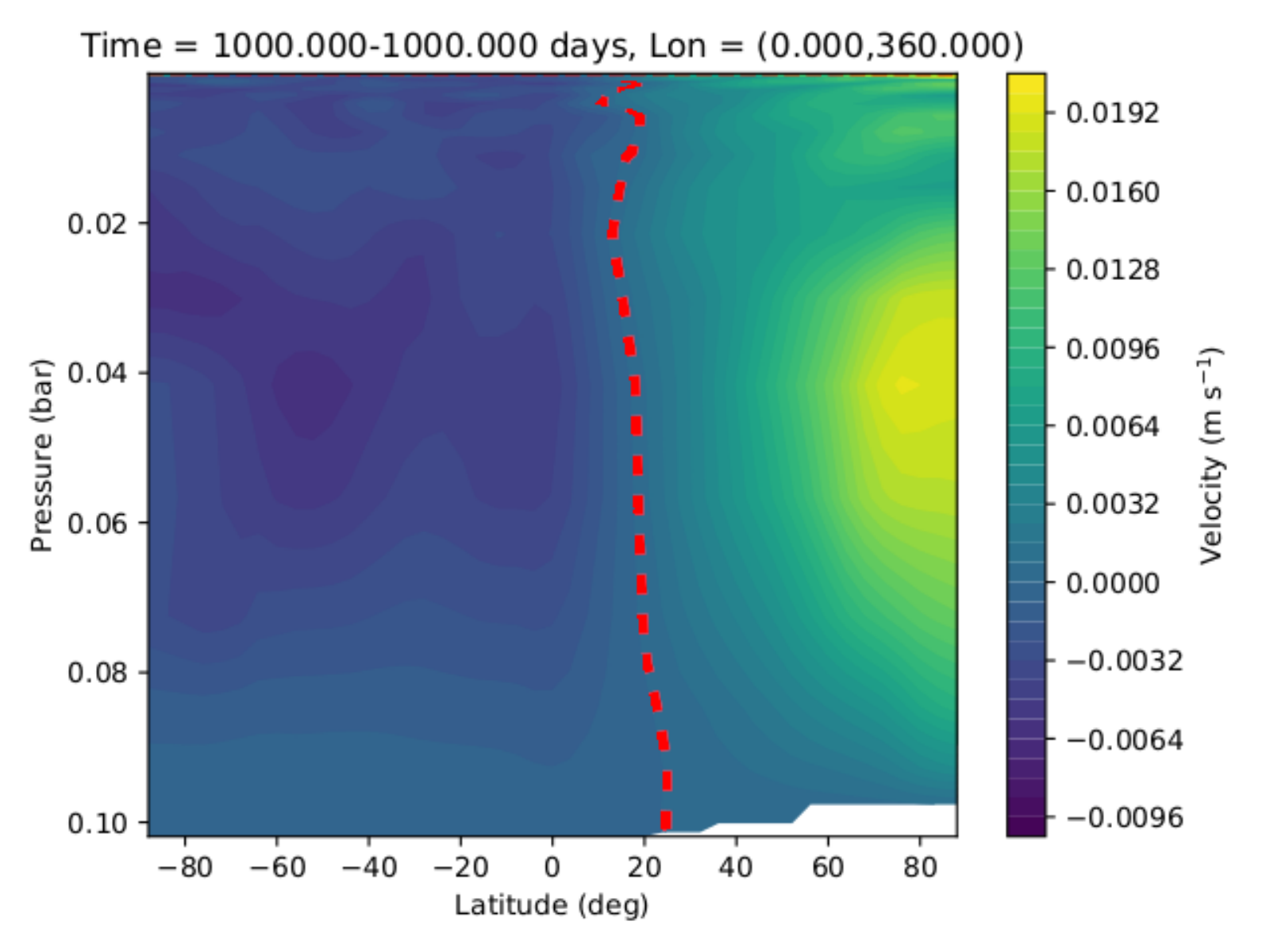} \\
\includegraphics[height=0.34\textwidth,trim = 0.0cm 0.0cm 0.0cm 0.0cm,clip]{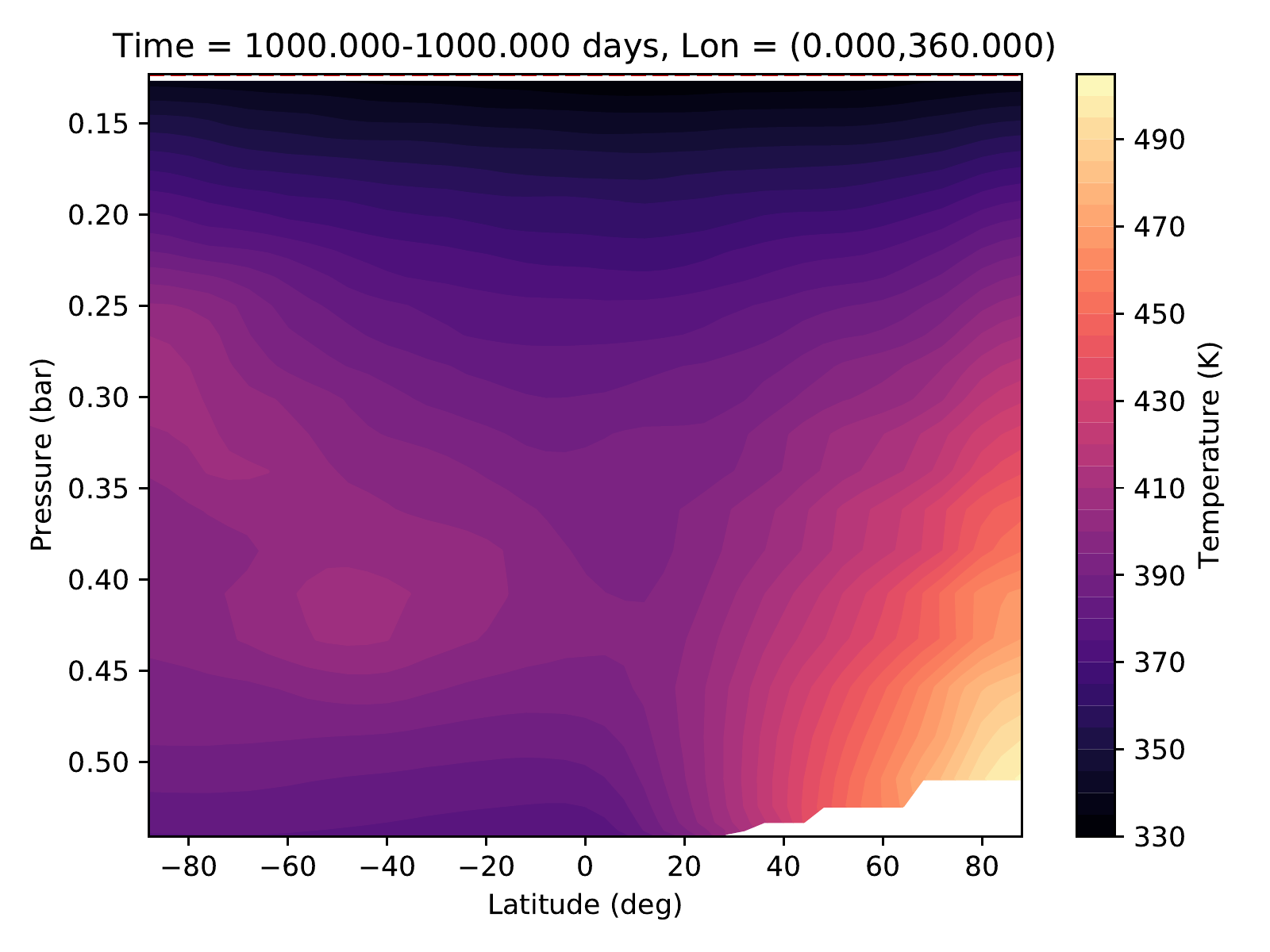}
\includegraphics[height=0.34\textwidth,trim = 0.0cm 0.0cm 0.0cm 0.0cm,clip]{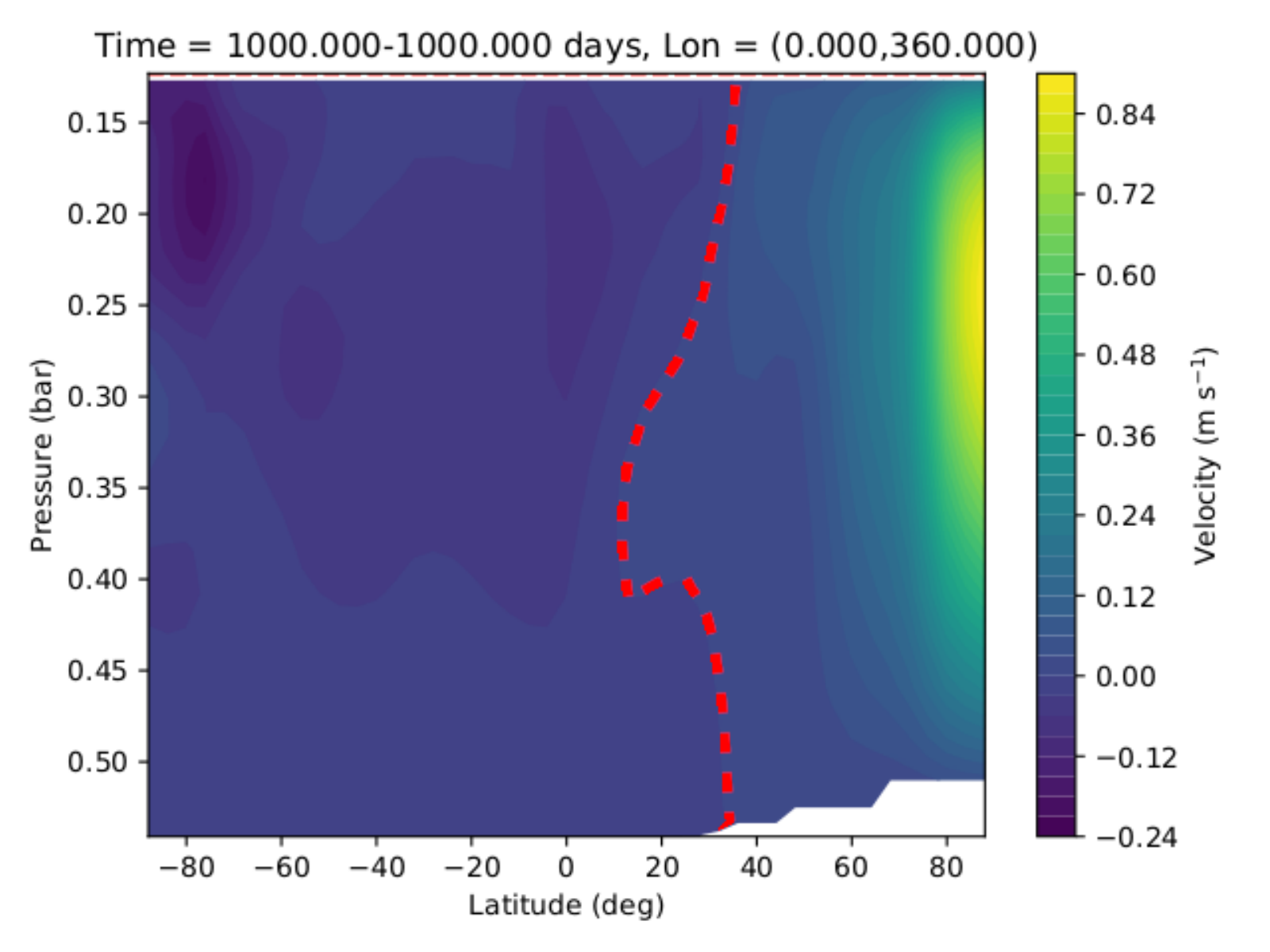} 
   \caption{\rec{Instantaneous snapshots of} zonally-averaged temperature (left panels) and vertical wind speed distributions (right panels) as functions of latitude and pressure with north pole at substellar point. {\it Top:} 0.1-bar $\carbondiox$-dominated atmosphere at $\Teq = 279$~K \citep[][Fig.~2]{Wordsworth2015}. {\it Bottom:} 0.5~bar $\nitrogen$-dominated atmosphere at $\Teq = 400$~K \citep[][Fig.~4]{KA2016}. Latitude is given in degrees (horizontal axis). \rec{This coordinate does not correspond to the usual latitude but to a latitude coordinate in the system of spherical coordinates where the north pole is the substellar point and the south pole the anti-stellar point. The associated colatitude is the stellar zenithal angle in this system of coordinates. Pressure is given} in bars (vertical axis), temperature in K, and vertical velocity in \units{m~s^{-1}}. The white area at the substellar point indicates the region where surface pressure is lower than the spherically averaged surface pressure due to the updraft. \rec{The dotted red line indicates the transition between rising and subsiding flows, where the vertical velocity is zero.}}  
       \label{fig:W2015fig2_KAfig4_thor}%
\end{figure*}

We illustrate the slow rotators regime by performing simulations with the Open-source General Circulation Model \texttt{THOR} \citep[][]{Mendonca2016,Deitrick2020}, which solves the general non-hydrostatic equations within a spherical shell using the icosahedral grid. In these calculations, simple grey gas radiative transfers are used and two cases treated by early works are reproduced: a 0.1-bar $\carbondiox$-dominated atmosphere at $\Teq  = 279$~K \citep[][Fig.~2]{Wordsworth2015}, and a 0.5-bar $\nitrogen$-dominated atmosphere at $\Teq = 400$~K \citep[][Fig.~4]{KA2016}. Figure~\ref{fig:W2015fig2_KAfig4_thor} shows \rec{instantaneous snapshots of} the zonally-averaged temperature and vertical wind speeds in both cases. These quantities are plotted as functions of latitude and pressure in the reference frame where the north pole is located at the substellar point and the south pole at the anti-stellar point. \rec{We emphasise that this latitude coordinate is different from the usual latitude, which is defined from the planet axis of rotation.} As we observe that a steady cycle has been reached in both simulations after $1000$~days, snapshots are taken at this date. The plots particularly highlight the day-night temperature gradient (left panels) and the dayside convective cell (right panels), which is characterised by strong upward flows in the substellar region.

When the Rossby deformation radius becomes less \kc{than} the planet radius ($\Rossbyradnorm <1$), the circulation regime changes and the atmospheric dynamics starts developing super-rotation, that is strong eastward equatorial jets \citep[][]{SG2002,SP2011,TC2010,HV2011,Heng2011,Leconte2013,Mendonca2019}. As shown by \cite{SP2011}, the emergence of super-rotation is due to the formation of standing, planetary-scale equatorial Rossby and Kelvin waves \citep[i.e. waves restored by the Coriolis acceleration; see e.g.][]{LS1997}, which feed zonal flows by continually pumping angular momentum from the mid-latitudes towards the equator. The latitude width of the equatorial band where super-rotating flows are accelerated through this mechanism scales as the equatorial Rossby radius ($\Rossbyrad$), and thus decays as the planet angular velocity increases. As the Rossby and Kelvin waves responsible for the super-rotation are caused by the latitudinal variations in radiative heating, the strength of equatorial jets increases with the day-night temperature difference \citep[][]{SP2011}. 

The speed of gravity waves that intervenes in the non-dimensional Rossby deformation length depends on the atmospheric structure. Denoting by $\ggravi$ the surface gravity, $\Rgp$ the ideal gas constant, and $\Mmolatm$ the mean molecular weight of the atmosphere, we first introduce the specific gas constant $\Rspec \define \Rgp / \Mmolatm$ and the pressure scale height, given as a function of temperature~$\temperature$ by

\begin{equation}
\Hpress \define \frac{\Rspec \temperature}{\ggravi}. 
\label{Hpress}
\end{equation}

Gravity waves are restored by the Archimedean force associated with the fluid buoyancy \citep[][]{GZ2008}. The strength of this force is quantified by the Brunt-\rec{V\"ais\"al\"a} frequency $\BVfreq$, which, in a dry stably stratified atmosphere, is given by

\begin{equation} 
\BVfreq^2 = \frac{\ggravi}{\temperature} \left( \frac{\ggravi}{\Cp} + \DD{\temperature}{\zz} \right),
\label{BVfreq}
\end{equation}

\noindent where $\Cp$ designates the heat capacity per unit mass of the gas at constant pressure, and $\zz$ the altitude. In terms of $\Hpress$ and $\BVfreq$, the typical speed of gravity waves is simply written as $\cwave  \sim \Hpress \BVfreq$ \citep[e.g.][]{KA2015,Leconte2013}. Thus, substituting these parameters by the above expressions (\eqs{Hpress}{BVfreq}) in \eq{Ronorm} and assuming an isothermal atmosphere yields

\begin{equation}
\Rossbyradnorm = \sqrt{ \frac{\Rspec \temperature^{1/2}}{2 \spinrate \Rpla \Cp^{1/2}}} .
\end{equation}

In the present work, we only consider the heat advected by a day-night stellar and anti-stellar circulation, which corresponds to the asymptotic regime of slowly rotating planets ($\Rossbyradnorm \gg 1$). We do not include other mechanisms of day-night heat transport, such as that due to the propagation of gravity waves, which was already studied in early works \citep[][]{KA2015,KA2016}. Thus, in \sect{sec:advection}, a rough approximation of the day-night heat flow is derived from a scale analysis in the heat engine framework \citep[][]{KA2016} using for wind speeds the prescription given by \cite{KK2018} in the case where circulation is balancing Rayleigh drag. As predicted by dimensional analyses \citep[e.g.][Sect.~3.2.1]{Leconte2013}, this heat flow is scaled by the ratio $\trad / \tadv$ between the radiative and advective timescales \citep[e.g.][]{SG2002}

\begin{equation}
\begin{array}{ll}
 \trad \define \dfrac{\psurf \Cp}{\ggravi \sigmaSB \Teq^3}, & \mbox{and} \ \tadv \define \dfrac{\Rpla}{\Vadv},
\end{array}
\label{trad_tadv}
\end{equation}

\noindent where $\psurf$ designates the atmospheric surface pressure and $\Vadv$ the typical speed of stellar and anti-stellar advective flows. This accounts for the fact that the heat transported by weak flows ($\trad/\tadv \ll 1$) is radiated towards space before being advected to the nightside. Conversely, strong flows ($\trad / \tadv \gg 1$) efficiently transport heat since the advected fluid parcels reach the nightside before being radiatively cooled. \rec{In the present work, the scaling law chosen for the advected heat flux is different from that used to describe thermal exchanges with the surface on the dayside in order to emphasise the distinction between the two involved mechanisms in the general case. These mechanisms could nevertheless be linked to each other by assuming the same scalings for the associated heat fluxes.}

Finally, we adopt dry thermodynamics as a first convenient step to set the basis of the model, ignoring thereby the role played by an ocean or moisture on the day-night heat redistribution. We note however that this role may be non-negligible. \cite{Edson2011} showed for instance that the presence of a slab ocean considerably affects the stability of the planet's atmosphere with respect to collapse, by inducing an additional heat transport by latent heat flux in the atmosphere and by heat diffusion in the ocean itself. Water vapour also leads to the formation of clouds, which affects the albedo of the atmosphere as well as its thermal structure and general circulation \citep[][]{Leconte2013}. While not taken into account in the present work, these aspects may be included in the model in future studies. 

\section{A two-layer grey radiative model}
\label{sec:radiative_model}

\kc{Radiative transfer is} the first building block of the model. We establish analytically in this section the vertical profiles of net radiative fluxes in the short- and longwave for the isothermal atmosphere. These profiles provide the integrated transfer functions that are used to derive the atmospheric and surface power budget equations. 

\subsection{Two-stream analytic solution}

In the two-stream, dual-band, and gray-gas approximations (see \sect{ssec:radiative_transfers}), the longwave and shortwave fluxes decouple and only propagate upwards or downwards. As it is governed by the same set of equations, the shortwave case is readily deduced from the longwave case that we treat hereafter. The longwave total and net fluxes are defined as $\Ftotlw \define \Fuplw + \Fdownlw$ and $\Fnetlw \define \Fuplw - \Fdownlw$, respectively. They are governed by the two-stream Schwarzschild equations, which may be written as the system of first order partial differential equations \citep[][]{Heng2017}

\begin{align}
&  \dd{\Ftotlw}{\optdepthlw} = \frac{1}{\betalw} \Fnetlw, \\
& \dd{\Fnetlw}{\optdepthlw} = \betalw \left( \Ftotlw - 2 \sigmaSB \temp^4 \right),
\end{align}

\noindent where $\dd{}{\XX}$ designates the partial derivative with respect to $\XX$, $\optdepthlw$ the optical depth of the atmosphere in the longwave, and $\betalw$ the scattering parameter. The optical depth characterises the optical thickness of the atmosphere and varies between $0$ (top of the atmosphere) and $\taugrlw$ (planet's surface). The scattering parameter indicates the fraction of absorbed flux with respect to the scattered component and takes its values between $0$ and $1$ \citep[$\betalw = 1$ corresponds to pure absorption and $\betalw = 0$ to pure scattering; see e.g.][Sect.~4.3]{Heng2017}. Eliminating the total flux in the preceding system yields the second order ordinary differential equation  

\begin{equation}
\DDn{\Fnetlw}{\optdepthlw}{2} -  \Fnetlw = - 2 \betalw  \DD{ \left(  \sigmaSB \temp^4 \right)}{\optdepthlw},
\label{rt_eq_order2}
\end{equation}

\noindent the notation $\DD{}{\XX}$ designating the total derivative \rec{of some quantity $\XX$}. 

As the isothermal approximation is assumed, the atmospheric temperature profile is approximated by the constant $\Tatm$. In this particular case, the right-hand member of \eq{rt_eq_order2} vanishes, which leads to the analytic expressions of the total and net fluxes,

\begin{align}
\label{Ftotlw}
\Ftotlw  = & \ \Aintlw \expo{\optdepthlw} + \Bintlw \expo{- \optdepthlw} +  2 \sigmaSB \Tatm^4, \\ 
\Fnetlw = & \ \betalw \left(  \Aintlw \expo{\optdepthlw} - \Bintlw \expo{- \optdepthlw} \right), 
\end{align}

\noindent and of the upwelling and downwelling fluxes, 

\begin{align}
\Fuplw = & \ \Aintlw \gamlwplus \expo{\optdepthlw} + \Bintlw \gamlwminus \expo{- \optdepthlw} + \sigmaSB \Tatm^4, \\
\label{Fdownlw}
\Fdownlw = & \ \Aintlw \gamlwminus \expo{\optdepthlw} + \Bintlw \gamlwplus \expo{- \optdepthlw} +   \sigmaSB \Tatm^4,
\end{align}

\noindent where $\Aintlw$ and $\Bintlw$ are the two integration constants of the solution, and $\gamlwminus$ and $\gamlwplus$ the coupling coefficients defined by 

\begin{equation}
\gamlwpm \define \frac{1}{2} \left( 1 \pm \betalw \right).
\end{equation}

The parameters $\gamlwpm$ characterise the coupling of radiative fluxes due to scattering. In the case of pure absorption ($\betalw = 1$), there is no coupling ($\gamlwminus = 0$ and $\gamlwplus = 1$). Conversely, in the case of pure scattering ($\betalw = 0$), the coupling is strong ($\gamlwminus = \gamlwplus = 1/2$). Similar expressions may be derived for the shortwave band by assuming $\Tatm = 0$ in \eqsto{Ftotlw}{Fdownlw}. The shortwave fluxes ($\Ftotsw$, $\Fnetsw$, $\Fupsw$, and $\Fdownsw$) are parametrised by the shortwave optical depth $\optdepthsw$ (such that $0 \leq \optdepthsw \leq \taugrsw$, $\taugrsw$ being the shortwave optical depth at the planet's surface), scattering parameter $\betasw $, coupling coefficients $\gamswpm \define \left( 1 \pm \betasw \right)/2$, and integration constants $\Aintsw$ and $\Bintsw$.

\begin{figure}[htb]
   \centering
   \includegraphics[width=0.48\textwidth,trim = 1.5cm 3cm 1cm 2cm,clip]{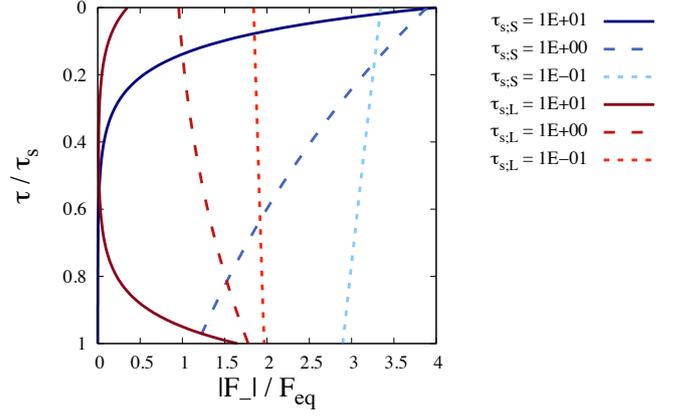}
      \caption{Net radiative fluxes in the short- and longwave (horizontal axis) as functions of optical depths (vertical axis) in case of isothermal atmosphere. Fluxes profiles are computed from the analytic formulae given by \eqs{Fnetsw_isoT}{Fnetlw_isoT} and normalised by the black body equilibrium flux $\Feq \define \sigmaSB \Teq^4$ (the equilibrium temperature $\Teq$ being given by \eq{Teq}). Optical depths are normalised by their value at the planet's surface, namely $\taugrsw$ for the shortwave and $\taugrlw$ for the longwave. Shortwave (blue lines) and longwave (red lines) net fluxes are plotted for various optical thicknesses, the optical depths at the planet's surface taking the values $\taugrsw = 0.1,1,10$ (from light to dark blue lines) and $\taugrlw = 0.1,1,10$ (from light to dark red lines). Pure absorption is assumed ($\betalw = \betasw = 1$). The incident, atmospheric, and surface fluxes are arbitrarily set to $\Finc = 4 \Feq$, $\Fatm  = 0.7 \Feq $, and $\Fsurf = 2 \Feq$, respectively.  }
       \label{fig:Fnet_tau}%
\end{figure}

The four integration constants of the solution are determined by assuming two boundary conditions for each \kc{wavelength} domain. At the top of the atmosphere, the shortwave downwelling flux is the stellar incident flux $\Finc$ and the longwave downwelling flux is zero because of the absence of thermal source at the infinity, which is written as $\Fdownsw = \Finc$ and $\Fdownlw = 0$. At the planet surface, a fraction of the shortwave downwelling flux is reflected while the remaining part is absorbed by the ground and re-emitted in the longwave. Denoting by $\Asurfsw$ the albedo of the planet's surface in the shortwave, and $ \Fsurf $ the flux emitted by the planet's surface in the longwave, these conditions are expressed as $\Fupsw = \Asurfsw \Fdownsw$ and $\Fuplw = \Fsurf$. We thus end up with the net fluxes in the short- and longwave,

\begin{equation}
\Fnetsw = - \Finc \usw \frac{ \left( \gamswminus - \Asurfsw \gamswplus \right) \ftranssw  \expo{\optdepthsw} + \left( \gamswplus - \Asurfsw \gamswminus \right) \ftranssw^{-1} \expo{-\optdepthsw} }{\gamswplus \left( \gamswplus - \Asurfsw \gamswminus \right) \ftranssw^{-1} - \gamswminus \left( \gamswminus - \Asurfsw \gamswplus \right) \ftranssw},
\label{Fnetsw_isoT}
\end{equation}

\noindent and
 
\begin{align}
\label{Fnetlw_isoT}
\Fnetlw = & \  \frac{1}{2} \Fatm \ulw \frac{\left( \gamlwminus \ftranslw - \gamlwplus \right) \expo{\optdepthlw} + \left( \gamlwplus \ftranslw^{-1} - \gamlwminus \right) \expo{- \optdepthlw}  }{\left( \gamlwplus \right)^2 \ftranslw^{-1} - \left( \gamlwminus \right)^2 \ftranslw}   \nonumber \\ 
 	      & +  \Fsurf \ulw \frac{ \gamlwplus \expo{\optdepthlw} + \gamlwminus \expo{-\optdepthlw} }{\left( \gamlwplus \right)^2 \ftranslw^{-1} - \left( \gamlwminus \right)^2 \ftranslw}. 
\end{align}

\noindent where we have introduced \kc{the atmospheric and surface fluxes, $\Fatm$ and $\Fsurf$,} and the vertically-integrated transmission functions

\begin{equation}
\begin{array}{ll}
\ftranslw \define \expo{- \taugrlw }, & \mbox{and} \ \ftranssw \define \expo{- \taugrsw}. 
\end{array}
\end{equation}

The vertical profiles of the short- and longwave net fluxes given by \eqs{Fnetsw_isoT}{Fnetlw_isoT} are plotted in \fig{fig:Fnet_tau} for various atmospheric \kc{optical depths}. One may observe here that increasing the opacity tends to confine the absorption region in the upper atmosphere in the shortwave band, while it tends to neutralise the net flux within the atmosphere in the longwave.

\subsection{Surface and atmospheric energy budgets}

The derived analytic solution is now used to establish the local energy budgets of the planet's surface and atmosphere at thermodynamical equilibrium. By considering the upper ($\optdepthlw = \optdepthsw = 0$) and lower ($\optdepthlw = \taugrlw$ and $\optdepthsw = \taugrsw$) boundaries, we obtain the fluxes radiated towards space at the top of the atmosphere and towards the atmosphere at the planet's surface, respectively. These fluxes are expressed as 

\begin{equation}
\left. \Fnetlw \right|_{\optdepthlw = 0} =   \ulw \frac{ \frac{1}{2} \Fatm \left(\gamlwplus \ftranslw^{-1} + \gamlwminus \ftranslw -1 \right) +  \Fsurf}{ \left(\gamlwplus \right)^2 \ftranslw^{-1} - \left( \gamlwminus \right)^2 \ftranslw} ,
\end{equation}

\begin{align}
\left. \Fnetlw \right|_{\optdepthlw = \taugrlw}= & \   - \frac{1}{2} \Fatm \ulw \frac{  \gamlwplus \ftranslw^{-1} + \gamlwminus \ftranslw -1 }{ \left(\gamlwplus \right)^2 \ftranslw^{-1} - \left( \gamlwminus \right)^2 \ftranslw}  \\
 & +  \Fsurf \ulw \frac{   \gamlwplus \ftranslw^{-1} + \gamlwminus \ftranslw }{ \left(\gamlwplus \right)^2 \ftranslw^{-1} - \left( \gamlwminus \right)^2 \ftranslw},
\end{align}

\begin{equation}
\left. \Fnetsw \right|_{\optdepthsw = 0}= -   \frac{ \Finc \usw \left[ \left( \gamswplus - \Asurfsw \gamswminus \right) \ftranssw^{-1} + \left( \gamswminus - \Asurfsw \gamswplus \right) \ftranssw \right] }{\gamswplus \left( \gamswplus - \Asurfsw  \gamswminus \right) \ftranssw^{-1} -  \gamswminus \left( \gamswminus - \Asurfsw \gamswplus \right) \ftranssw},
\end{equation}

\noindent  and

\begin{equation}
\left. \Fnetsw \right|_{\optdepthsw = \taugrsw} = - \frac{ \Finc \usw \left( 1 - \Asurfsw \right) }{\gamswplus \left( \gamswplus - \Asurfsw  \gamswminus \right) \ftranssw^{-1} -  \gamswminus \left( \gamswminus - \Asurfsw \gamswplus \right) \ftranssw},
\end{equation}

\noindent which brings out the coefficient parameterising the surface and atmospheric energy balances,

\begin{align}
\left. \Fnetlw \right|_{\optdepthlw= 0} = & \ \Alw \Fatm  + \left( 1 - \Agrlw \right) \Fsurf , \\
\left. \Fnetlw \right|_{\optdepthlw = \taugrlw} = & - \Alw \Fatm + \Klw \Fsurf, \\
\left. \Fnetsw \right|_{\optdepthsw = 0}   = & - \Ksw \Finc,  \\
\left. \Fnetsw \right|_{\optdepthlw = \taugrsw}  = & - \left( 1 - \Agrsw \right) \Finc ,
\end{align}

\noindent with $\Alw \define \Klw + \Agrlw - 1$ (similarly, in the short wavelength domain, we may define $\Asw  \define \Ksw + \Agrsw - 1$). These coefficients are expressed as 

\begin{equation}
\Alw = \ulw \frac{\gamlwplus \ftranslw^{-1} + \gamlwminus \ftranslw -1}{\left( \gamlwplus \right)^2 \ftranslw^{-1} - \left( \gamlwminus \right)^2 \ftranslw},
\end{equation}

\begin{equation}
\Asw = \usw \frac{\left( \gamswminus - \Asurfsw \gamswplus \right) \ftranssw + \left(  \gamswplus - \Asurfsw \gamswminus \right) \ftranssw^{-1} + \Asurfsw - 1}{\gamswplus \left( \gamswplus - \Asurfsw  \gamswminus \right) \ftranssw^{-1} -  \gamswminus \left( \gamswminus - \Asurfsw \gamswplus \right) \ftranssw},
\end{equation}

\begin{equation}
\Klw = \ulw \frac{\gamlwplus \ftranslw^{-1} + \gamlwminus \ftranslw}{\left( \gamlwplus \right)^2 \ftranslw^{-1} - \left( \gamlwminus \right)^2 \ftranslw},
\end{equation}

\begin{equation}
\Ksw = \usw \frac{\left( \gamswminus  - \Asurfsw \gamswplus \right) \ftranssw + \left( \gamswplus - \Asurfsw \gamswminus \right) \ftranssw^{-1}}{\gamswplus \left( \gamswplus - \Asurfsw  \gamswminus \right) \ftranssw^{-1} -  \gamswminus \left( \gamswminus - \Asurfsw \gamswplus \right) \ftranssw},
\end{equation}

\begin{equation}
\Agrlw = \frac{\left(\gamlwplus \right)^2 \ftranslw^{-1} - \left( \gamlwminus \right)^2 \ftranslw - \ulw}{\left(\gamlwplus \right)^2 \ftranslw^{-1} - \left( \gamlwminus \right)^2 \ftranslw},
\end{equation}

\noindent and

\begin{align}
\Agrsw = & \ \frac{\gamswplus \left( \gamswplus - \Asurfsw \gamswminus \right) \ftranssw^{-1} - \gamswminus \left( \gamswminus - \Asurfsw \gamswplus \right) \ftranssw }{\gamswplus \left( \gamswplus - \Asurfsw \gamswminus \right) \ftranssw^{-1} - \gamswminus \left( \gamswminus - \Asurfsw \gamswplus \right) \ftranssw} \\
 & - \frac{\left( 1 - \Asurfsw \right) \usw}{\gamswplus \left( \gamswplus - \Asurfsw \gamswminus \right) \ftranssw^{-1} - \gamswminus \left( \gamswminus - \Asurfsw \gamswplus \right) \ftranssw}. \nonumber
\end{align}

\noindent Thus, introducing the heat power per unit area due to surface turbulent exchanges on the dayside ($\Dconv$) and day-night advection ($\Dadv$) -- specified in \sect{sec:sensible_heating} and \sect{sec:advection}, respectively -- the local energy balances of the planet's surface and atmosphere are formulated as 

\begin{align}
\label{budget_surf_day}
\left( 1 - \Agrsw \right) \Finc + \Alw \Fatm - \Klw \Fsurf - \Dconv = 0 & \ \ \mbox{(surf.; day),}  \\
\label{budget_atm_day}
\Asw \Finc - 2 \Alw \Fatm + \Alw \Fsurf - \Dadv + \Dconv = 0 & \ \ \mbox{(atm.; day),} \\
\label{budget_surf_nitght}
 \Alw \Fatm - \Klw \Fsurf = 0 & \ \  \mbox{(surf.; night),} \\
\label{budget_atm_night}
 - 2 \Alw \Fatm + \Alw \Fsurf + \Dadv  = 0 & \ \  \mbox{(atm.; night).}
\end{align}

\noindent  We note that \eqs{budget_surf_day}{budget_atm_day} are the counterparts of Eqs.~(14) and (15) in \cite{Wordsworth2015}, respectively. \kc{Following \cite{Pierrehumbert2011} and \cite{Wordsworth2015}, the notation $\Dconv$ refers to 'sensible heat flux', which designates an energy flux corresponding to a change of temperature of the fluid\footnote{\kc{Sensible heat is thus distinct from latent heat, which designates the energy associated with the change of phase of a condensable substance \citep[][Sect.~6.3]{Pierrehumbert2011}.}}.}

As illustrated by \fig{fig:model_overview}, the dayside and nightside hemispheres of the planet are characterised by different surface and atmospheric temperatures ($\Tday$, $\Taday$, $\Tnight$, and $\Tanight$). This leads to average the preceding energy budgets over each of the two hemispheres. Hence, we introduce the dayside and nightside hemisphere-averaged surface fluxes, denoted by $\Bday$ and $\Bnight$, the blackbody fluxes radiated by the atmosphere, $\Baday$ and  $\Banight$, and the averaged heat flows, these quantities being defined by \citep[see][Eqs.~(16) and (17)]{Wordsworth2015}

\begin{equation}
\begin{array}{ll}
\Bday \define \sigmaSB \Tday^4, & \Bnight \define \sigmaSB \Tnight^4,  \\[0.2cm]
 \Baday \define \sigmaSB \Taday^4, & \Banight \define \sigmaSB \Tanight^4,  \\[0.2cm]
 \meanDconv \define \dfrac{\integ{\Dconv}{\unitsurf}{\iday}{}}{2 \pi \Rpla^2}, &  \meanDadv \define \dfrac{\integ{\Dadv}{\unitsurf}{\iday}{}}{2 \pi \Rpla^2} .
\end{array}
\end{equation}

Besides, the incident stellar flux is given as a function of the stellar zenithal angle $\zenithphi$ by  

\begin{equation}
\begin{array}{ll}
	\Finc \define \Fstar \cos \zenithphi & \mbox{if} \  0 \leq \zenithphi \leq \dfrac{\pi}{2} , \\ 
	\Finc \define 0 & \mbox{if} \ \dfrac{\pi}{2} < \zenithphi \leq \pi.
\end{array}
\end{equation}

\noindent Thus, averaged over the dayside and nightside hemispheres, the system of \eqsto{budget_surf_day}{budget_atm_night} becomes

\begin{align}
\label{mean_budget_surf_day}
\frac{1}{2} \left( 1 - \Agrsw \right) \Fstar + \Alw \Baday - \Klw \Bday - \meanDconv = 0, &   \\
\label{mean_budget_atm_day}
\frac{1}{2} \Asw \Fstar - 2 \Alw \Baday + \Alw \Bday - \meanDadv + \meanDconv = 0, & \\
\label{mean_budget_surf_nitght}
 \Alw \Banight - \Klw \Bnight= 0, & \\
\label{mean_budget_atm_night}
 - 2 \Alw \Banight + \Alw \Bnight + \meanDadv  = 0 .&
\end{align}

\noindent \kc{This set of equations is the counterpart of Eqs.~(21-23) in \cite{Wordsworth2015} extended outside of the optically thin limit and with an additional energy flux between the dayside and nightside hemispheres ($\meanDadv$).}

\begin{figure*}[htb]
   \centering
    \begin{flushleft}
   \hspace{0.17\textwidth} $\taugrsw = 10^{-4}$
  \hspace{0.17\textwidth} $\taugrsw = 10^{-2}$
\hspace{0.17\textwidth}  $\taugrsw = 10^{0}$ \\
 \end{flushleft}
  \vspace{-0.2cm}
    \raisebox{0.5cm}{\includegraphics[width=0.023\textwidth]{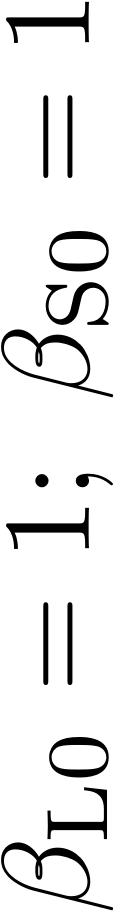}}
    \hspace{0.2cm}
    \includegraphics[width=0.023\textwidth,trim = 3.2cm 4cm 23.3cm 2cm,clip]{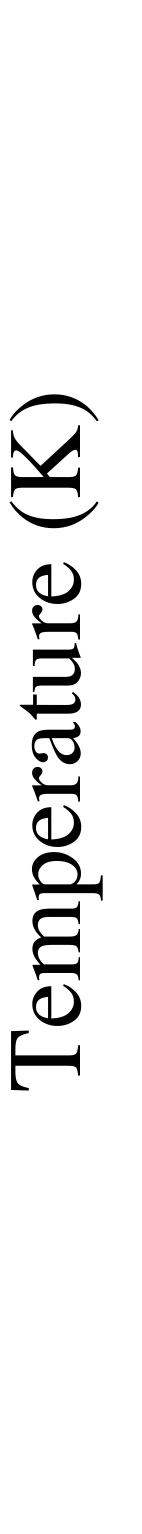} 
    \includegraphics[width=0.28\textwidth,trim = 5.0cm 4cm 5.5cm 2cm,clip]{auclair-desrotour_fig4b} 
   \includegraphics[width=0.28\textwidth,trim = 5.0cm 4cm 5.5cm 2cm,clip]{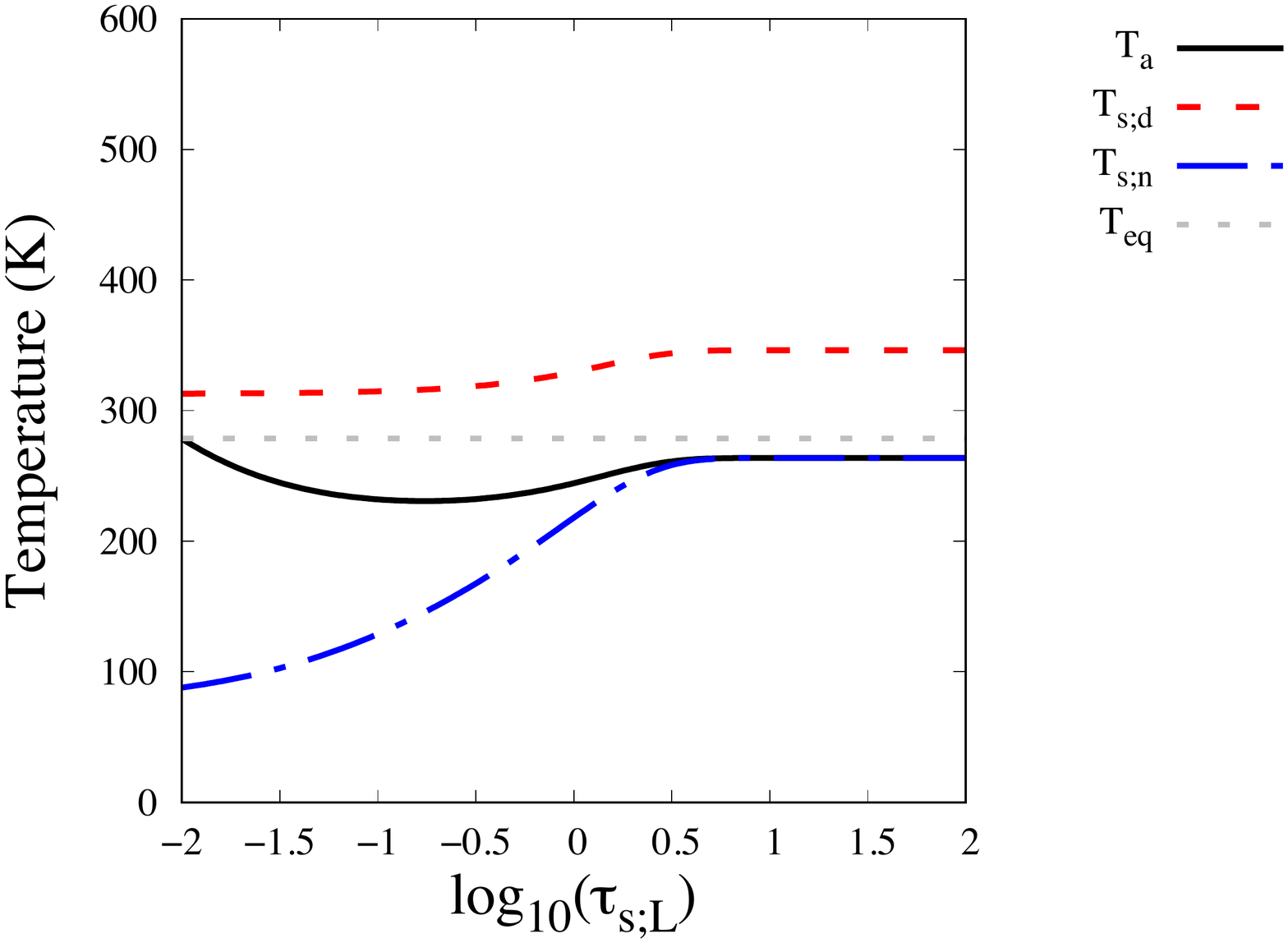} 
   \includegraphics[width=0.28\textwidth,trim = 5.0cm 4cm 5.5cm 2cm,clip]{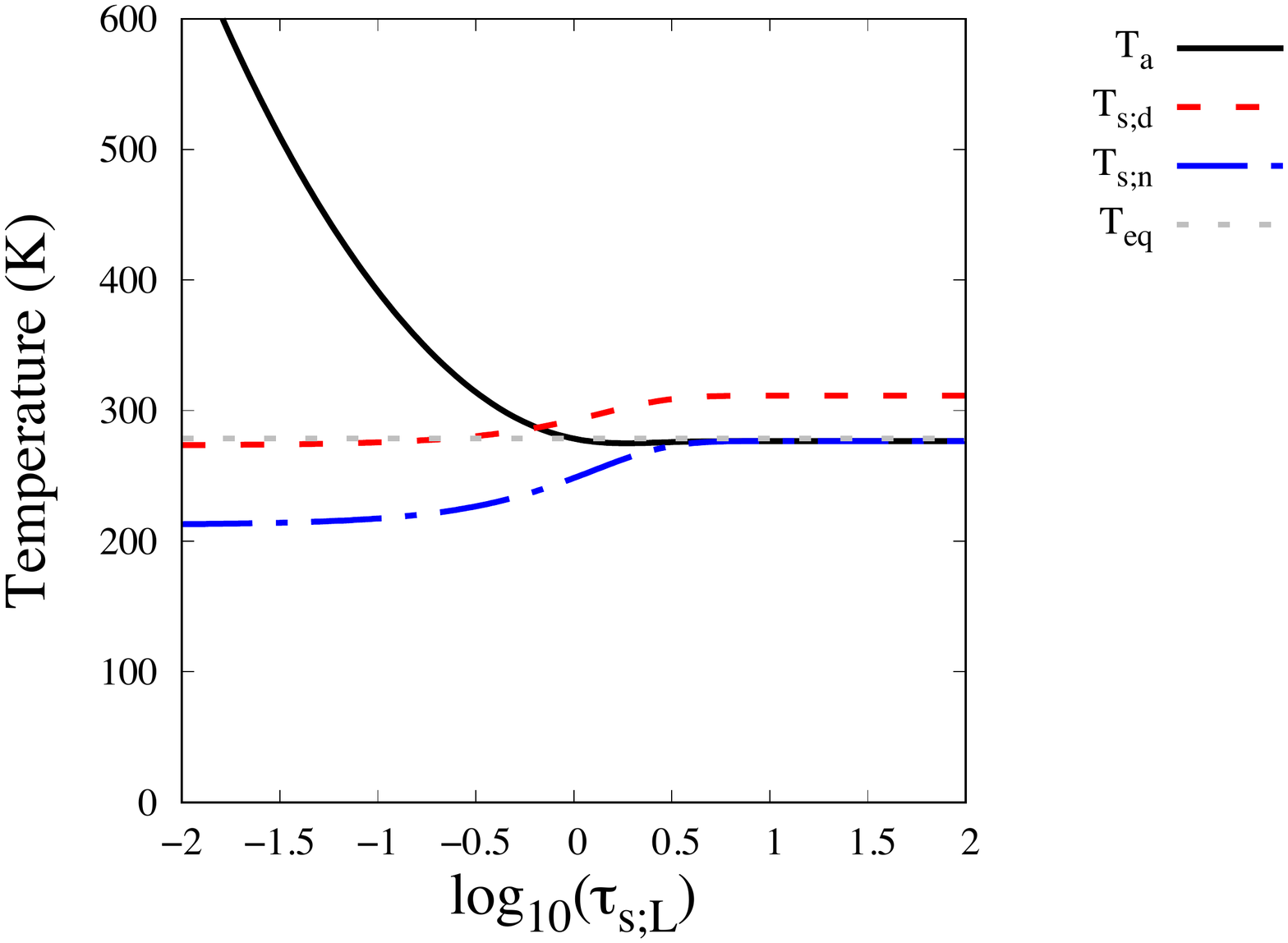}
    \includegraphics[width=0.07\textwidth,trim = 22.6cm 8cm 2cm 2cm,clip]{auclair-desrotour_fig4d}  \\
    \hspace{-0.07\textwidth}
    \raisebox{0.5cm}{\includegraphics[width=0.023\textwidth]{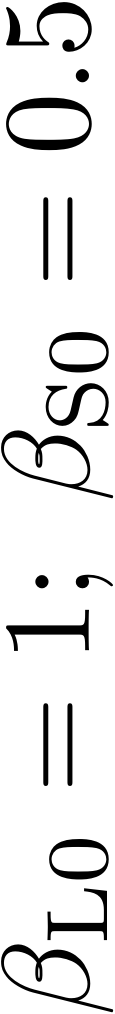}}
    \hspace{0.2cm}
    \includegraphics[width=0.023\textwidth,trim = 3.2cm 4cm 23.3cm 2cm,clip]{auclair-desrotour_fig4b} 
     \includegraphics[width=0.28\textwidth,trim = 5.0cm 4cm 5.5cm 2cm,clip]{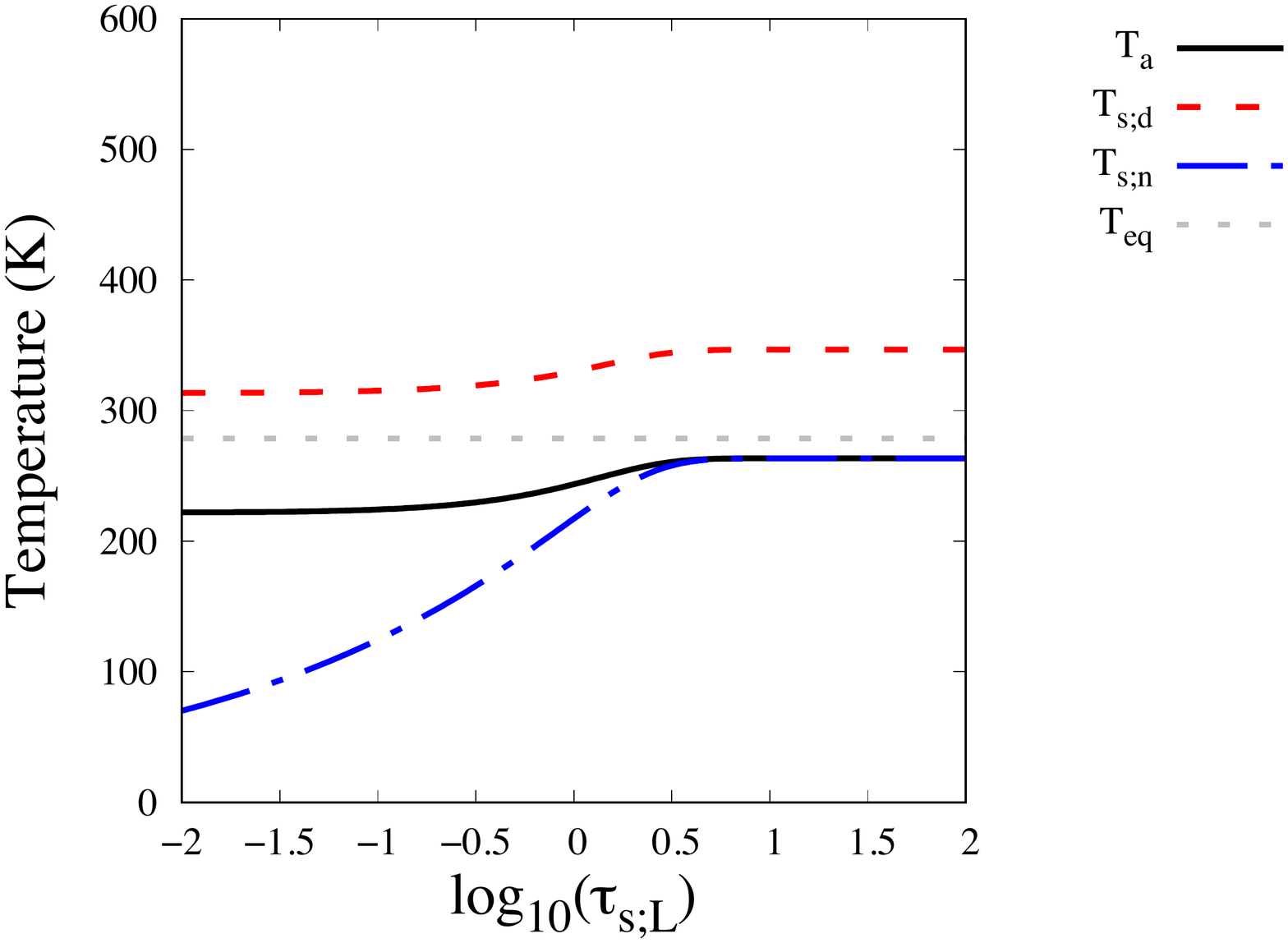} 
   \includegraphics[width=0.28\textwidth,trim = 5.0cm 4cm 5.5cm 2cm,clip]{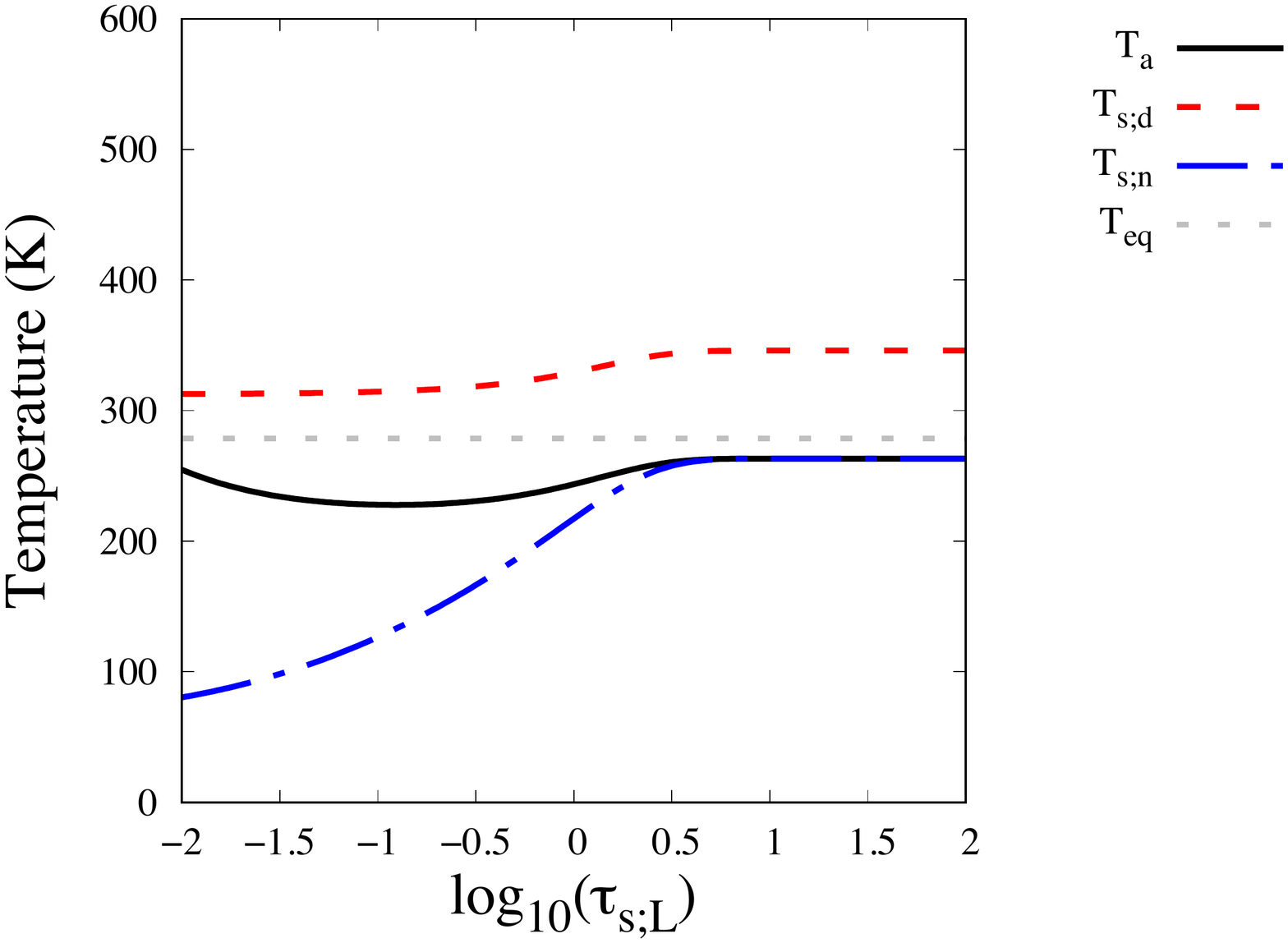} 
   \includegraphics[width=0.28\textwidth,trim = 5.0cm 4cm 5.5cm 2cm,clip]{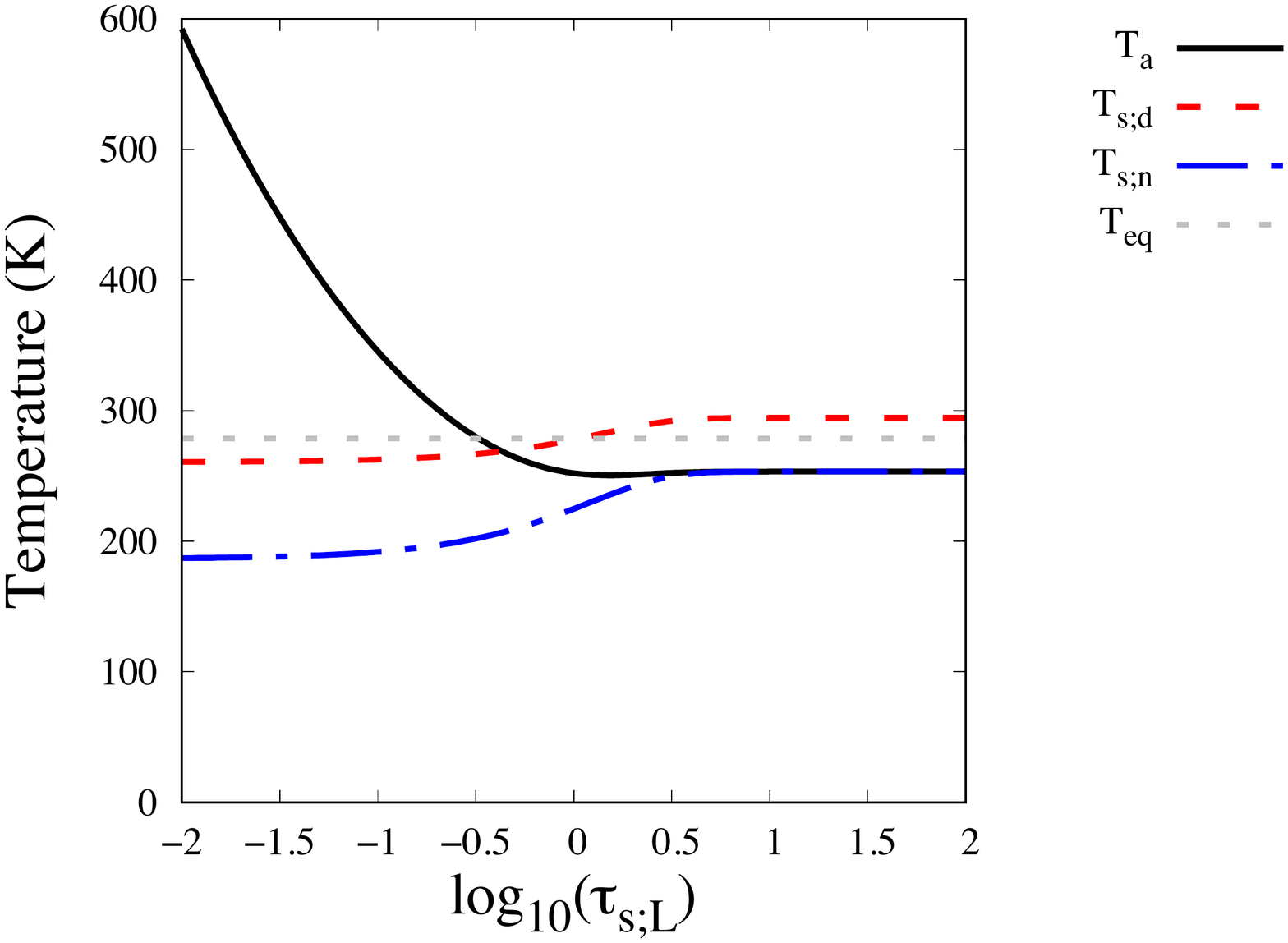}\\
   \hspace{-0.07\textwidth}
     \raisebox{0.5cm}{\includegraphics[width=0.023\textwidth]{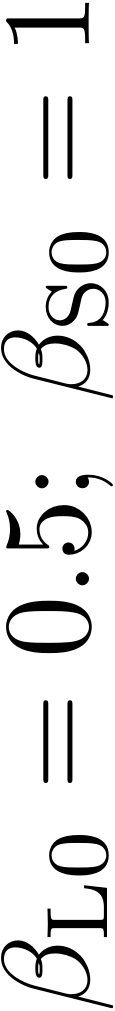}}
    \hspace{0.2cm}
    \includegraphics[width=0.023\textwidth,trim = 3.2cm 4cm 23.3cm 2cm,clip]{auclair-desrotour_fig4b} 
     \includegraphics[width=0.28\textwidth,trim = 5.0cm 4cm 5.5cm 2cm,clip]{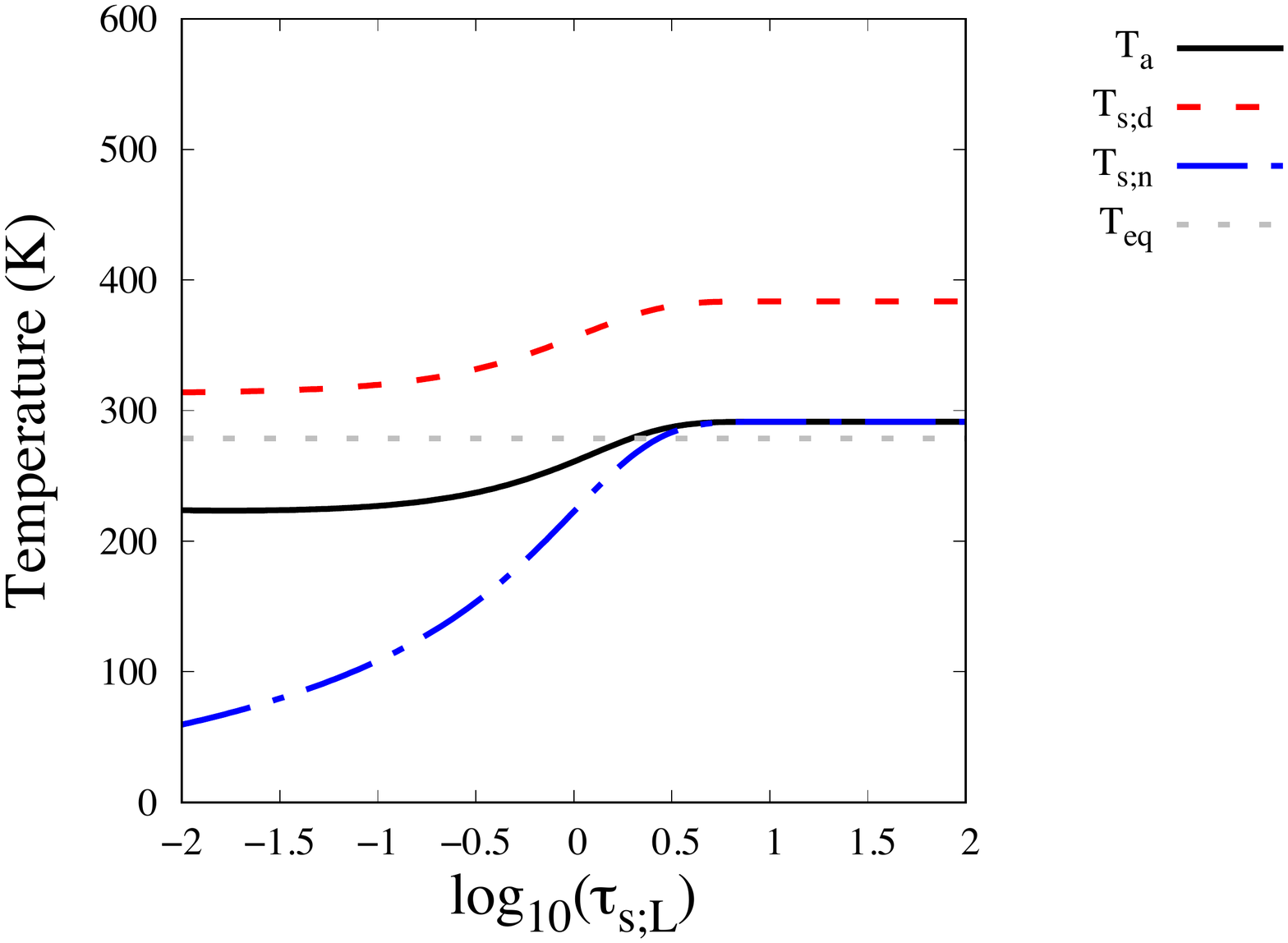} 
   \includegraphics[width=0.28\textwidth,trim = 5.0cm 4cm 5.5cm 2cm,clip]{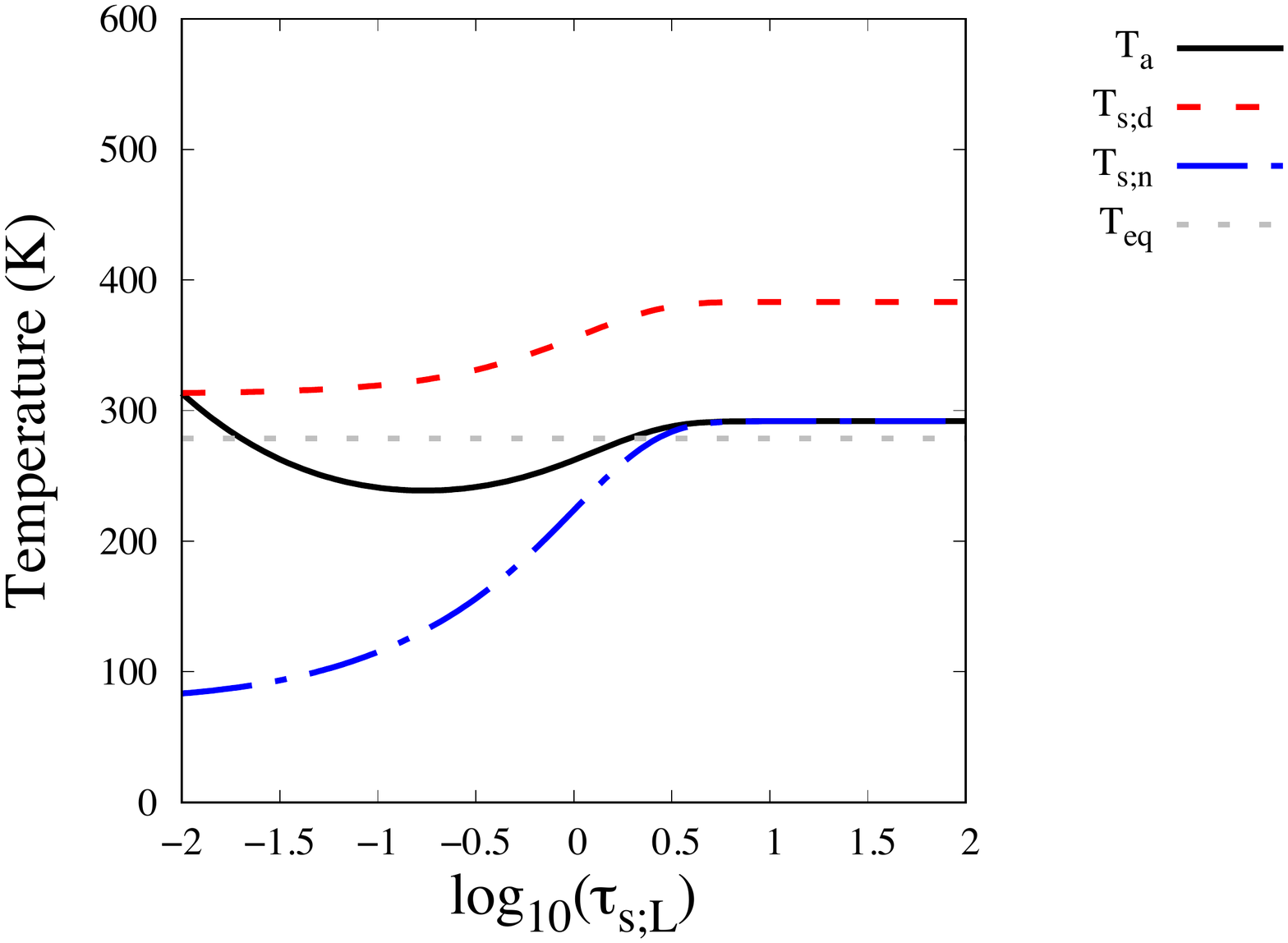} 
   \includegraphics[width=0.28\textwidth,trim = 5.0cm 4cm 5.5cm 2cm,clip]{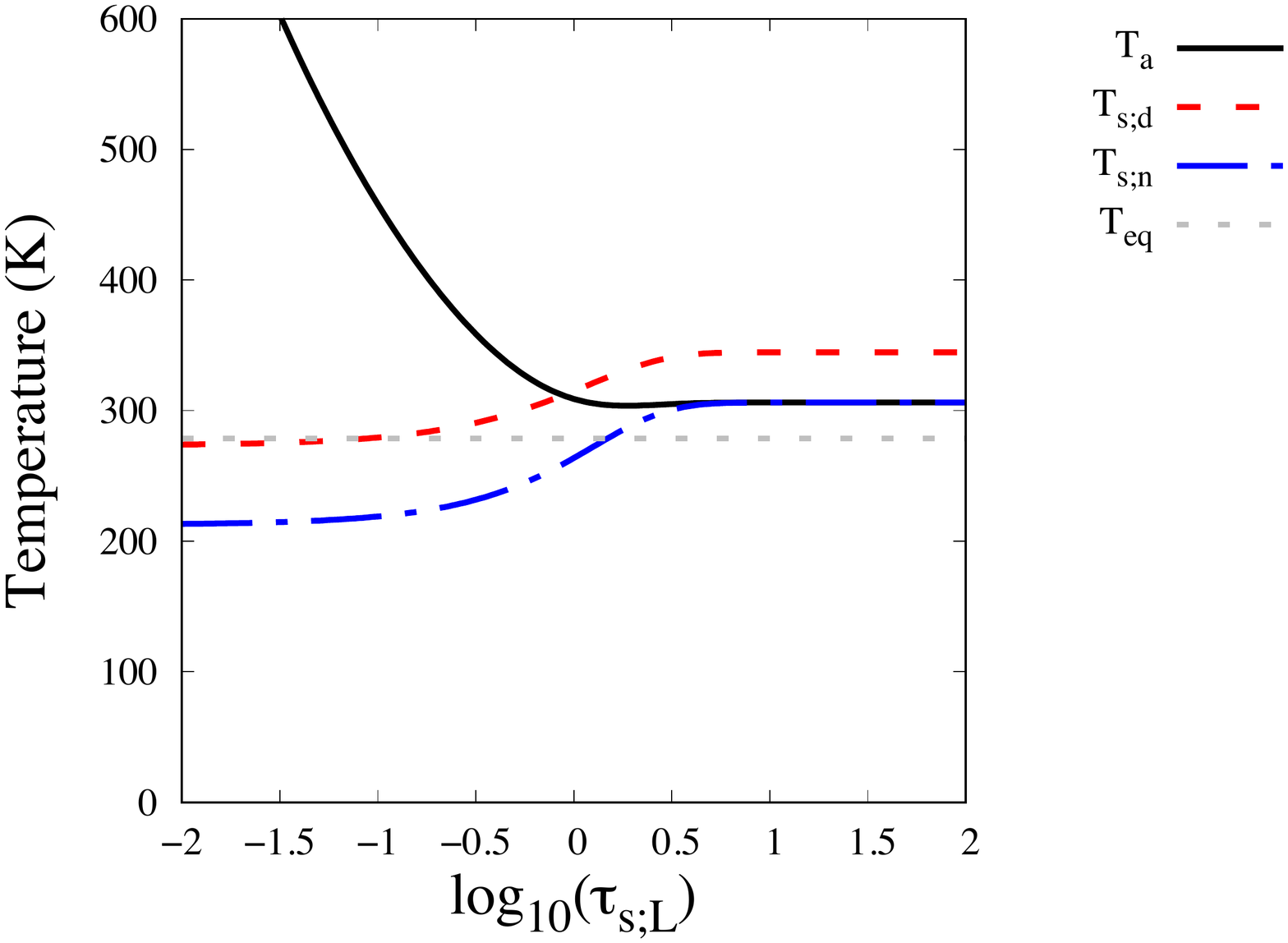}\\
    \hspace{-1cm}
   \includegraphics[width=0.28\textwidth,trim = 5.0cm 2cm 5.5cm 17.6cm,clip]{auclair-desrotour_fig4h} 
    \includegraphics[width=0.28\textwidth,trim = 5.0cm 2cm 5.5cm 17.6cm,clip]{auclair-desrotour_fig4h} 
     \includegraphics[width=0.28\textwidth,trim = 5.0cm 2cm 5.5cm 17.6cm,clip]{auclair-desrotour_fig4h} 
   \caption{Atmospheric, dayside and nightside surface temperatures in pure radiative regime. Temperatures are plotted as functions of the logarithm of $\taugrlw$ in the case of pure absorption (top panels), scattering in the shortwave (middle panels) and scattering in the longwave (bottom panels). {\it Left:} Quasi-transparence in the shortwave ($\taugrsw = 10^{-4}$). {\it Middle:} Small shortwave opacity ($\taugrsw = 10^{-2}$). Large shortwave opacity ($\taugrsw = 10^{0}$). The atmospheric temperature (solid black line), dayside surface temperature (dashed red line), nightside surface temperature (dashed blue line), and black-body equilibrium temperature (dotted grey line) are plotted using \eqsfour{Tarad}{Tsdrad}{Tsnrad}{Teq}, respectively. Parameters values: $\Asurfsw = 0.2$, and $\Fstar = 1366$\units{W m^{-2}}. }
       \label{fig:temp_taurad}%
\end{figure*}

\subsection{Purely radiative case}
\label{ssec:radiative_case}

Before investigating the role played by sensible heating and advection, it is worth examining the purely radiative case. For simplification, we place ourselves in the regime treated by \cite{Wordsworth2015}, where day-night heat transfers are strong enough to homogenise the atmospheric temperature ($\Taday = \Tanight = \Tatm$ and $\Baday = \Banight = \Batm$). In this framework, the sensible heating is neglected (\kc{$\meanDconv = 0$}), and the advected heat flow does not intervene in the equilibrium, meaning that the dayside and nightside atmospheric budgets may be combined into a globally averaged atmospheric budget,

\begin{equation}
\frac{1}{2} \Asw \Fstar - 4 \Alw \Baday + \Alw \left( \Bday + \Bnight \right)  = 0.
\label{global_averaged_atm_budget}
\end{equation}

\noindent Solving this equation together with \eqs{mean_budget_surf_day}{mean_budget_surf_nitght} yields the surface and atmospheric equilibrium temperatures

\begin{equation}
\Tatm = \Teq  \left[ \frac{\Alw \left( 1 - \Agrsw \right) + \Klw \Asw }{\Alw \left( 2 \Klw - \Alw \right)}  \right]^{\frac{1}{4}}, 
\label{Tarad}
\end{equation}

\begin{equation}
\Tday = \Teq \left[ \frac{ \left( 4 \Klw - \Alw \right) \left( 1 - \Agrsw \right) + \Klw \Asw }{ \Klw \left( 2 \Klw - \Alw \right) }  \right]^{\frac{1}{4}}, 
\label{Tsdrad}
\end{equation}

\begin{equation}
\Tnight = \Teq \left[ \frac{\Alw \left( 1 - \Agrsw \right) + \Klw \Asw }{\Klw \left( 2 \Klw - \Alw \right)}  \right]^{\frac{1}{4}},
\label{Tsnrad}
\end{equation}

\noindent which capture the non-linear dependence of \kc{radiative transfer} on the atmospheric optical thickness in the general case, and highlight asymptotic regimes. 

Assuming pure absorption in the longwave ($\betalw = 1$) and \kc{transparency} in the shortwave ($\taugrsw = 0$), we recover the behaviour derived by \cite{Wordsworth2015} in the optically thin limit ($\ftranslw \approx 1 - \taugrlw$). Conversely, as the optical depth increases, both the atmosphere and surface temperatures converge towards the black body temperature, as showed by \tab{tab:radiative_asymp}. We note that the surface temperature of rocky planets in the optically thick limit are expected to be higher than that predicted by the model, because the atmospheric structure cannot be approximated by an isothermal temperature profile in this limit. For instance, Venus' troposphere is characterised by a strong temperature gradient \citep[e.g.][]{Seiff1985}, which is far closer to the adiabatic profile ($\BVfreq = 0$) than to the isothermal profile. Consequently, \kc{Venus's} mean surface temperature is $\Tsurf \approx 740$\units{K} whereas its blackbody equilibrium temperature in the absence of albedo is $\Teq = 327.3$~K (we have taken the value $\Fstar = 2601.3$\units{W~m^{-2}} for the Solar bolometric flux).

\begin{table}[h]
\centering
 \textsf{\caption{\label{tab:radiative_asymp} Optically thin and thick asymptotic limits in the pure radiative (dayside sensible heating and day-night heat transfers ignored) and pure absorption approximations (no scattering). We adopt the notations $\optdepthgen \ll 1$ and $\optdepthgen \gg 1$ to designate the optically thin ($\taugrsw \ll 1$; $\taugrlw \ll 1$) and thick ($\taugrsw \gg 1$; $\taugrlw \gg1$) limits, respectively. The case treated by \cite{Wordsworth2015} is obtained by setting $\taugrsw = 0$ in the optically thin limit.}}
\begin{small}
    \begin{tabular}{ c c c}
      \hline
      \hline
      \textsc{Parameter} & \textsc{Thin ($\optdepthgen \ll 1$)}  & \textsc{Thick ($\optdepthgen \gg 1$)}  \\ 
      \hline 
      $\Alw$ & $\optdepthlw$ & $1$ \\
      $\Klw$ & $1$ & $1$ \\
      $\Agrlw$ & $\optdepthlw$ & $1$ \\
      \hline 
      $\Asw$ & $\left( 1 + \Asurfsw \right) \optdepthsw $ & $1$ \\
      $\Ksw$ & $1 - \Asurfsw$ & $1$ \\
      $\Agrsw$ & $\Asurfsw$ & $1$ \\
      \hline 
      $\Tatm$ & $ \Teq \left[ \frac{\left( 1 - \Asurfsw \right) }{2} \left( 1 + \frac{1 + \Asurfsw}{1 - \Asurfsw} \frac{\taugrsw}{\taugrlw} \right) \right]^{\frac{1}{4}} $ & $ \Teq $ \\
      $\Tday$ & $\Teq \left[ 2 \left( 1 - \Asurfsw \right) \right]^{\frac{1}{4}}$ & $ \Teq$ \\
      $\Tnight$ & $ \Teq \left[ \frac{ \left( 1 - \Asurfsw \right) \taugrlw}{2 } \left( 1 + \frac{1 + \Asurfsw}{1 - \Asurfsw} \frac{\taugrsw}{\taugrlw} \right) \right]^{\frac{1}{4}} $ & $ \Teq$ \\
      \hline
    \end{tabular}
\end{small}
 \end{table}

The effect that absorption and scattering have on the steady state is explored in \fig{fig:temp_taurad}, where the atmospheric and surface temperatures are plotted against the longwave opacity for various scattering parameters and shortwave opacities. We first consider the quasi-transparent limit in the shortwave ($\taugrsw = 10^{-4}$), which corresponds to $\taugrsw \ll \taugrlw$. As the longwave optical thickness increases, the equilibrium state switches from the optically thin regime, where $\Tnight \scale \taugrlw^{1/4} $, to the optically thick one, where the three temperatures reach a plateau. The \kc{interplay} between short- and longwave \kc{optical depths} defines \kc{another} dimension in the parameter space. When $\taugrlw \lesssim \taugrsw$, the atmosphere has to increase its temperature to evacuate the power absorbed in the shortwave. As a consequence, the atmospheric temperature increases as the optical thickness in the longwave decays, which is the behaviour observed in \fig{fig:temp_taurad} (right panels). 

Scattering in the shortwave and in the longwave have opposite effects on temperatures. When scattering in the shortwave is present ($\betasw  =  0.5$, middle panels), it induces an anti-greenhouse cooling. A fraction of the incident stellar flux is scattered back to space, meaning that the planet absorbs less energy than in the case of pure absorption. As a consequence, the atmosphere and surface temperatures are smaller. Conversely, the presence of scattering in the longwave (\kc{$\betalw = 0.5$}, bottom panels) generates an effect known as scattering greenhouse effect \citep[e.g.][Sect.~4.7, p.~71]{Heng2017}. Thermal emissions of the surface and atmosphere towards space are scattered back to the planet, leading to the observed increase of the atmospheric, dayside and nightside surface temperatures with respect to the case of pure absorption.

\section{Inclusion of sensible heating}
\label{sec:sensible_heating}

\begin{figure}[htb]
   \centering
   \includegraphics[width=0.45\textwidth,trim = 6.0cm 0cm 0cm 3.0cm,clip]{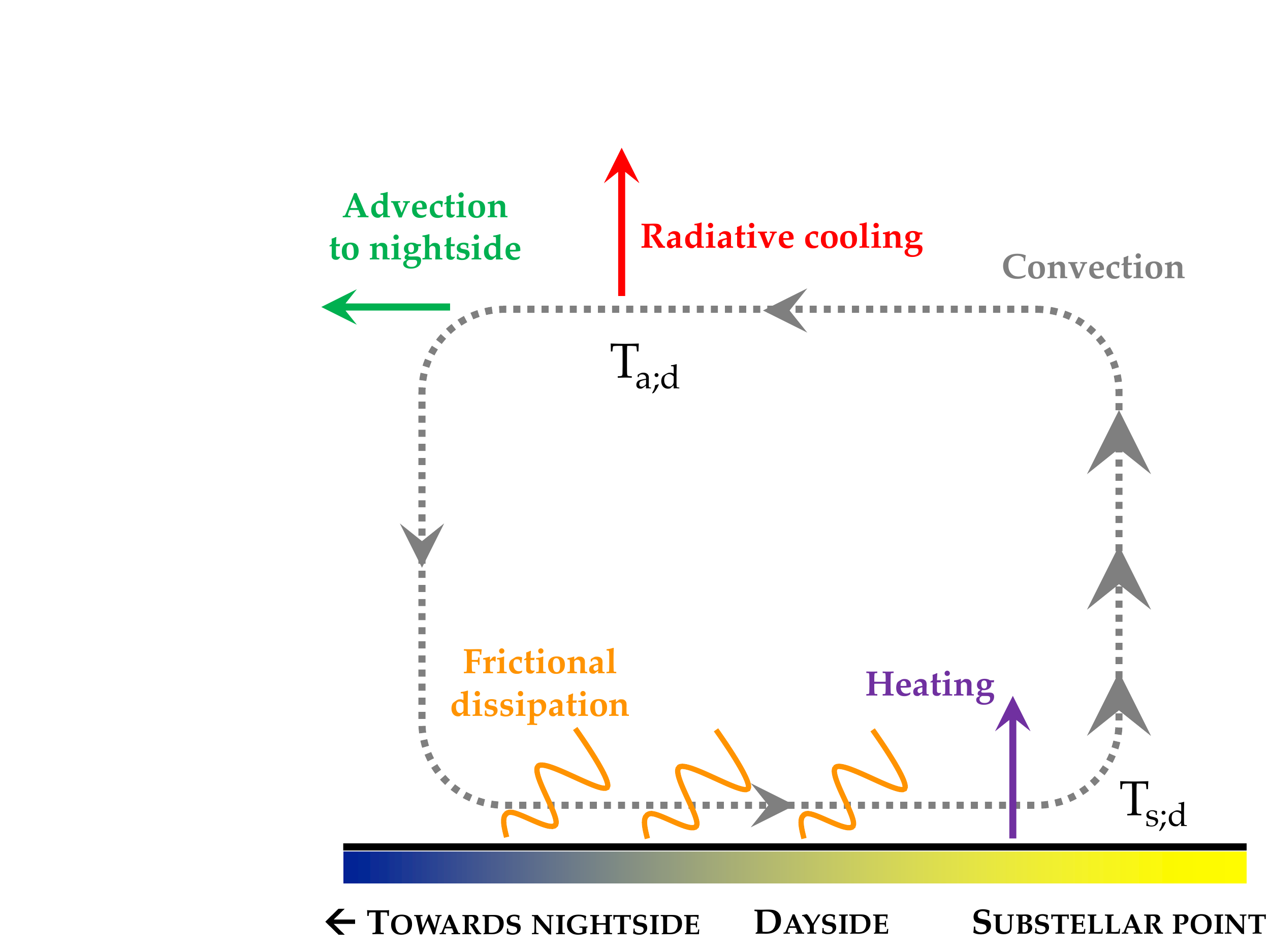}
      \caption{A diagram of the dayside atmospheric heat engine. The heat engine is driven by dayside heating and cooling to space. A fluid parcel works against friction in the boundary layer, which limits the speed of the flow.}
       \label{fig:convection_cell}%
\end{figure}

As a second step, the dayside sensible heating is included in the model. While we still assume a uniform atmospheric temperature ($\Taday =\Tanight = \Tatm$), the surface turbulent heat flux \kc{$\meanDconv$} is now expressed as a function of the dayside surface and atmospheric temperatures. We introduce first the expression of turbulent heat exchanges between the surface and the atmosphere within the planetary boundary layer (PBL), and second the scaling law of the horizontal velocity parametrising this flux, \kc{following} \cite{KA2016}, where the atmospheric circulation is treated as a heat engine (\fig{fig:convection_cell}). 

\subsection{Scaling of turbulent heat exchanges}
\label{ssec:scaling_sensible_heating}

The turbulent heat exchanges due to convection in the dayside PBL may be quantified in the framework of the mixing length theory \citep[][${\rm 5^{th}}$ edition, Sect.~8.3.5, p.~268]{Holton1973}, as detailed in \append{app:turbulent_heat_exchanges}. The hemisphere-averaged sensible heat flux is written as \citep[e.g.][Eq.~(6.11), p. 396]{Pierrehumbert2011}

\begin{equation}
\meanDconv = \Cd \Cp \rhoatm \Vconv \left(\Tday - \Taday \right),
\label{Fturb}
\end{equation}

\noindent where $\rhoaday = \psurf / \left( \Rspec \Taday \right)$ is the surface density, $\Cd$ the bulk drag coefficient accounting for the strength of friction, and $\Vconv$ the typical horizontal wind speed quantifying the strength of the convective cell that develops in the vicinity of the substellar point. If the ground is warmer than air, heat is carried away from the ground ($\meanDconv > 0$) at a rate proportional to the temperature difference. Conversely, if the ground is cooler than the air, the atmosphere warms the ground ($\meanDconv<0$). \rec{We note that the convective cell vanishes in case of stable stratification, which implies $\Vconv = 0$ and $\meanDconv = 0$. Thus, we ignore the sensible flux on the nightside. On the dayside however, although radiative transfer is described in the isothermal atmosphere approximation for simplification, we assume that the planetary boundary layer is convective, so that $\meanDconv \neq 0$ in the general case.  }

The bulk drag coefficient weakly depends on the surface and boundary layer properties. In the case of a neutrally buoyant boundary layer, following the empirical scaling law proposed by von K\'arm\'an for the mixing length (\append{app:turbulent_heat_exchanges}), it is expressed as a function of the surface roughness length $\roughheight$, the thickness of the turbulent boundary layer $\zsurf$, and the von K\'arm\'an constant $\karman \approx 0.4$, as \citep[e.g.][Eq.~(7)]{Esau2004}

\begin{equation}
\Cd = \left[ \frac{\karman}{\ln \left( \zsurf / \roughheight \right) } \right]^{2}. 
\label{bulk_coeff}
\end{equation}

\noindent The bulk drag coefficient is of order $\Cd \sim 10^{-3}$ \citep[a typical value of $\Cd$ over oceans is $\Cd \approx 1.5 \times 10^{-3}$; e.g.][Sect.~8.3.1]{Holton1973}, although it may be larger over rough ground. In the present work, this parameter is set to $\Cd = 3.4 \times 10^{-3}$, which is the value used by \cite{Wordsworth2015} and corresponding to $\roughheight \sim 10^{-2}$~m and $\zsurf \sim 10$~m. 

The characteristic horizontal wind speed varies with the surface and atmospheric temperatures. To take this dependence into account, \cite{Wordsworth2015} derived a relationship between $\Vconv$, $\Tatm$, and $\Tday$ from the thermodynamic equation by assuming the Weak Temperature Gradient (WTG) approximation, that is considering that winds are driven by small horizontal temperature gradients. This equation (Eq. (43) in the article) was combined with \eq{Fturb} to close the system of equations derived from hemisphere-averaged energy budgets. Later, by performing a series of numerical simulations using \kc{a} GCM, \cite{KA2016} showed that the atmospheric circulations of rocky planets resemble heat engines. Particularly, they established that the heat engine theory better fit horizontal wind speeds than the relationship derived by \cite{Wordsworth2015} in the explored region of the parameter space. We thus choose to follow here the prescription proposed by \cite{KA2016}. 

In this framework, the dayside convection is idealised as a single overturning cell between the substellar point and cooler regions. This cell is driven by the surface-atmosphere temperature difference, and the heat engine is fed by the absorption of radiative fluxes. The atmosphere absorbs heat near the dayside surface at the temperature $\Tday$ (hot reservoir) and emits it to space at the temperature $\Taday$ (cold reservoir) in the upper regions of the atmosphere. During the cycle, a fluid parcel works against friction in the boundary layer, which is the place where the energy is dissipated. This work is formulated as $\work = \Cd \rhoaday \Vconv^3$ \citep{BE1998,KA2016}. In the case of an isentropic cycle (i.e. composed of adiabatic reversible processes solely), $\work$ is related to the amount of power per unit area available to drive atmospheric motion $\Qconv$ by Carnot's theorem \rec{\citep[e.g.][]{Schroeder2000}}

\begin{equation}
\work = \effthermconv \Qconv, 
\end{equation}

\noindent the factor $\effthermconv \define \left( \Tday - \Taday \right) / \Tday$ being the atmosphere's thermodynamic efficiency. The typical horizontal wind speed of the convective cell is thus expressed as 

\begin{equation}
\Vconv = \effconv \left( \frac{ \effthermconv \Qconv}{\Cd \rhoaday} \right)^{\frac{1}{3}} \rec{,}
\label{Vconv}
\end{equation}

\noindent where we have introduced the efficiency coefficient $\effconv$ accounting for the non-isentropic nature of the thermodynamic cycle. For an isentropic cycle, $\effconv = 1$, which corresponds to the idealised Carnot's heat engine. In reality, the thermodynamic cycle is not isentropic because additional dissipative processes, such as diffusion, induce an irreversible production of entropy, which decreases the efficiency of the heat engine. As a consequence, the effective amount of energy available to drive atmospheric motion is smaller than its theoretical estimate, and $\effconv < 1$. \cite{KA2016} hence noted that the typical wind speeds given by their GCM simulations could be twice smaller than those predicted by the theory, that is $\effconv = 1/2$.

The power per unit area available to drive atmospheric motion ($\Qconv$) has to be specified from an ad hoc prescription. As it was benchmarked against GCM simulations, we \kc{follow} that proposed by \cite{KA2016}, which is formulated as 

\begin{equation}
\Qconv = 2 \Feq \Ksw \left( 1 - \expo{- \taugrlw} \right), 
\label{Qconv}
\end{equation}

\noindent with $\Feq \define \sigmaSB \Teq^4$, the black body equilibrium temperature $\Teq$ being defined in \eq{Teq}.

The first factor of the expression given by \eq{Qconv} is just the hemisphere-averaged power received by the planet on the dayside. The second factor is the vertically-integrated atmospheric transmission function in the shortwave. The third factor is \kc{a function of optical depth} scaling the fraction of flux emitted by the surface in the longwave that is absorbed by the atmosphere. Assuming $\taugrlw \ll 1$ leads to $1 - \expo{- \taugrlw} \approx \taugrlw$ and we recover the usual transfer function in the optically thin limit \citep[][]{Pierrehumbert2011}.

\begin{figure*}[htb]
   \centering
 \begin{flushleft}
   \hspace{0.17\textwidth} $\taugrlw = 10^{-2}$
  \hspace{0.17\textwidth} $\taugrlw = 10^{-1}$
\hspace{0.17\textwidth}  $\taugrlw = 10^{0}$ \\
 \end{flushleft}
 \vspace{-0.2cm}
    \includegraphics[height=0.26\textwidth,trim = 3.2cm 2.5cm 5.5cm 2.0cm,clip]{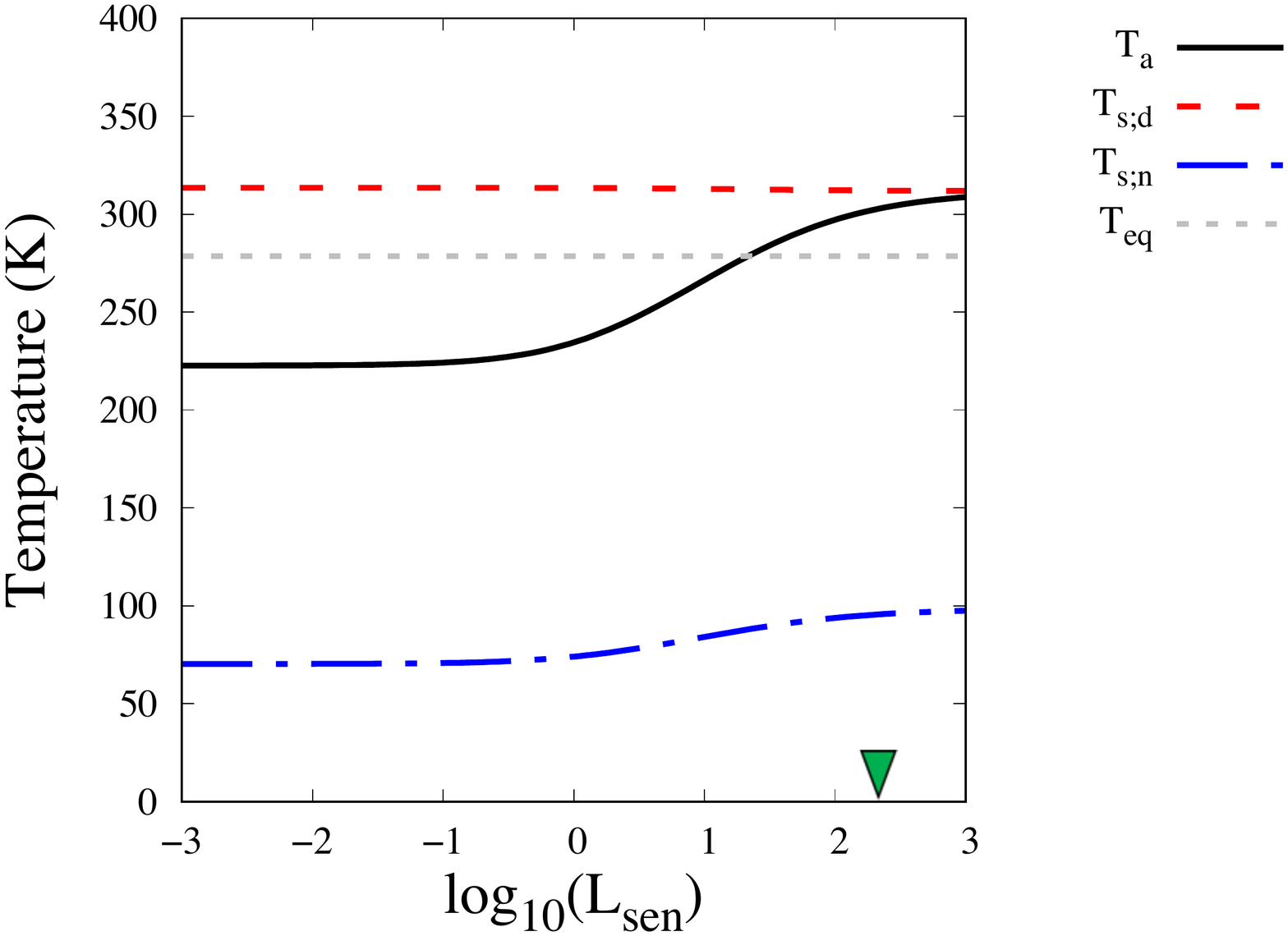} 
   \includegraphics[height=0.26\textwidth,trim = 5.0cm 2.5cm 5.5cm 2cm,clip]{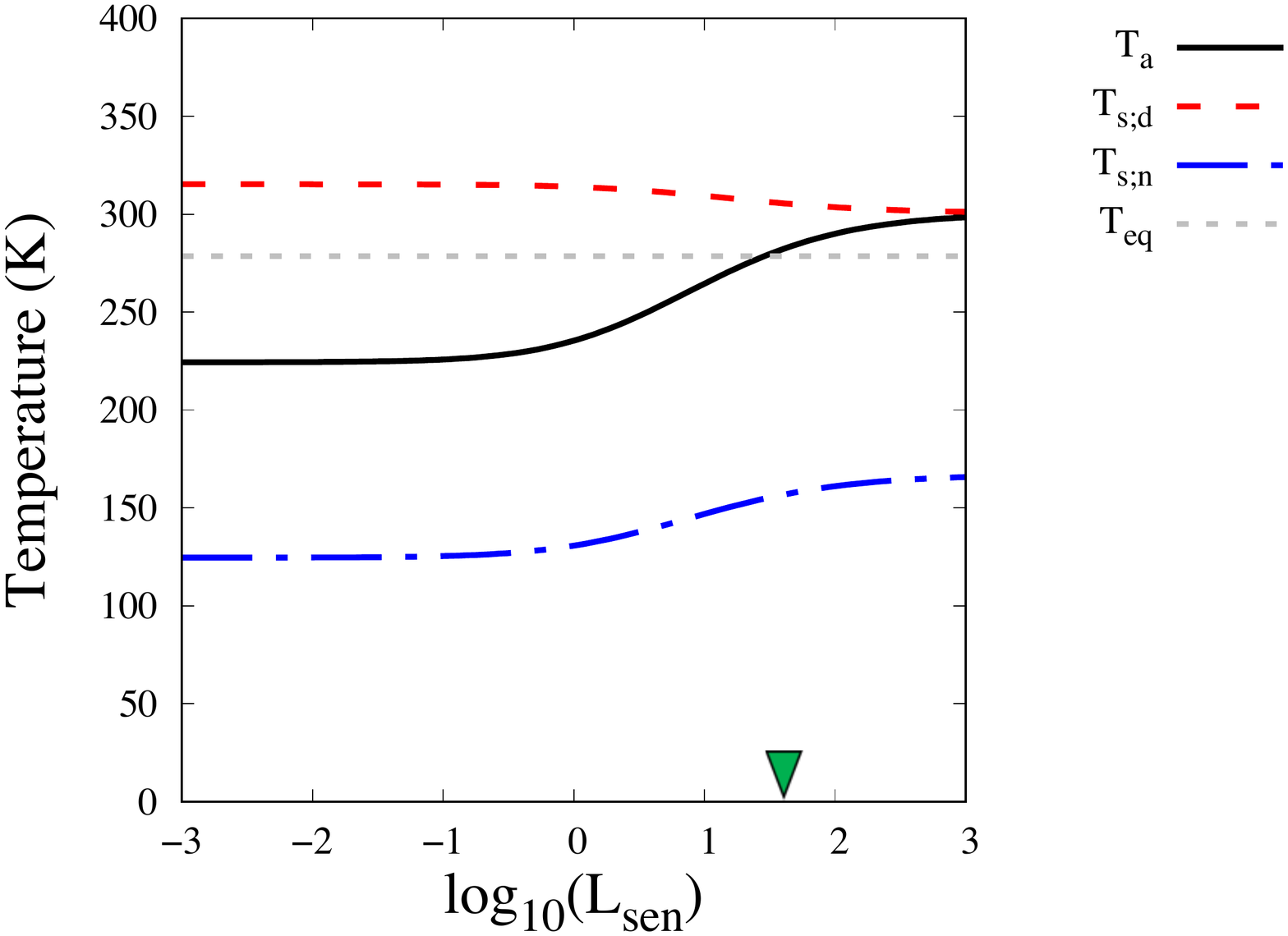} 
   \includegraphics[height=0.26\textwidth,trim = 5.0cm 2.5cm 5.5cm 2cm,clip]{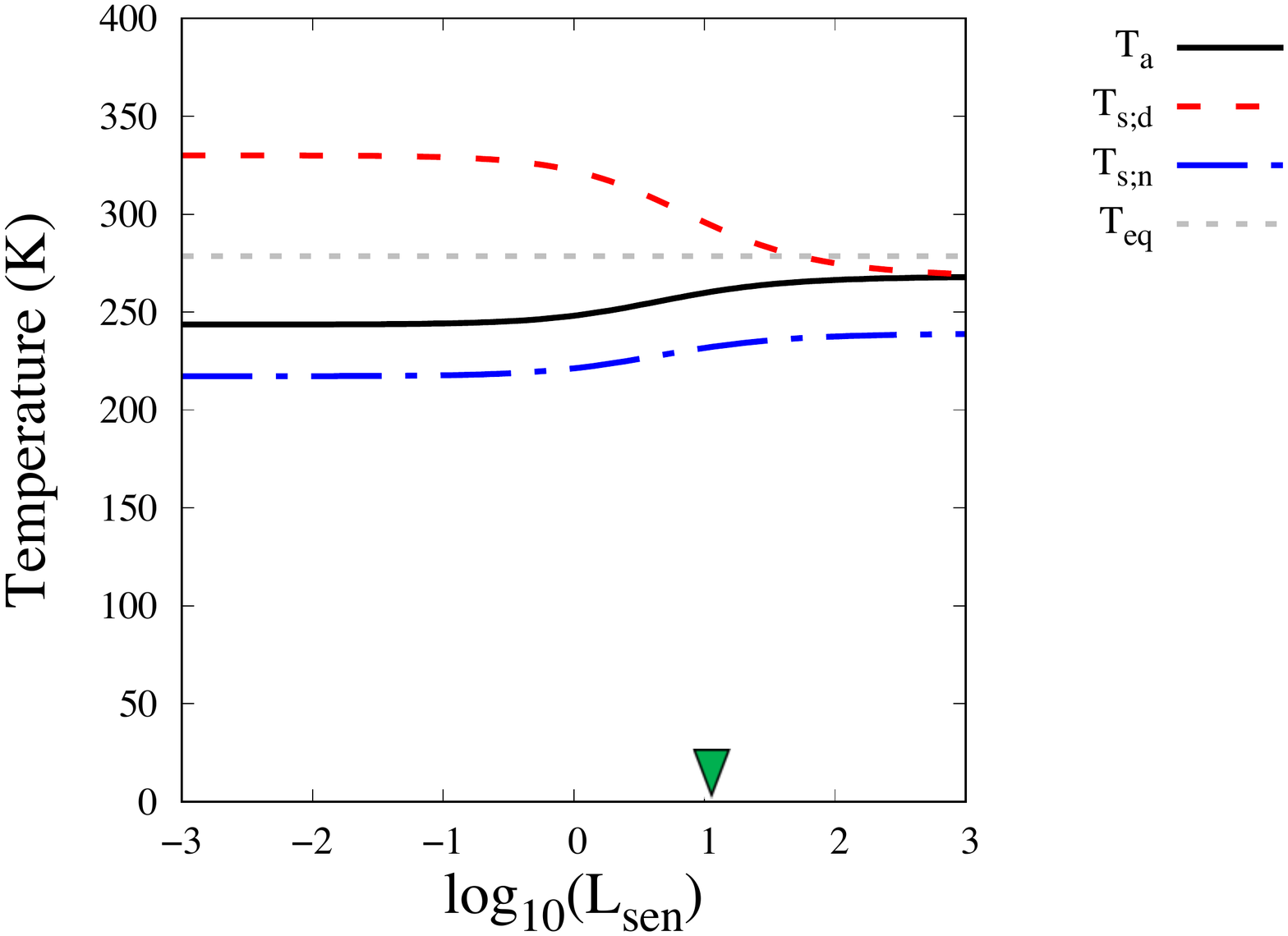}
    \includegraphics[width=0.07\textwidth,trim = 22.6cm 8cm 2cm 2cm,clip]{auclair-desrotour_fig6c}     
   \caption{Atmospheric, dayside and nightside surface temperatures as functions of logarithm of control parameter $\Lconv$. {\it Left:} $\taugrlw = 0.01$ (quasi-transparent). {\it Middle:} $\taugrlw = 0.1$ (small optical thickness). {\it Right:} $\taugrlw = 1$ (Earth-like optical thickness). In all cases, the atmosphere is quasi-transparent in the shortwave ($\taugrsw = 10^{-4}$), pure absorption is assumed ($\betalw = \betasw = 1$), $\Asurfsw = 0.2$, and $\Fstar = 1366$\units{W m^{-2}}. The atmospheric temperature is designated by the solid black line, the dayside surface temperature by the dashed red line, the nightside surface temperature by the dotted blue line, \rec{and the black body equilibrium temperature (\eq{Teq}) by the dotted grey line. The value of $\Lconv$ for a 1-bar $\carbondiox$-dominated atmosphere is indicated by the green triangle.}}
       \label{fig:temp_Lconv}%
\end{figure*}

\subsection{Role played by sensible heating}
\label{ssec:role_sensible_heating}

The formulation of the sensible heat flux given by \eqs{Fturb}{Vconv} closes the system formed by energy budget equations and allows us to determine the state of equilibrium of the system. It is convenient here to adopt the formalism of \cite{Wordsworth2015} since we follow the same approach as this early work. Thus, introducing the normalised temperature, radiative flux, and sensible heat flux, 

\begin{equation}
\begin{array}{lll}
\Tnorm \define \dfrac{\Taday}{\Tday}, & \Fnorm \define \dfrac{\Fstar}{2 \sigmaSB \Tday^4}, & \mbox{and} \ \Dconvnorm \define \dfrac{\meanDconv}{\Alw \sigmaSB \Tday^4},
\end{array}
\label{Tnorm_Fnorm}
\end{equation}

\noindent we transform \eqsthree{mean_budget_surf_day}{global_averaged_atm_budget}{Fturb} into 

\begin{align}
\label{thick_eq1}
\frac{\Klw}{\Alw} = & \ \frac{1 - \Agrsw}{\Alw} \Fnorm + \Tnorm^4 - \Dconvnorm , \\ 
\label{thick_eq2}
\left( 4 - \frac{\Alw}{\Klw} \right) \Tnorm^4 - 1 = & \ \frac{\Asw}{\Alw} \Fnorm + \Dconvnorm , \\
\label{thick_eq3}
\Dconvnorm = & \displaystyle \ \Lconv \Fnorm^{11/12} \Tnorm^{-2/3} \left( 1 - \Tnorm \right)^{4/3},
\end{align}

\noindent where the dimensionless parameter $\Lconv$ is defined as 

\begin{equation}
\Lconv \define \frac{2 \Cp \Cd \psurf \effconv}{\Alw \Rspec \Fstar} \left[ \frac{\Qconv \Rspec }{\Cd \psurf} \left( \frac{\Fstar}{2 \sigmaSB} \right)^{\frac{1}{4}} \right]^{\frac{1}{3}}. 
\label{Lconv}
\end{equation}

The parameter $\Lconv$ controls the intensity of sensible heating with respect to radiative heating on the dayside. The case $\Lconv = 0$ corresponds to pure radiative equilibrium (no turbulent heat exchanges), while $\Lconv \gg 1$ indicates a very strong turbulent mixing within the boundary layer, which leads to $\Tatm = \Tday$. The system of \eqsto{thick_eq1}{thick_eq3} is finally reduced to one single equation. By combining \eqs{thick_eq1}{thick_eq2} we first express the normalised stellar flux as a function of the normalised temperature,

\begin{equation}
\Fnorm = \frac{\Alw}{\Ksw} \left[ \left( 3 - \frac{\Alw}{\Klw} \right) \Tnorm^4 + \frac{\Klw}{\Alw} - 1  \right] \rec{.}
\label{Fnorm_Tnorm_thick2}
\end{equation}

\noindent Substituted in \eqs{thick_eq2}{thick_eq3}, this expression of $\Fnorm$ yields

\begin{align}
\label{eqsingle_noadv}
\left( 4 - \frac{\Alw}{\Klw} \right) \Tnorm^4 - 1 - \frac{\Asw}{\Ksw} \left[ \left( 3 - \frac{\Alw}{\Klw} \right) \Tnorm^4 + \frac{\Klw}{\Alw} - 1  \right] & \\
 -  \Lconv \left( \frac{\Alw}{\Ksw}  \right)^{\frac{11}{12}}  \left[ \left( 3 - \frac{\Alw}{\Klw} \right) \Tnorm^4 + \frac{\Klw}{\Alw} - 1  \right]^{\frac{11}{12}} \frac{\left( 1 - \Tnorm \right)^{\frac{4}{3}}}{\Tnorm^{\frac{2}{3}}} &   = 0 , \nonumber
\end{align}

\noindent which is the counterpart of Eq.~(45) in \cite{Wordsworth2015} with the heat engine approach proposed by \cite{KA2016}. This equation can be solved by using a combination of the dichotomy and secant methods \citep[e.g.][p.~449]{press2007numerical}. The obtained $\Tnorm$ is then substituted in \eq{Fnorm_Tnorm_thick2} and \eqsto{thick_eq1}{thick_eq3} to compute successively $\Fnorm$, $\Dconvnorm$, the atmospheric temperature, and the dayside and nightside surface temperatures. \kc{Numerical calculations are performed using \texttt{TRIP} \citep[][]{GL2011}.}

Figure~\ref{fig:temp_Lconv} shows the evolution of the atmospheric temperature ($\Tatm$), and the dayside ($\Tday$) and nightside ($\Tnight$) surface temperatures with the atmospheric opacity in the longwave -- quantified by $\taugrlw$ -- and the control parameter of sensible heating $\Lconv$. \kc{In this example,} pure absorption is assumed ($\betalw = \betasw = 1$), the atmosphere is quasi-transparent in the shortwave ($\taugrsw = 10^{-4}$), and parameters are set to $\Asurfsw = 0.2$ and $\Fstar = 1366$\units{W~m^{-2}}, which is the stellar flux received by the Earth. \rec{The value of $\Lconv$ in the typical case of a 1-bar $\carbondiox$-dominated atmosphere (green triangle) is estimated by taking $\psurf = 1$~bar, $\Rspec = 189$\units{J~kg^{-1}~K^{-1}}, $\Cp = 650 $\units{J~kg^{-1}~K^{-1}}, $\Cd = 3.4 \times 10^{-3}$, and $\effconv = 1/2$. }

 The figure highlights the role played by sensible heating. As $\Lconv$ increases, the thermodynamic state of equilibrium switches from the purely radiative regime ($\Lconv \ll1$) to a strongly convective regime ($\Lconv \gg 1$), where atmospheric and dayside surface temperatures are homogenised by sensible exchanges. \rec{Both in the optically thin ($\taugrlw \ll 1$) and thick ($\taugrlw \gg 1$) limits,} we retrieve for $\Tnorm$ a similar evolution with $\Lconv$ as that shown by Fig.~7 of \cite{Wordsworth2015}, which indicates that the two different prescriptions used to estimate the typical horizontal wind speed (heat engine and scaling analysis using the thermodynamic equation) give comparable results. As expected, increasing the atmospheric opacity tends to accentuate the greenhouse effect, and thus to increase the nightside surface temperature.

\section{Inclusion of large-scale advection}
\label{sec:advection}

\begin{figure}[htb]
   \centering
   \includegraphics[width=0.45\textwidth,trim = 6.0cm 0cm 0cm 3.0cm,clip]{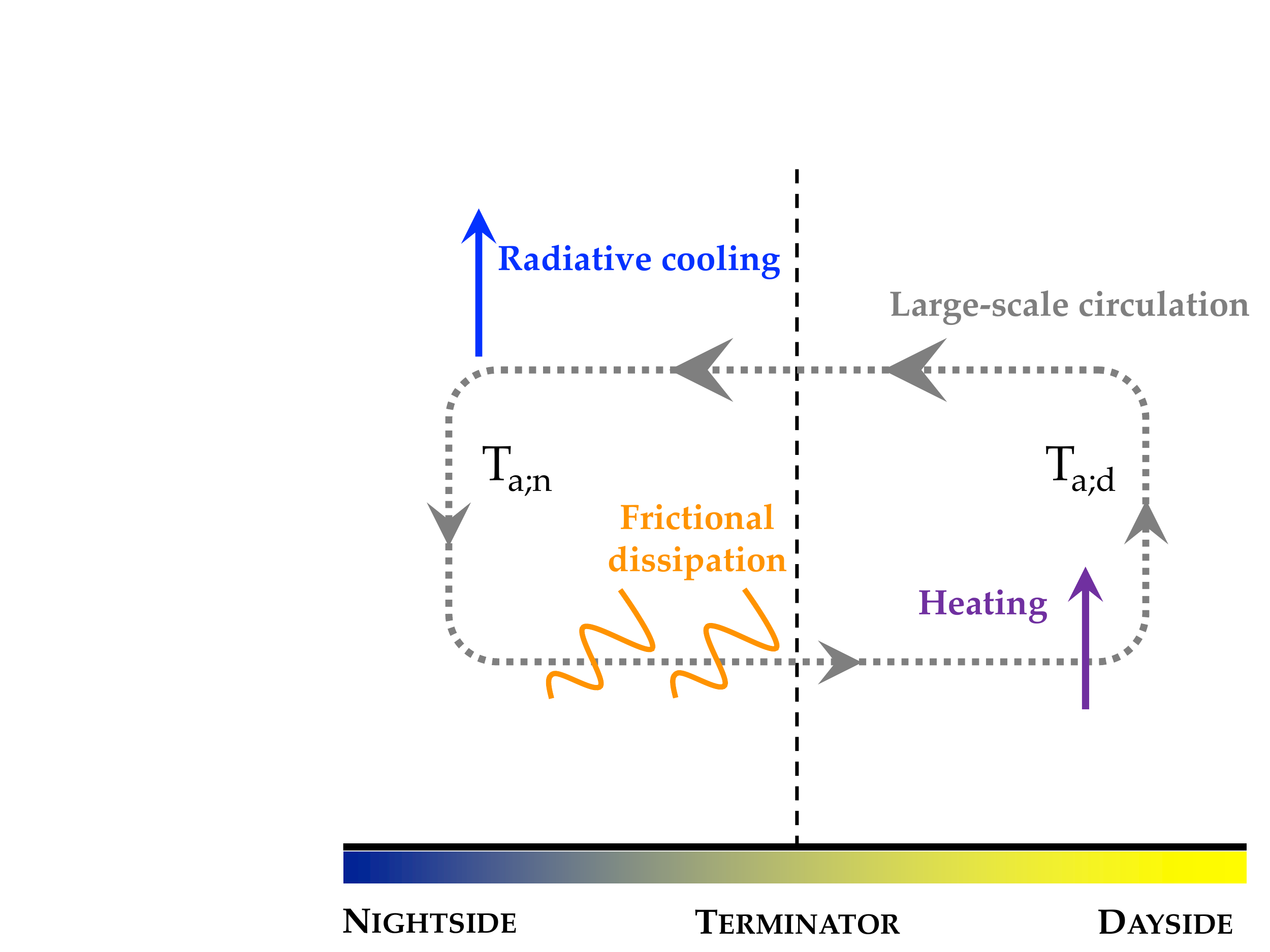}
      \caption{A diagram of large-scale stellar and anti-stellar atmospheric circulation as planetary heat engine. The heat engine is driven by dayside heating and nightside cooling to space. In this study, the atmospheric circulation is balancing Rayleigh drag, which causes frictional energy dissipation.}
       \label{fig:advection_cell}%
\end{figure}

The last physical ingredient that has to be introduced in the model is a mechanism responsible for heat transport from the dayside to the nightside. As discussed in \sect{ssec:heat_transport_circulation}, heat can be advected by large-scale atmospheric flows or transported by gravity waves \citep[e.g.][]{KA2015,KA2016}, which propagate in stably stratified fluid layers \citep[][]{GZ2008}. In the present work, we assume that heat transport is due to atmospheric circulation solely and we focus on the regime of slow rotators ($\Rossbyradnorm \gg 1$), where mean flows are symmetric with respect to the axis connecting the stellar and anti-stellar points. From now on, we relax the uniform temperature approximation and consider that $\Taday \neq \Tanight$ in the general case. First, we derive a scaling of the heat flux associated with advection $\Dadv$ as a function of the dayside and nightside atmospheric temperatures by using the heat engine theory (\fig{fig:advection_cell}). Second, we use this expression to derive and solve the equilibrium equation in the general case. 

\subsection{Scaling of the advected heat flux}

To quantify the atmospheric heat transport, we proceed to a standard scale analysis \citep[e.g.][Sect.~9.2.3, p.~610]{Pierrehumbert2011}. The heat per unit mass contained in a fluid \kc{parcel} starting from the dayside is $\Cp \Taday$. When the \kc{parcel} moves to the nightside, it decays to $\Cp \Tanight$. The energy released by the \kc{parcel} thus corresponds to $\Cp \left( \Taday - \Tanight \right)$. Moreover, the mass flow that goes across the annulus separating the day- and nightside is proportional to the air column mass $\psurf/\ggravi$, to the size of the annulus $2 \pi \Rpla$, and to a typical wind speed $\Vadv$. According to these considerations, the hemisphere-averaged heat flux due to atmospheric circulation may be written as  

\begin{equation}
\meanDadv = \effcirc \frac{\psurf \Cp }{\ggravi \Rpla} \Vadv \left( \Taday - \Tanight \right),
\label{meanDadv}
\end{equation}

\noindent where $\effcirc$ is an efficiency coefficient related to the three-dimensional geometry of the flow. The wind speed $\Vadv$ has now to be specified as a function of the atmospheric temperatures and of the system parameters. \rec{In order to clearly separate the mechanisms responsible for day-night heat transport on the one hand, and for sensible heat exchanges with the surface on the other hand, we choose to adopt a scaling law for the advection velocity different from that used to parametrise the convective cell (\eq{Vconv}). We note however that the two mechanisms could be linked to each other by adopting the same scaling for both velocities.}

\cite{KK2018} showed that the large-scale atmospheric circulation of hot Jupiters can be modelled as planetary heat engines and derived the typical wind speeds of mean flows in this framework. The reasoning applied to hot Jupiters in their work holds for rocky planets, and is very similar to that followed in \sect{ssec:scaling_sensible_heating} to derive the expression of the sensible heating flux used in the model (\eq{thick_eq3}). Here, the hot and cold reservoirs of the heat engine are the dayside and nightside parts of the atmosphere, and the temperature gradient driving the flow is thus horizontal instead of being vertical. In steady state, the work generated by differential heating and cooling is balanced by dissipation of kinetic energy, which determines the typical wind speed of the flow. In the present study, we assume that the atmospheric circulation is balancing Rayleigh drag \recc{in the terminator region}. As a consequence, wind speeds scale as \citep[e.g.][Eq.~(11)]{KK2018}

\begin{equation}
\Vadv = \effadv \left( \effthermadv \Qadv \frac{\tdrag \ggravi }{\psurf} \right)^{\frac{1}{2}},
\label{Vadv}
\end{equation}

\noindent where $\effthermadv \define \left( \Taday - \Tanight \right) / \Taday $ designates Carnot's thermodynamic efficiency for the large-scale circulation, $\effadv$ the efficiency coefficient associated with the non-isentropic nature of the cycle, $\Qadv$ the \rec{theoretical} amount of power available to drive atmospheric motion, given by \eq{Qconv}, and $\tdrag$ the drag timescale, that is the timescale over which winds are linearly damped by Rayleigh drag. Substituting $\Vadv$ by \eq{Vadv} in \eq{meanDadv}, introducing the total efficiency factor $\efftot  \define \effcirc \effadv $, and extracting the dependence on the dayside and nightside atmospheric temperatures yields

\begin{equation}
\meanDadv = \efftot \frac{\psurf \Cp }{\ggravi \Rpla} \left(\Qadv \frac{\tdrag \ggravi }{\psurf} \right)^{\frac{1}{2}} \Taday \left( 1 - \frac{\Tanight}{\Taday} \right)^{\frac{3}{2}},
\label{meanDadv1}
\end{equation}

\noindent where we retrieve the scaling $\meanDadv / \Feq \scale \trad / \tadv$ given by early studies \citep[e.g.][]{Leconte2013}. We remind \kc{ourselves} here that $\trad$ and $\tadv$ are the radiative and advective timescales defined by \eq{trad_tadv}. The ratio $\trad / \tadv$ quantifies the ability of the stellar and anti-stellar circulation to transport heat from the dayside to the nightside. The case $\trad / \tadv = + \infty$ corresponds to the regime treated by \cite{Wordsworth2015} and \sect{sec:sensible_heating} of the present work, where advection is so strong that the atmosphere is horizontally well mixed and its temperature homogenised. Conversely, if $\trad / \tadv \ll 1$, advection is not efficient in warming the nightside, which leads to strong day-night temperature gradients. \rec{We note that the drag timescale ($\tdrag$) and the efficiency factor ($\efftot$) appearing in \eq{meanDadv1} are unknown a priori. Both these parameters and the scaling law used for $\meanDadv$ should eventually be constrained with the help of more sophisticated models solving the coupled momentum and thermodynamic equations, such as GCMs typically. }

\subsection{Steady state equation in the general case}

We can now write the system of equations describing the steady state in the general case, where \kc{radiative transfer}, sensible heating and large-scale circulation are taken into account (\fig{fig:model_overview}). In coherence with the formalism employed in \sect{ssec:role_sensible_heating}, we introduce the normalised temperature and heat flux,

\begin{equation}
\begin{array}{ll}
\Tanorm \define \dfrac{\Tanight}{\Taday}, & \mbox{and} \  \Dadvnorm \define \dfrac{\meanDadv}{\Alw \sigmaSB \Tday^4}. 
\end{array}
\end{equation}

The equations \eqsthree{mean_budget_surf_day}{global_averaged_atm_budget}{mean_budget_atm_night} thus become

\begin{align}
\label{thickgen_eq1}
\frac{\Klw}{\Alw} = & \ \frac{1 - \Agrsw}{\Alw} \Fnorm + \Tnorm^4 - \Dconvnorm , \\ 
\label{thickgen_eq2}
\left( 4 - \frac{\Alw}{\Klw} \right) \Tnorm^4 - 1 = & \ \frac{\Asw}{\Alw} \Fnorm + \Dconvnorm -  \Dadvnorm, \\
\label{thickgen_eq3}
- \left( 2 - \frac{\Alw}{\Klw} \right) \Tanorm^4 \Tnorm^4 + \Dadvnorm  = & \ 0, 
\end{align}

\noindent where the normalised sensible heating and advection fluxes are expressed as 

\begin{align}
\label{thickgen_eq4}
\Dconvnorm = & \displaystyle \ \Lconv \Fnorm^{11/12} \Tnorm^{-2/3} \left( 1 - \Tnorm \right)^{4/3}, \\
\label{thickgen_eq5}
\Dadvnorm = & \displaystyle \ \Ladv \Tnorm \Fnorm^{3/4} \left( 1 - \Tanorm \right)^{3/2}. 
\end{align}

\noindent In the preceding equation, we have introduced the non-dimensional parameter 

\begin{equation}
\Ladv \define  \efftot \frac{\psurf \Cp }{\ggravi \Rpla \sigmaSB } \left( \frac{2 \sigmaSB}{\Fstar} \right)^{\frac{3}{4}} \left(\Qadv \frac{\tdrag \ggravi } {\psurf} \right)^{\frac{1}{2}} ,
\label{Ladv}
\end{equation}

\noindent which controls the intensity of heating due to atmospheric circulation with respect to radiative cooling. This parameter is complementary with $\Lconv$, defined by \eq{Lconv}, which controls the intensity of sensible heating with respect to radiative cooling \kc{on dayside}. 

\begin{figure}[htb]
   \centering
   \includegraphics[width=0.48\textwidth,trim = 1.5cm 3cm 1cm 2cm,clip]{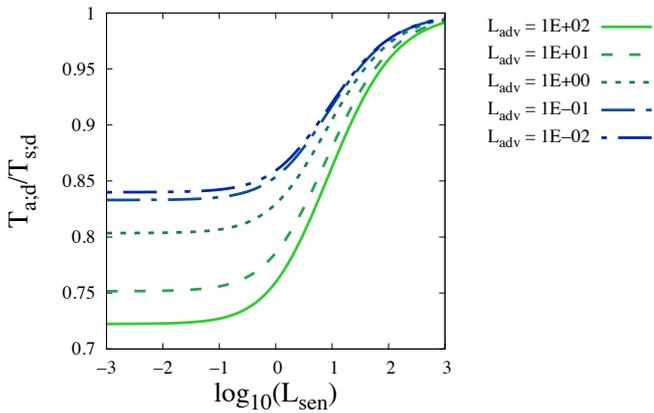}
      \caption{Normalised temperature $\Tnorm = \Taday / \Tday$ as function of logarithm of parameter $\Lconv$ for various values of $\Ladv$. The parameter $\Ladv$, which controls advective heat transport by the stellar and anti-stellar circulation, varies in the range $\logdix \left( \Ladv \right) = -2, \ldots , 2$ (from dark blue to green).  The atmosphere is quasi-transparent in the shortwave ($\taugrsw = 10^{-4}$), pure absorption is assumed ($\betalw = \betasw = 1$), $\Asurfsw = 0.2$, and $\Fstar = 1366$\units{W m^{-2}}, similarly as in \fig{fig:temp_Lconv}. Here, the atmosphere is optically thin in the longwave: $\taugrlw = 0.1$, which corresponds to the middle panels of \fig{fig:temp_Lconv}.    }
       \label{fig:tempnorm_Ladv}%
\end{figure}

\begin{figure*}[htb]
   \centering
    \begin{flushleft}
   \hspace{0.20\textwidth} $\Ladv = 10^{-2}$
  \hspace{0.17\textwidth} $\Ladv = 10^{0}$
\hspace{0.17\textwidth}  $ \Ladv = 10^{2}$ 
 \end{flushleft}
  \vspace{-0.2cm}
    \raisebox{4.5cm}{$\taugrlw = 0.1$}
    \hspace{-0.2cm}
    \includegraphics[width=0.023\textwidth,trim = 3.2cm 4cm 23.3cm 2cm,clip]{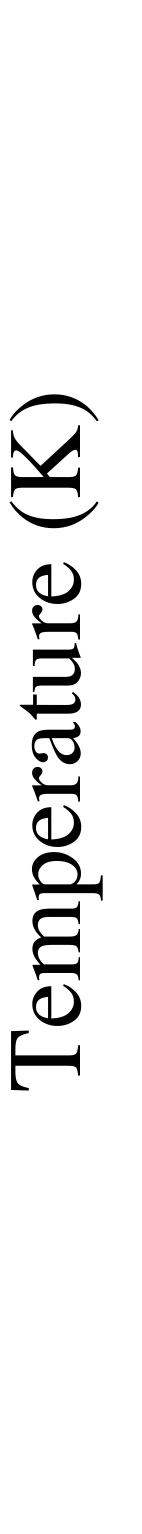} 
    \includegraphics[width=0.26\textwidth,trim= 5.0cm 4cm 7.0cm 2cm,clip]{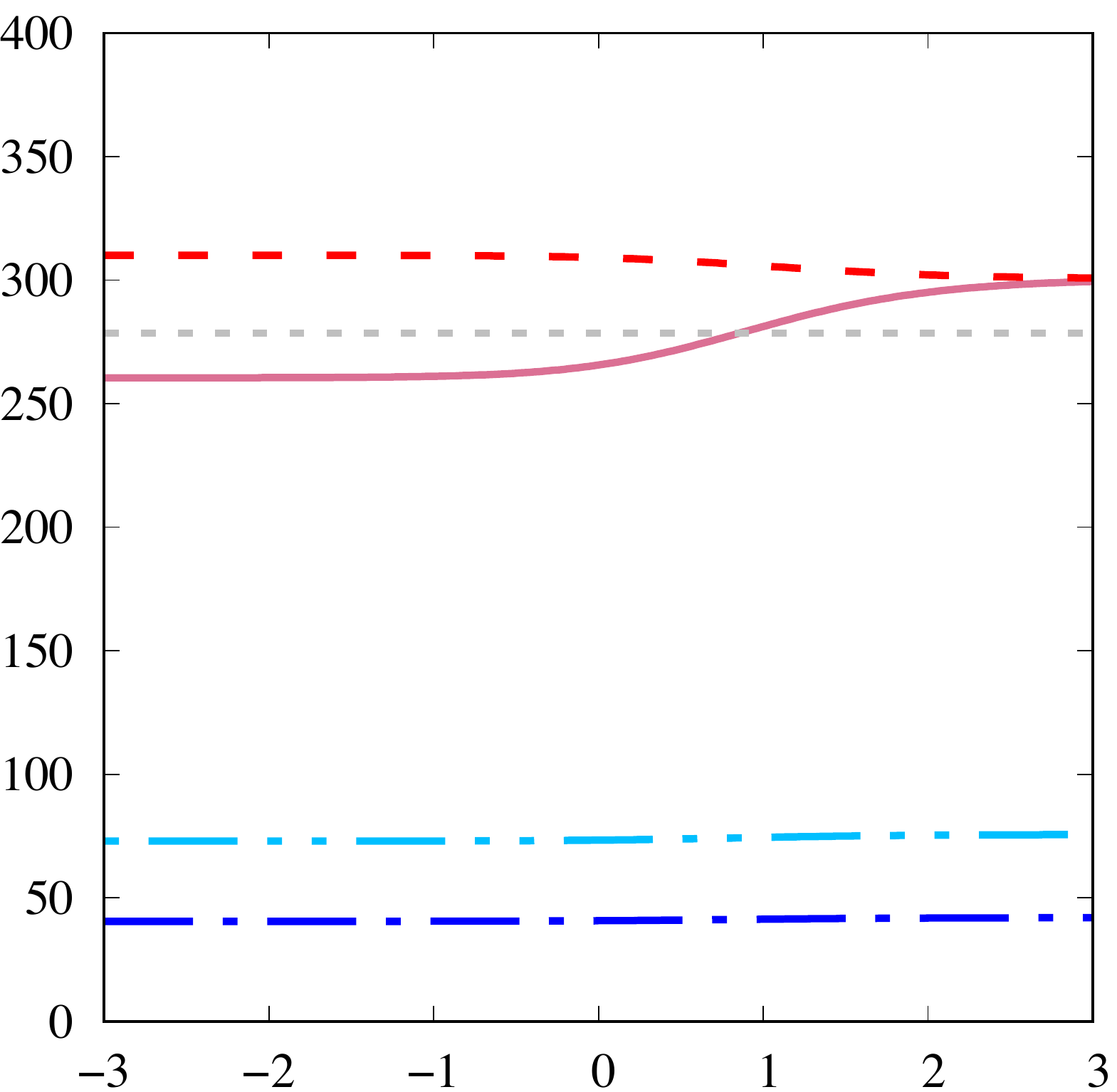} 
   \includegraphics[width=0.26\textwidth,trim = 5.0cm 4cm 7.0cm 2cm,clip]{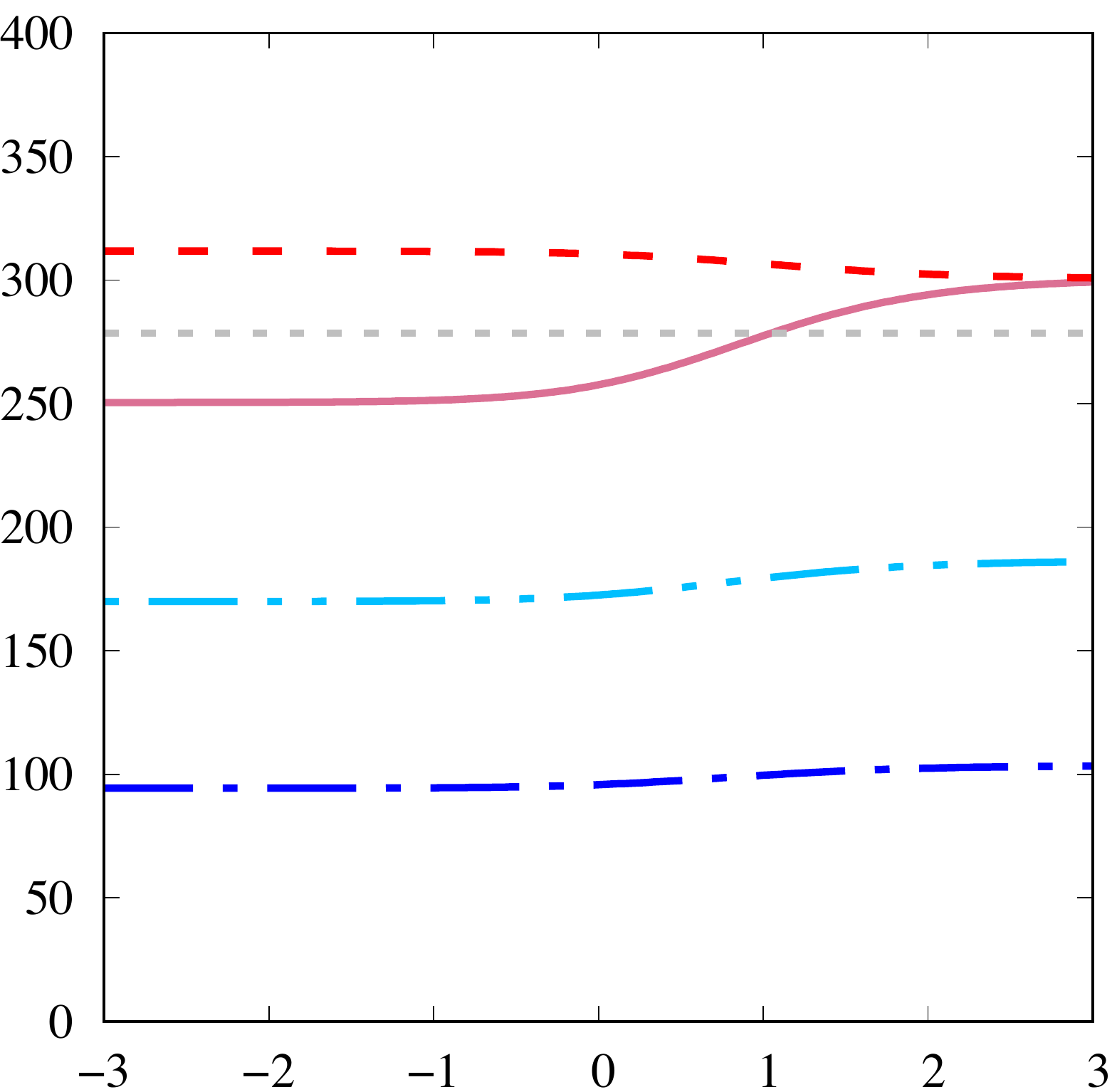} 
   \includegraphics[width=0.26\textwidth,trim = 5.0cm 4cm 7.0cm 2cm,clip]{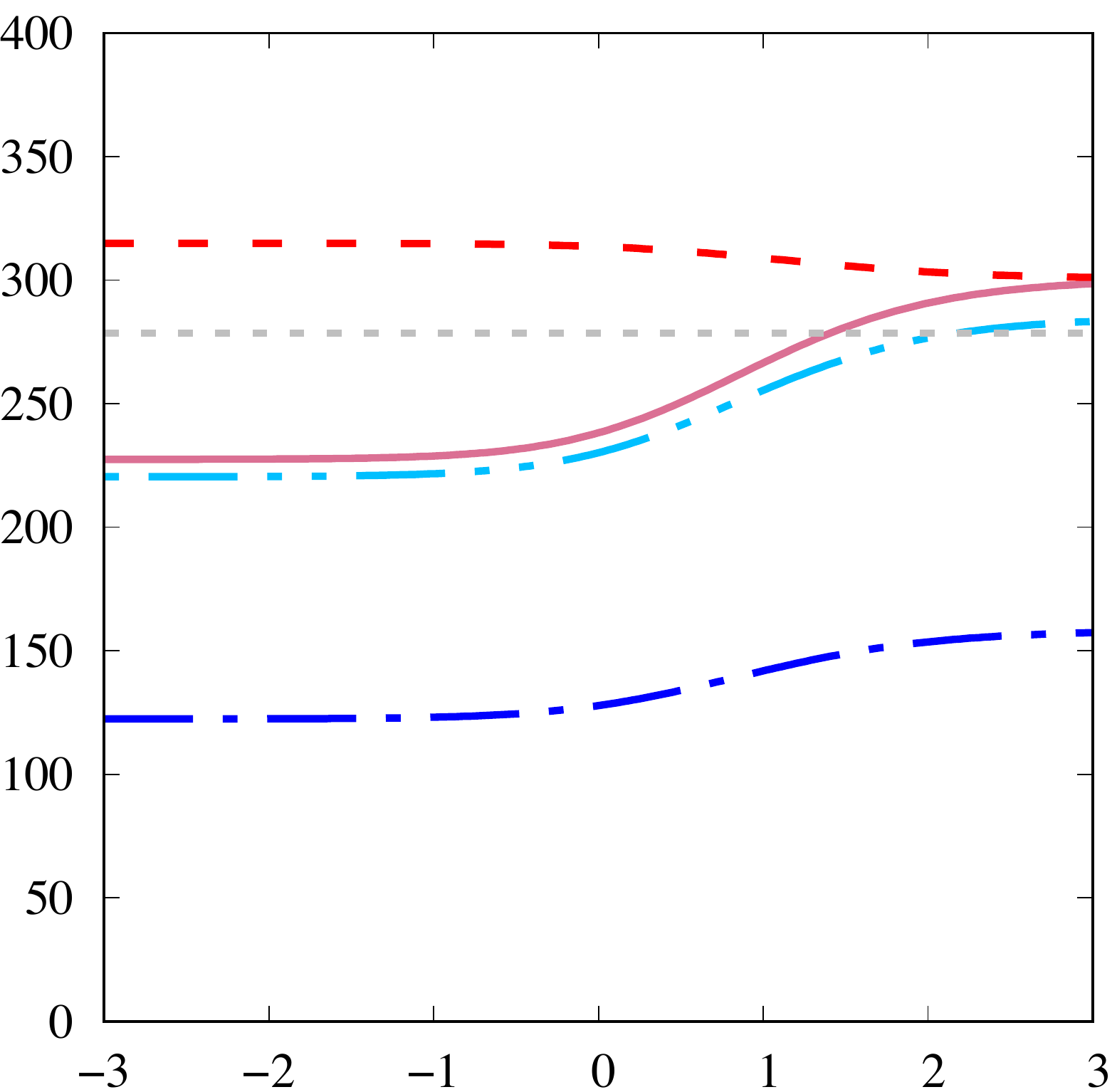}
    \includegraphics[width=0.07\textwidth,trim = 22.6cm 8cm 2cm 2cm,clip]{auclair-desrotour_fig9d}  \\
    \hspace{-0.07\textwidth}
    \raisebox{4.5cm}{$\taugrlw = 1.0$}
    \hspace{-0.2cm}
    \includegraphics[width=0.023\textwidth,trim = 3.2cm 4cm 23.3cm 2cm,clip]{auclair-desrotour_fig9a} 
     \includegraphics[width=0.26\textwidth,,trim =  5.0cm 2.5cm 7.0cm 2cm,clip]{auclair-desrotour_fig9a} 
   \includegraphics[width=0.26\textwidth,trim = 5.0cm 2.5cm 7.0cm 2cm,clip]{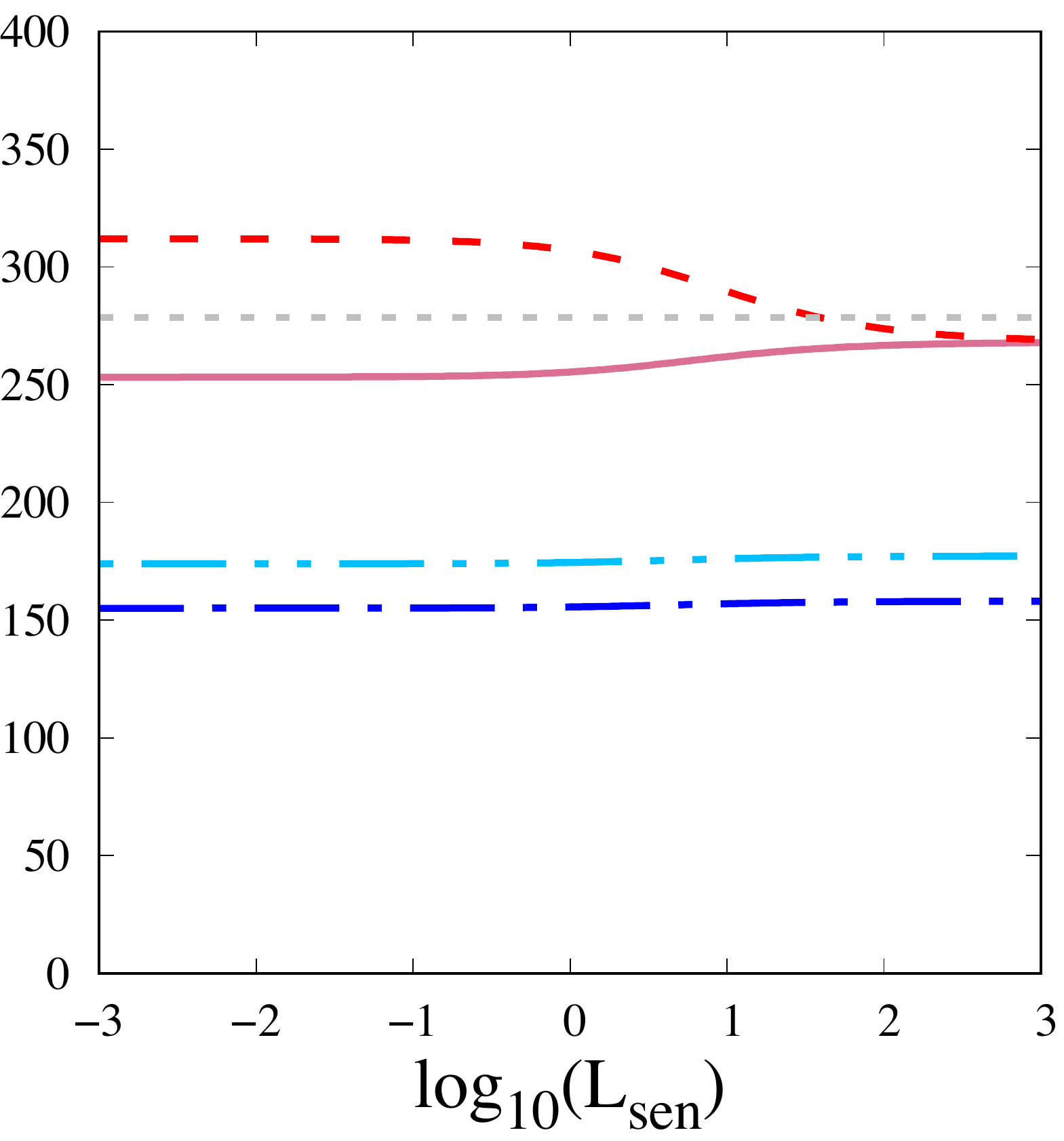} 
   \includegraphics[width=0.26\textwidth,trim = 5.0cm 2.5cm 7.0cm 2cm,clip]{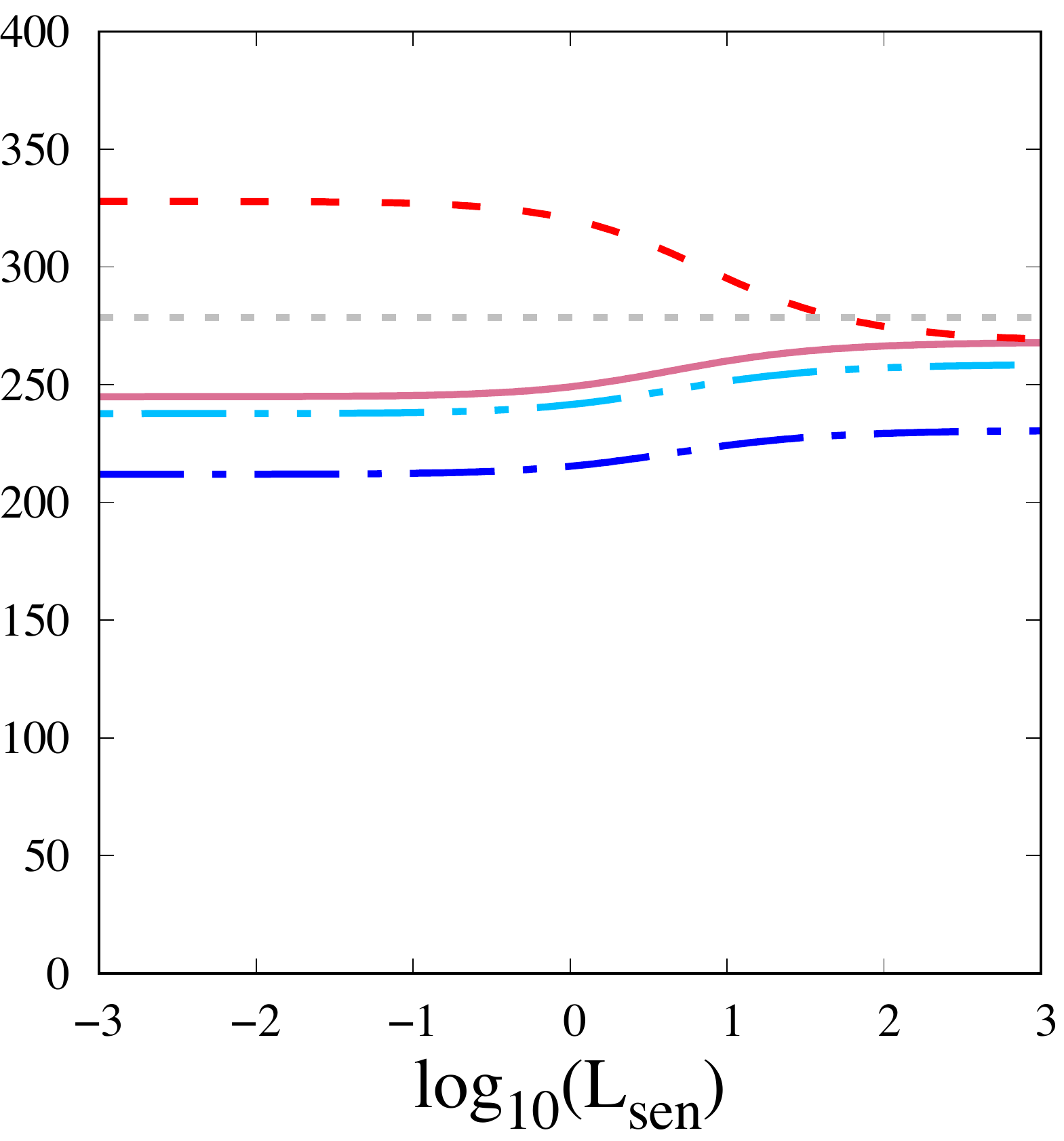}\\
   \caption{Atmospheric, dayside and nightside surface temperatures as functions of control parameters $\Lconv$ and $\Ladv$ for optically thin ($\taugrlw = 0.1$, top panels) and thick ($\taugrlw = 1.0$, bottom panels) atmospheres in longwave. {\it Left:} $\Ladv = 0.01$ (weak circulation). {\it Middle:} $\Ladv = 1.0$ (medium-strength circulation). {\it Right:} $\Ladv = 100$ (strong circulation). In all cases, the atmosphere is quasi-transparent in the shortwave ($\taugrsw = 10^{-4}$), pure absorption is assumed ($\betalw = \betasw = 1$), $\Asurfsw = 0.2$, and $\Fstar = 1366$\units{W m^{-2}}, similarly as in \fig{fig:temp_Lconv}. The dayside atmospheric (solid pink line) and surface (dashed red line) temperatures, as well as the nightside atmospheric (dotted cyan line) and surface (dashed blue line) temperatures are plotted as a function of the logarithm of $\Lconv$. The notation $\logdix$ designates the decimal logarithm.}
       \label{fig:temp_Ladv}%
\end{figure*}

The system of \eqsto{thickgen_eq1}{thickgen_eq5} may be reduced to one single equation by proceeding similarly as in \sect{ssec:role_sensible_heating}. First, $\Fnorm $ is expressed as a function of $\Tanorm$ and $\Tnorm$ by substituting $\Dadvnorm$ by \eq{thickgen_eq5} in \eq{thickgen_eq3},

\begin{equation}
\Fnorm = \left( \frac{2 - \frac{\Alw}{\Klw} }{\Ladv} \right)^{\frac{4}{3}} \frac{  \Tanorm^{16/3} \Tnorm^4 }{\left( 1 - \Tanorm \right)^2}. 
\label{Fnorm_Tnorm_Tanorm}
\end{equation}

Second, \eqs{thickgen_eq1}{thickgen_eq2} are combined to eliminate $\Dconvnorm$, and the resulting equation is combined with \eq{thickgen_eq5} to eliminate $\Dadvnorm$. This yields the expression of $\Tnorm$ as a function of $\Tanorm$,

\begin{equation}
\Tnorm =  \left( \frac{ \displaystyle \frac{\Klw}{\Alw} - 1}{ \displaystyle \Tanorm^4 \left[  \frac{\Ksw}{\Alw} \left( \frac{2 - \frac{\Alw}{\Klw} }{\Ladv} \right)^{\frac{4}{3}} \frac{\Tanorm^{\frac{4}{3}}}{\left( 1 - \Tanorm \right)^2} + \frac{\Alw}{\Klw}-2  \right] - 1 } \right)^{\frac{1}{4}}. 
\label{Tnorm_Tanorm}
\end{equation}

Third, $\Tnorm$ is substituted by \eq{Tnorm_Tanorm} in \eq{Fnorm_Tnorm_Tanorm}, and both $\Tnorm$ and $\Fnorm$ are substituted by their expressions as functions of $\Tanorm$ in the normalised fluxes (\eqs{thickgen_eq4}{thickgen_eq5}), which allows us to reduce the system of \eqsto{thickgen_eq1}{thickgen_eq5} to a single equation of $\Tanorm$. By using \eq{thickgen_eq1}, we get 

\begin{align}
\label{eqsingle_gen}
 \left[ \frac{1 - \Agrsw}{\Alw} \left( \frac{2 - \frac{\Alw}{\Klw} }{\Ladv} \right)^{\frac{4}{3}} \frac{  \Tanorm^{16/3}  }{\left( 1 - \Tanorm \right)^2} + 1 \right] \Tnorm^4 \left( \Tanorm \right) - \frac{\Klw}{\Alw} & \\
      - \Lconv  \left( \frac{2 - \frac{\Alw}{\Klw} }{\Ladv} \right)^{\frac{11}{9}} \left(  \frac{\Tanorm^{\frac{8}{3}}}{1 - \Tanorm} \right)^{\frac{11}{6}} \left[ 1 - \Tnorm \left( \Tanorm \right) \right]^{\frac{4}{3}} \Tnorm^3 \left( \Tanorm \right)  & =  0.  \nonumber
\end{align}

This equation can be solved using a combination of the dichotomy and secant methods \citep[][]{press2007numerical}, as previously done for \eq{eqsingle_noadv}. However, we note that the denominator of \eq{Tnorm_Tanorm} is negative if $\Tanorm$ is less than a minimum value $\Tanorminf$, meaning that $\Tnorm$ is not defined in this range. \kc{This minimum value simply results from the fact that the advected heat flux cannot be greater than the heat absorbed on the dayside hemisphere. As the heat transport by large-scale advection increases as $\Tanorm$ decays (see \eq{thickgen_eq5}), a maximum of $\Dadvnorm$ corresponds to a minimum of $\Tanorm$, which is $\Tanorminf$. We note that $\Tanorminf \ll 1 $ if the efficiency of advection is low ($\Ladv \ll1$), while \rec{$\Tanorminf \approx 1$} if the efficiency is high ($\Ladv \gg 1$).} \kc{As a consequence, one has to preliminarily} calculate $\Tanorminf$ by finding the zero of the denominator of $\Tnorm$ in \eq{Tnorm_Tanorm}. As a second step, the solution of \eq{eqsingle_gen} is sought within the interval $\Tanorminf< \Tanorm \leq 1$ considering the fact that the nightside atmospheric temperature cannot be greater than the dayside atmospheric temperature in this model. Although there is no evidence for the existence and unicity of solutions from an analytical point of view, we verify them numerically in the studied range of the parameter space (see \append{app:solving_equation}). 

However, as $\Ladv \rightarrow + \infty$, the dayside and nightside atmospheric temperatures tend to join together, meaning that  very small temperature variations lead to huge variations of the flux associated with atmospheric circulation ($\Dadvnorm$). In other words, the day-night atmospheric temperature difference do not contain information any more in the strong circulation limit, and one should drop \eqs{thickgen_eq3}{thickgen_eq5}, which corresponds to the case treated in \sect{ssec:role_sensible_heating} and by \cite{Wordsworth2015}. This degeneracy has repercussions on the numerical \kc{solution} of \eq{eqsingle_gen}, which cannot be performed beyond a certain upper bound of $\Ladv \gg 1$ (typically $\Ladv \sim 10^{2} - 10^{3}$). We also note that the logarithm of the distance to $\Tanorminf$ is a better appropriate coordinate than $\Tanorm$ itself in the root-finding procedure because of the strong variation of the function given by \eq{eqsingle_gen} in the vicinity of $\Tanorminf$ (see \fig{fig:function_Tanorm}). 

Figure~\ref{fig:tempnorm_Ladv} shows the evolution of the normalised temperature $\Tnorm = \Taday / \Tday$ (\eq{Tnorm_Fnorm}) as a function of the parameters controlling sensible heating ($\Lconv$) and large-scale heat transport ($\Ladv$) in the optically thin case of \fig{fig:temp_Lconv} ($\taugrlw = 0.1$, middle panels). We observe here the transition between the asymptotic limits of weak ($\Ladv \ll 1$) and strong ($\Ladv \gg 1$) atmospheric circulations. In the regime of strong advection, the solution \kc{resembles that obtained by \cite{Wordsworth2015} (see Fig.~7, top panel)}. Conversely, while the circulation weakens, a larger amount of heat remains on the planet's dayside. As a consequence, the temperature difference between the surface and the atmosphere decays. 

In \fig{fig:temp_Ladv}, the four temperatures of the system in steady state are plotted as functions of $\Lconv$ for $\Ladv = 0.01,1,100$ in the optically thin ($\taugrlw = 0.1$) and thick ($\taugrlw = 1.0$) regimes. This figure shows that the nightside surface temperature strongly depends on $\Ladv$, while it is not very sensitive to $\Lconv$. As discussed above, increasing the efficiency of heat transport by the atmospheric circulation leads the nightside atmospheric and surface temperature to increase. In the regime of strong circulation ($\Ladv \gg 1$), the atmospheric nightside and dayside temperatures join together and we recover the results obtained in \fig{fig:temp_Lconv}, where $\Ladv = + \infty$. The dependence of the nightside temperature on the atmospheric circulation is increased by greenhouse effect. In the optically thick case the temperature evolution between the cases $\Ladv = 0.01 $ and $\Ladv = 100$ is larger than in the optically thick one.

\section{Atmospheric stability}
\label{sec:atmospheric_stability}

The analytical developments made in the preceding sections allow us to characterise the stability of the atmosphere, that is its ability to conserve its greenhouse gases and, more generally, to remain unchanged in composition. In the following, we use the model (\eq{eqsingle_gen}) in parallel with that proposed by \cite{Wordsworth2015} (Eq.~(45) of the article) to quantify the effects of the different processes participating to global heat redistribution on atmospheric stability. 

The collapse is triggered by the condensation of one of the gases present in the atmospheric mixture, which occurs from the moment that the temperature of the air goes below the condensation temperature of the gas. The gas then condensates and forms an ice sheet at the planet's surface in the cold trap. This modifies the radiative properties of the atmosphere, as well as the atmospheric circulation, which are both coupled to the atmospheric gas mixture. In the standard case of greenhouse gases, collapse acts as a positive feedback by leading the atmosphere to cool down, and favouring thereby the condensation of other gases. 

In the simplified approach of the present work, the complex dynamics of the collapse is not treated and the cold trap is assumed to be the whole nightside surface. We thus consider that, because of atmospheric circulation, a fluid \kc{parcel} will end up cooling down at the nightside surface temperature, and we do not investigate the timescale of the process consequently. Under these hypotheses, the atmospheric stability is simply determined by the ratio between the nightside surface temperature and the lowest condensation temperature among the greenhouse gases present in the mixture, following early analytical studies \citep[][]{Wordsworth2015,KA2016}.

In order to better understand the role played by the different involved processes, we use as a reference case the case studied by \cite{Wordsworth2015}, which is an Earth-sized planet hosting a $\carbondiox$-dominated atmosphere. In this case, the atmosphere is assumed  to be stable with respect to atmospheric collapse if $\Tnight > \Tcondcarbdiox \left( \psurf \right) $, the condensation temperature of $\carbondiox$ in~K being given, below the triple point ($\pressure < 5.18 \times 10^{5}$\units{Pa}), by \citep[][]{Fanale1982,Wordsworth2010b,Wordsworth2015}

\begin{equation}
\Tcondcarbdiox = \dfrac{3167.8}{23.23 - \ln \left( 0.01 \pressure \right) },
\label{TcondCO2_1}
\end{equation}

\noindent and, beyond the triple point ($\pressure \geq 5.18 \times 10^{5}$\units{Pa}), by 

\begin{equation}
\Tcondcarbdiox = 684.2 - 92.3 \ln \left( \pressure \right) + 4.32 \ln^2 \left( \pressure \right).
\label{TcondCO2_2}
\end{equation}

To specify the relation between the short- and longwave optical thicknesses and surface pressure, we use a linear law of the form \citep[][Eq.~(12)]{Wordsworth2015}

\begin{equation}
\label{taugr}
\begin{array}{ll}
 \taugrsw  = \dfrac{2  \klw \psurf }{\ggravi \betasw} , & \mbox{and} \ \taugrlw  = \dfrac{2  \klw \psurf }{\ggravi \betalw},
\end{array}
\end{equation}

\noindent where $\ksw$ and $\klw$ are the supposed constant shortwave and longwave opacities, respectively. The longwave opacity is set to $\klw = 5 \times 10^{-5}$\units{m^2~kg^{-1}}, which is the value used by \cite{Wordsworth2015} in GCM simulations. This value corresponds to $\taugrlw \approx 1$ for a $1$-bar atmosphere. The shortwave opacity is set to $\ksw = 10^{-9}$\units{m^2~kg^{-1}}, that is a value corresponding to the quasi-transparent regime ($\ksw \ll \klw$). We note that pure absorption is assumed in the reference case ($\betasw = \betalw = 1$). Finally, the parameters characterising heat engines are specified from the observations made by \cite{KA2016} and \cite{KK2018} using GCM simulations. As mentioned above, the efficiency of the heat engine responsible for sensible heating was noticed to be twice smaller than the theoretical value \citep[][]{KA2016}. Thus, it is set to $\effconv = 0.5$. 


As the timescale of Rayleigh drag and the efficiency parameter of advection both scale the inter-hemispheric heat transport due to atmospheric circulation (see \eq{Ladv}), only one of these two parameters is needed to quantify the flux. Thus we \recc{arbitrarily} fix $\tdrag = 10$~days, and choose the efficiency parameter $\effadv$ so that $\Ladv \gg 1$, which corresponds to the asymptotic regime of the thermally homogenised atmosphere studied by \cite{Wordsworth2015}. The parameters values used in the reference case are gathered in \tab{tab:param_wordsworth2015}.

\begin{table}[h]
\centering
 \textsf{\caption{\label{tab:param_wordsworth2015} Values of parameters used in the reference case of the present work \citep[case treated by][]{Wordsworth2015}. The acronyms SW and LW are used in place of 'shortwave' and 'longwave', respectively. }}
\begin{small}
    \begin{tabular}{l l l}
      \hline
      \hline
      \textsc{Parameter} & \textsc{Symbol}  & \textsc{Value}  \\ 
      \hline
      \\[-0.2cm]
     \multicolumn{3}{c}{\textit{Planet characteristics}} \\
      Planet mass ($\Mearth$) & $\Mpla$ & 1.0  \\
      Planet radius ($\Rearth$) & $\Rpla$ & 1.0  \\[0.2cm]
      \multicolumn{3}{c}{\textit{Atmospheric properties}} \\
      Molecular mass (${\rm g~mol^{-1}}$) & $\Mmolatm$ & 44.01 \\
      Heat capacity per unit mass (${\rm J~kg^{-1}~K^{-1}}$) & $\Cp$ & $650$  \\
      SW mean opacity (${\rm m^2~kg^{-1}}$) & $\ksw$ & $10^{-9}$  \\
      LW mean opacity (${\rm m^2~kg^{-1}}$) & $\klw$ & $5 \times 10^{-5}$  \\
      SW scattering parameter & $\betasw$ & $1.0$ \\
      LW scattering parameter & $\betalw$ & $1.0$ \\[0.2cm]
      \multicolumn{3}{c}{\textit{Surface properties}} \\
      Surface albedo & $\Asurfsw$ & $0.2$ \\
      Bulk drag coefficient & $\Cd$ & $3.4 \times 10^{-3}$  \\
       \multicolumn{3}{c}{\textit{Parameters of the heat engines}} \\
       Efficiency parameter of sensible heating & $\effconv$ & $0.5$  \\
       Efficiency parameter of advection & $\effadv$ & $8 \times 10^{-3}$ \\
       Timescale of Rayleigh drag (days) & $\tdrag$ & 10 \\[0.2cm]
       \hline
    \end{tabular}
\end{small}
 \end{table}

\begin{figure*}[htb]
   \centering
       \begin{flushleft}
   \hspace{0.13\textwidth} 
   Radiative case \hspace{0.25\textwidth} General case \\
 \end{flushleft}
  \vspace{-0.2cm}
\hspace{-0.16\textwidth}
    \includegraphics[width=0.018\textwidth,trim = 0.5cm 2.5cm 33.cm 1.9cm,clip]{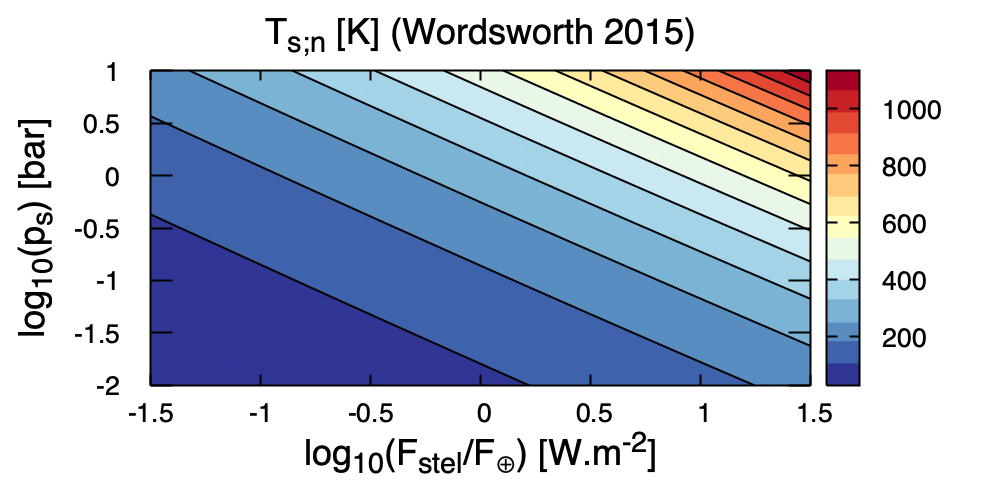} 
     \includegraphics[width=0.36\textwidth,trim = 2.5cm 2.5cm 2.cm 1.9cm,clip]{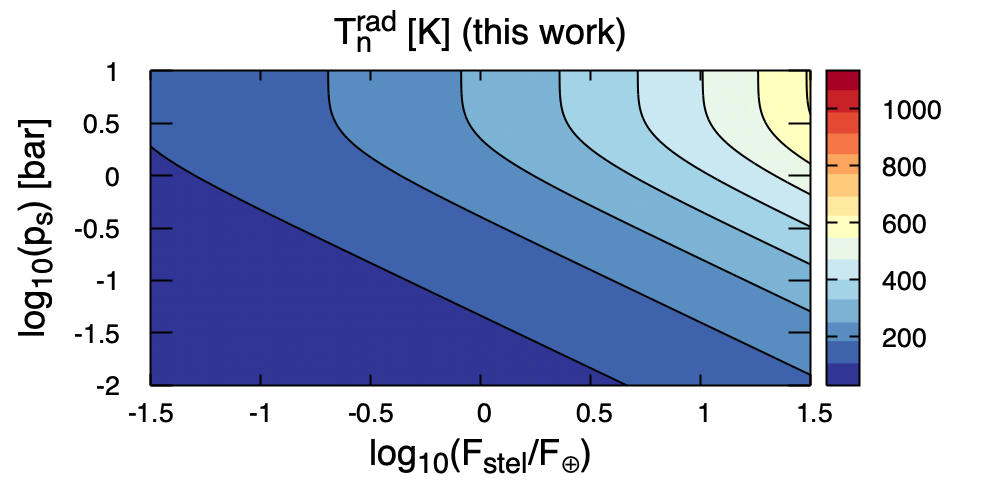} 
    \includegraphics[width=0.36\textwidth,trim = 2.5cm 2.5cm 2.cm 1.9cm,clip]{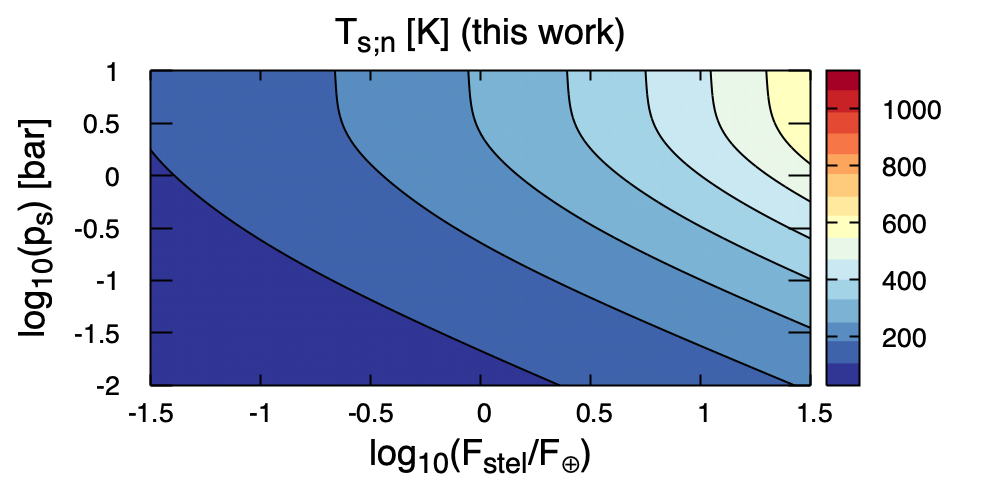} 
     \raisebox{1.5cm}{This work}
    \\
    \hspace{-0.18\textwidth}
 \includegraphics[width=0.018\textwidth,trim = 0.5cm 2.5cm 33.cm 1.9cm,clip]{auclair-desrotour_fig10a.png} 
  \includegraphics[width=0.36\textwidth,trim = 2.5cm 2.5cm 2.cm 1.9cm,clip]{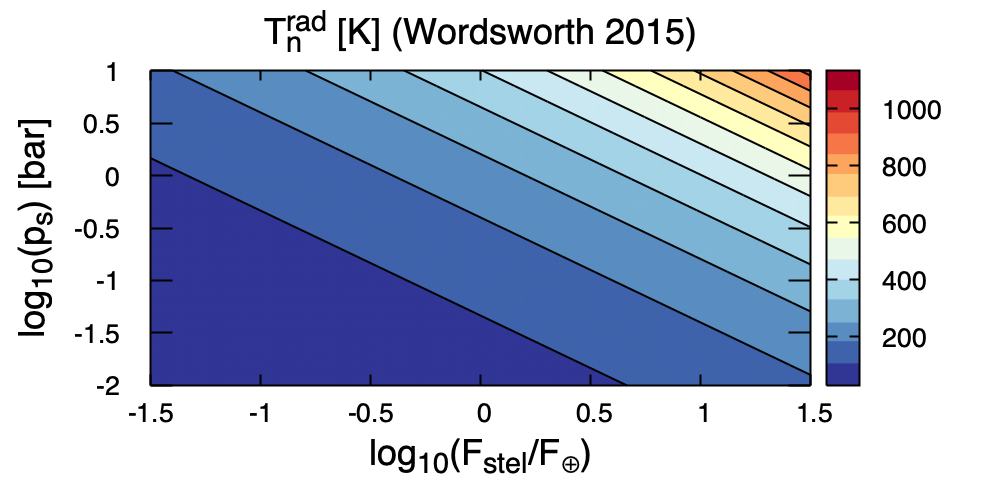} 
    \includegraphics[width=0.36\textwidth,trim = 2.5cm 2.5cm 2.cm 1.9cm,clip]{auclair-desrotour_fig10a.png} 
    \raisebox{1.5cm}{W2015}
    \\
   \includegraphics[width=0.018\textwidth,trim = 0.5cm 2.5cm 33.cm 1.9cm,clip]{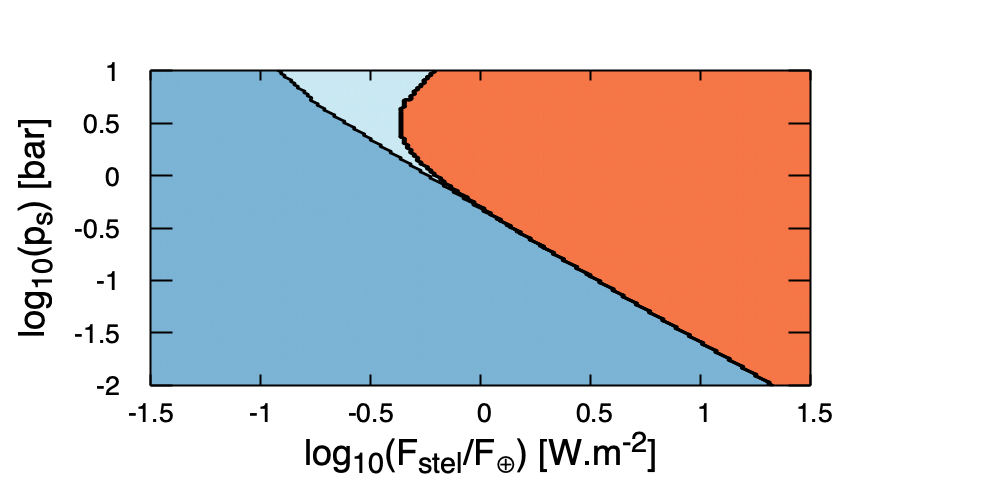} 
  \includegraphics[width=0.36\textwidth,trim = 2.5cm 2.5cm 2.cm 1.9cm,clip]{auclair-desrotour_fig10e.png} 
    \includegraphics[width=0.36\textwidth,trim = 2.5cm 2.5cm 2.cm 1.9cm,clip]{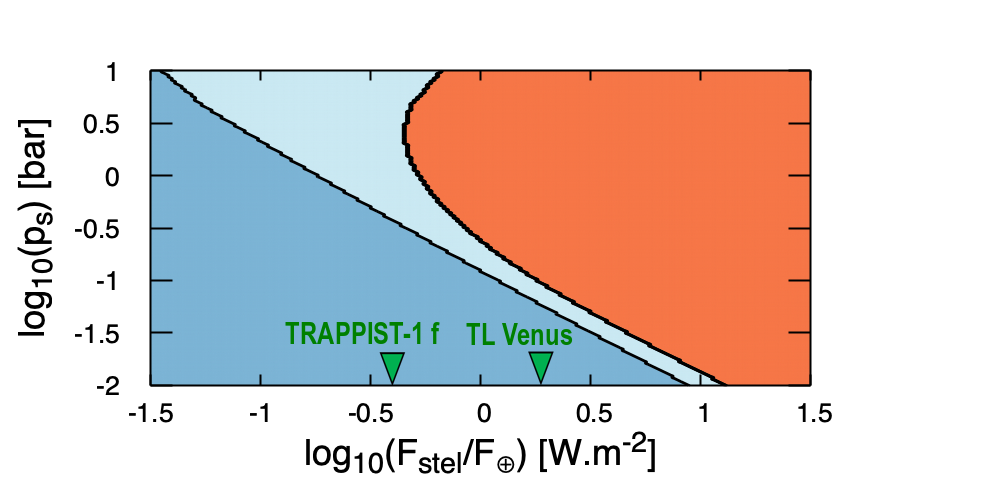}
     \includegraphics[width=0.23\textwidth,trim = 0cm 13cm 15cm 0.0cm,clip]{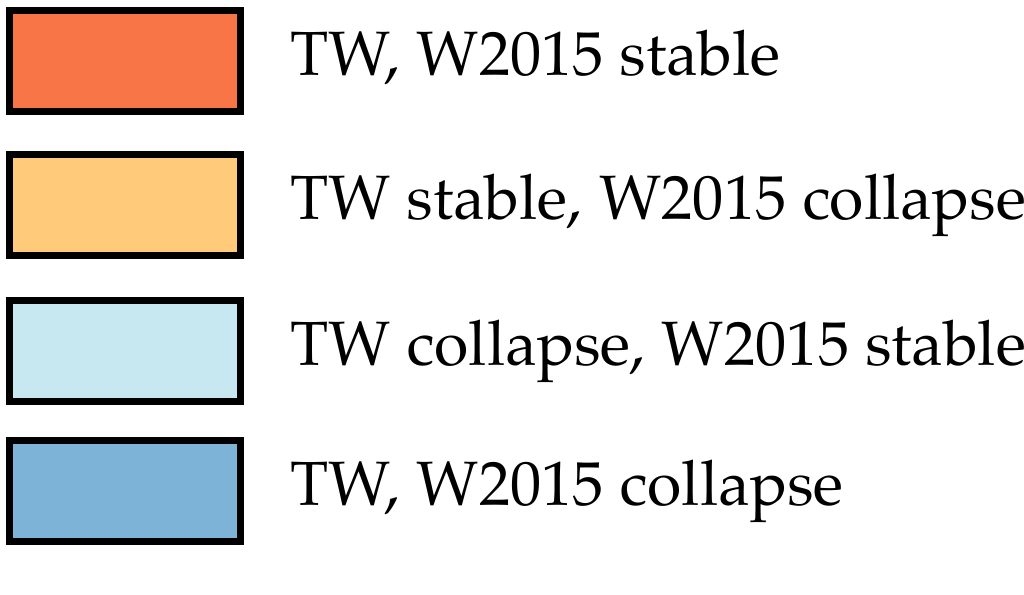} \\
     \vspace{-0.3cm}
    \begin{flushleft}
    \hspace{2.5cm}
    \small{$\logdix \left( \Fstar / \Fearth \right)$}  \hspace{4.8cm} \small{$\logdix \left( \Fstar / \Fearth \right)$}
    \end{flushleft}
       \vspace{-0.2cm}
   \caption{Nightside temperature (K) predicted by analytical theory in this work (top panels) and in \cite{Wordsworth2015} (middle panel), and associated comparative stability diagrams (bottom panels) as functions of logarithms of stellar flux (horizontal axis) and surface pressure (vertical axis). {\it Left:} Purely radiative case as defined in \sect{ssec:radiative_case} ($\Lconv = 0$, $\Ladv = + \infty$). {\it Right:} General case (i.e., with sensible heating and heat transport by stellar and anti-stellar atmospheric circulation). The stellar flux is normalised by the modern Earth's value $\Fearth = 1366$\units{W~m^{-2}}, and the surface pressure is given in bar. In stability diagrams, the acronyms 'TW' and 'W2015' are used for 'This work' and '\cite{Wordsworth2015}', respectively. Colours indicate the stability of the steady state predicted by the two models: both models predict stability (orange-red areas); the present model predicts stability while W2015 predicts collapse (orange areas); the present model predicts collapse while W2015 predicts stability (light blue areas); both models predict collapse (blue areas). \rec{The incident stellar fluxes of TRAPPIST-1 f and a hypothetical tidally locked Venus are designated by green triangles.} These plots are obtained by solving \eq{eqsingle_gen} and Eq.~(45) of \cite{Wordsworth2015} with parameters values given by \tab{tab:param_wordsworth2015}.  }
       \label{fig:map_stabilite_W2015}%
\end{figure*}

\subsection{Optical opacity favours collapse at low stellar fluxes}
\label{ssec:opacity_low_fluxes}

Our investigations start with the role played by longwave opacity on atmospheric collapse. By performing simulations with a GCM with correlated-k radiative transfer, \cite{Wordsworth2015} computed a \kc{diagram} of the atmospheric stability as a function of stellar flux and surface pressure in the reference case of the $\carbondiox$-dominated atmosphere for an Earth-sized planet and a $10~\Mearth$-super-Earth \citep[][Fig.~12]{Wordsworth2015}. This \kc{diagram} shows that the thresholds between stable and unstable regions of the parameter space predicted by GCM simulations and the analytic theory in the optically thin limit diverge from each other at low stellar fluxes, while they match well at high stellar fluxes. The divergence is patent around $\Fstar \approx 500 $\units{W~m^{-2}}, where the critical pressure increases drastically as the stellar flux decays. 

Following \cite{Wordsworth2015}, we interpret this feature as a limitation of the optically thin approximation assumed in this early work. In order to better understand it, we compute the nightside temperature both with Wordsworth's model (designated by the acronym 'W2015'), and our model ('This work'), which captures in a simplified way the non-linear dependence of the atmospheric thickness on the surface pressure. Figure~\ref{fig:map_stabilite_W2015} shows the results of these calculations in the purely radiative case ($\Lconv = 0$, $\Ladv = + \infty$) and in \recc{the case where sensible heating is introduced ($\Lconv \neq 0$), refereed to as 'General case'. As mentioned above, this later case corresponds to the asymptotic regime treated by \cite{Wordsworth2015}, where the day-night heat transport is efficient ($\Ladv \gg 1$), leading to $\Tanight \approx \Taday$. In this regime, the values chosen for the timescale of Rayleigh drag ($\tdrag$) and the efficiency parameter of advection ($\effadv$) do not affect the planet's thermal state of equilibrium and the associated atmospheric stability from the moment that $\Ladv \gg 1$. Besides, the computed nightside temperature and atmospheric stability correspond to upper bounds since they can only decay if the efficiency of the day-night heat transport is modified\footnote{\recc{If the value chosen for $\effadv$ were small enough, one would leave the asymptotic regime of efficient day-night heat transport and get in the transition regime showed by \figs{fig:tempnorm_Ladv}{fig:temp_Ladv}, which corresponds to a weaker heat transport, and thus leads to lower nightside temperatures and decreases the atmospheric stability.}}.} 

\recc{The values of parameters used in the two cases are given by \tab{tab:param_wordsworth2015}}. Nightside temperatures as well as the corresponding stability diagrams are plotted as function of the stellar flux and surface pressure in logarithmic scales. \rec{For comparison, the incident stellar fluxes of TRAPPIST-1~f \citep[][]{Gillon2017} and a hypothetical tidally locked Venus are indicated in the bottom right panel of the figure (green triangles).}

We first focus on stability diagrams (\fig{fig:map_stabilite_W2015}, bottom panels). Both in the purely radiative and in the general cases, the two models diverge at low stellar fluxes. While the evolution of the critical pressure with the stellar flux is well described by a power law in the optically thin limit (W2015), it undergoes a radical change of behaviour around $\Fstar \approx \Fearth$ when the non-linear dependence of the atmospheric optical thickness on surface pressure is taken into account (this work). This change may be explained by comparing the scalings of the nightside temperature with surface pressure given by the two models (\fig{fig:map_stabilite_W2015}, top and middle panels). Below $\psurf \approx 1$~bar, the atmospheric opacity is linear with $\klw$, as predicted by \eq{Tsnrad} in the purely radiative case (see \tab{tab:radiative_asymp}). Combining the fact that $\Tnight \scale \Teq \taugrlw^{1/4}$ in the optically thin limit with the expression of the condensation temperature of $\carbondiox$ given by \eq{TcondCO2_1} yields the scaling

\begin{equation}
\pcrit \sim \frac{4 \ggravi \sigmaSB \left( \Tcondcarbdioxc \right)^4 }{\klw \left( 1 - \Asurfsw \right) \Fstar} \scale \Fstar^{-1}. 
\label{pcrit_scaling}
\end{equation}

\noindent where $\Tcondcarbdioxc \define \Tcondcarbdiox \left( 100 \, \mbox{Pa} \right) = 136.37 $~K is the condensation temperature of carbon dioxide at $1$~mbar. We remark that this scaling law depends on the relation between the longwave optical thickness and pressure. In the case where this relation is a power law of the form $\taugrlw \scale \psurf^{\nexptau}$ with $\nexptau>0$, the preceding scaling becomes $\pcrit \scale \Fstar^{-1/\nexptau}$. For instance, if the opacity increases due to pressure broadening, \rec{then} $\nexptau = 2$ \citep[][]{RC2012,Pierrehumbert2011}, meaning that the dependence of the critical pressure on the stellar flux is weaker in this case than in the studied case. 

Beyond $\psurf \approx 1$~bar, the atmospheric opacity stops growing linearly with $\klw$. As a consequence, $\Tnight / \left( \Tnight \right)_{\rm rad.} \rightarrow 0$ as surface pressure increases, which leads to the observed dependence inversion. We note that the change of behaviour predicted by the model exceeds that derived from GCM simulations since the critical pressure tends to decay as surface pressure increases for $\psurf \gtrsim 3 $~bar. This may be related to the limitations of the isothermal approximation made to derive radiative transfer functions in \sect{sec:radiative_model}, which is not appropriate to model the structure of thick atmospheric layers as discussed in the case of Venus. In spite of these limitations, the isothermal approximation turns out to be sufficient to capture the effect of large longwave opacities on the atmospheric stability at low stellar fluxes, and recover the associated behaviour of the critical pressure highlighted by GCM simulations. 

\recc{Besides, we observe that the prediction of the model for atmospheric stability in Fig.~\ref{fig:map_stabilite_W2015} corresponds to an upper limit since stability diagrams were computed in the asymptotic regime previously treated by \cite{Wordsworth2015}, which maximises day-night heat transport ($\Ladv \gg 1$). Decreasing the efficiency parameter of advection would decrease the atmospheric stability and widen the region of the parameter space where collapse may occur as detailed in \sect{ssec:planet_size}. In the light of these considerations, the obtained stability diagrams (Fig.~\ref{fig:map_stabilite_W2015}, bottom panels) predict atmospheric collapse on TRAPPIST-1~f, which agrees with the results obtained by \cite{Wolf2017b} using 3D GCM simulations.}

\begin{figure*}[htb]
   \centering
       \begin{flushleft}
   \hspace{0.27\textwidth} 
   This work / Radiative case \hspace{0.17\textwidth} This work / W2015
 \end{flushleft}
  \vspace{-0.2cm}
  \raisebox{1.5cm}{ $\effconv = 10^{0~}$}
  \hspace{0.3cm}
    \includegraphics[width=0.018\textwidth,trim = 0.5cm 2.5cm 33.cm 1.9cm,clip]{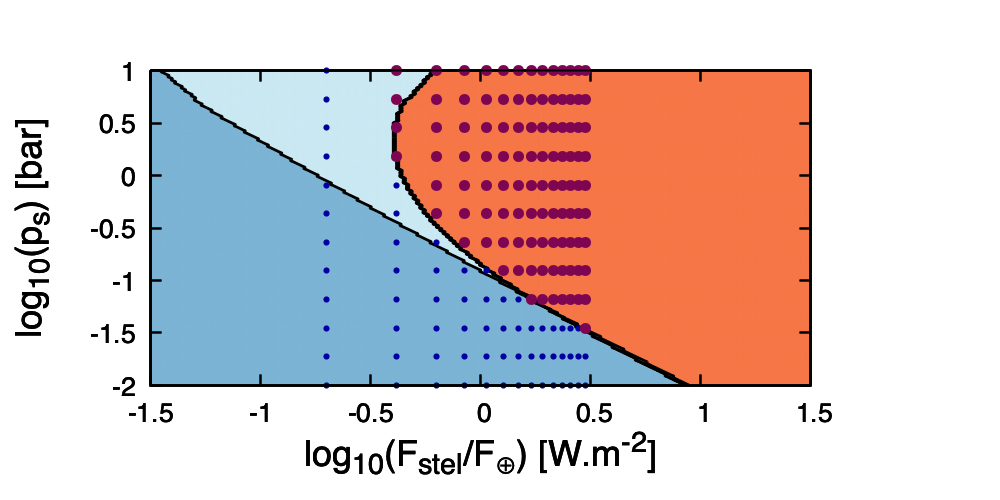} 
     \includegraphics[width=0.36\textwidth,trim = 2.5cm 2.5cm 2.cm 1.9cm,clip]{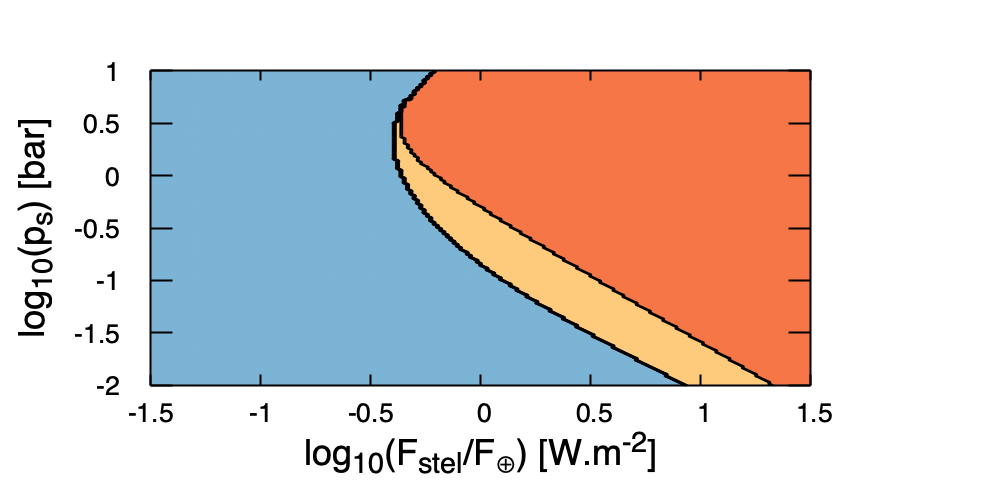} 
    \includegraphics[width=0.36\textwidth,trim = 2.5cm 2.5cm 2.cm 1.9cm,clip]{auclair-desrotour_fig11a.png} 
    \\
     \raisebox{1.5cm}{ $\effconv = 10^{-2}$}
      \hspace{0.2cm}
  \includegraphics[width=0.018\textwidth,trim = 0.5cm 2.5cm 33.cm 1.9cm,clip]{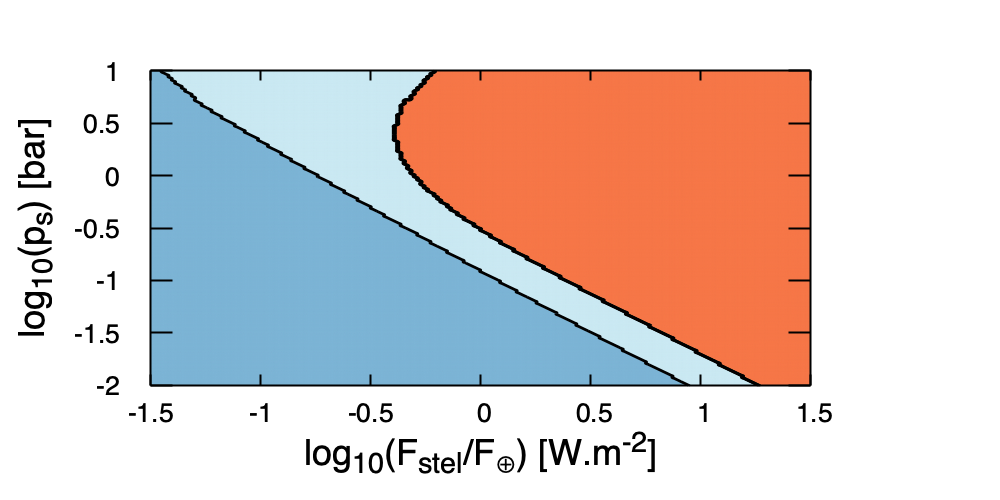} 
     \includegraphics[width=0.36\textwidth,trim = 2.5cm 2.5cm 2.cm 1.9cm,clip]{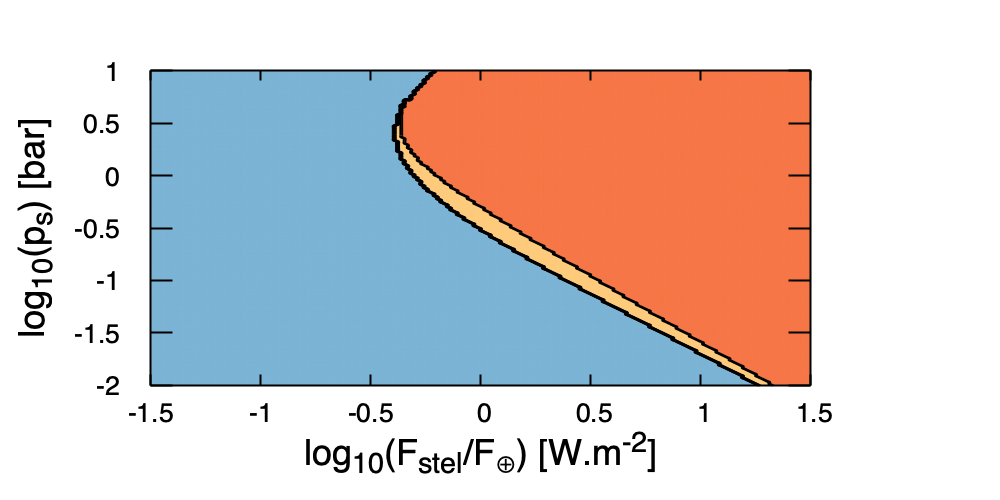} 
    \includegraphics[width=0.36\textwidth,trim = 2.5cm 2.5cm 2.cm 1.9cm,clip]{auclair-desrotour_fig11c.png} 
    \\
    \raisebox{1.5cm}{ $\effconv = 10^{-4}$}
     \hspace{0.2cm}
    \includegraphics[width=0.018\textwidth,trim = 0.5cm 2.5cm 33.cm 1.9cm,clip]{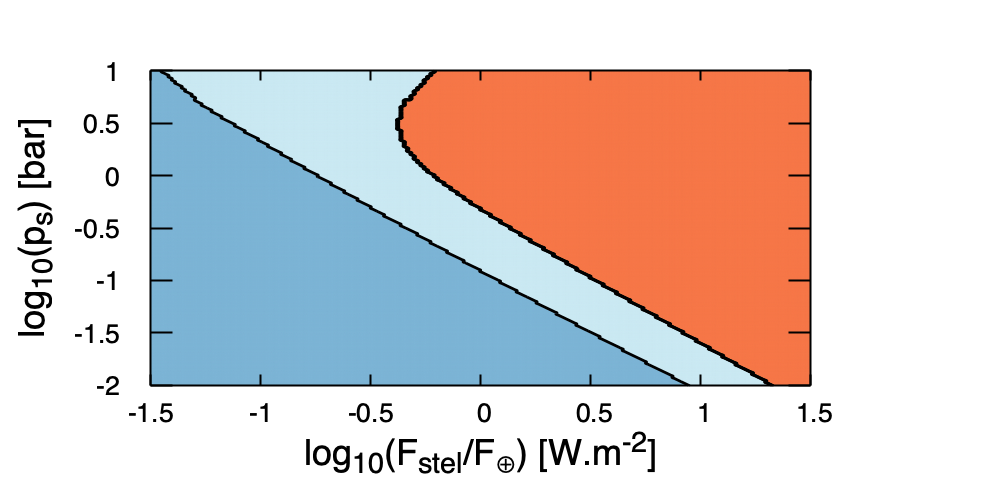} 
     \includegraphics[width=0.36\textwidth,trim = 2.5cm 2.5cm 2.cm 1.9cm,clip]{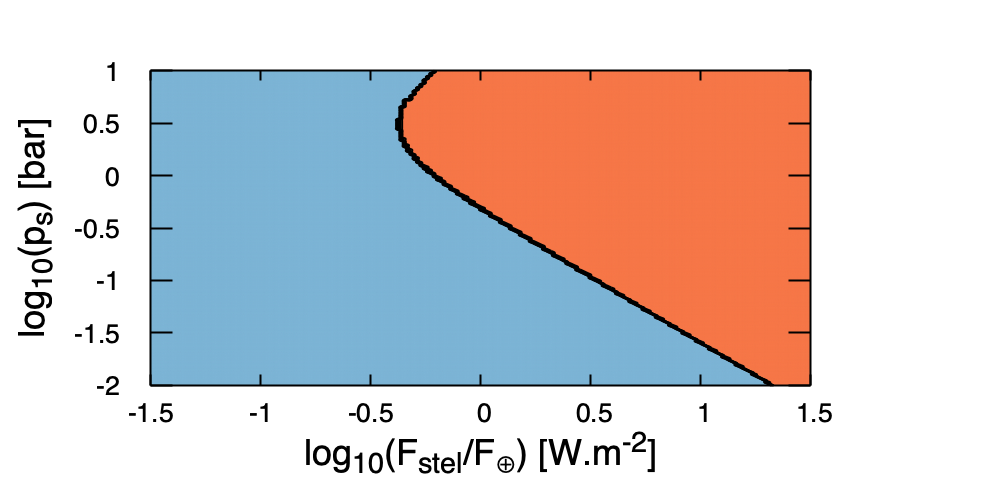} 
    \includegraphics[width=0.36\textwidth,trim = 2.5cm 2.5cm 2.cm 1.9cm,clip]{auclair-desrotour_fig11e.png} \\
     \vspace{-0.3cm}
    \begin{flushleft}
    \hspace{6.0cm}
    \small{$\logdix \left( \Fstar / \Fearth \right)$}  \hspace{4.5cm} \small{$\logdix \left( \Fstar / \Fearth \right)$} \\[0.2cm]
    \hspace{5.0cm}
    \includegraphics[width=0.23\textwidth,trim = 0cm 13.5cm 15cm 0.0cm,clip]{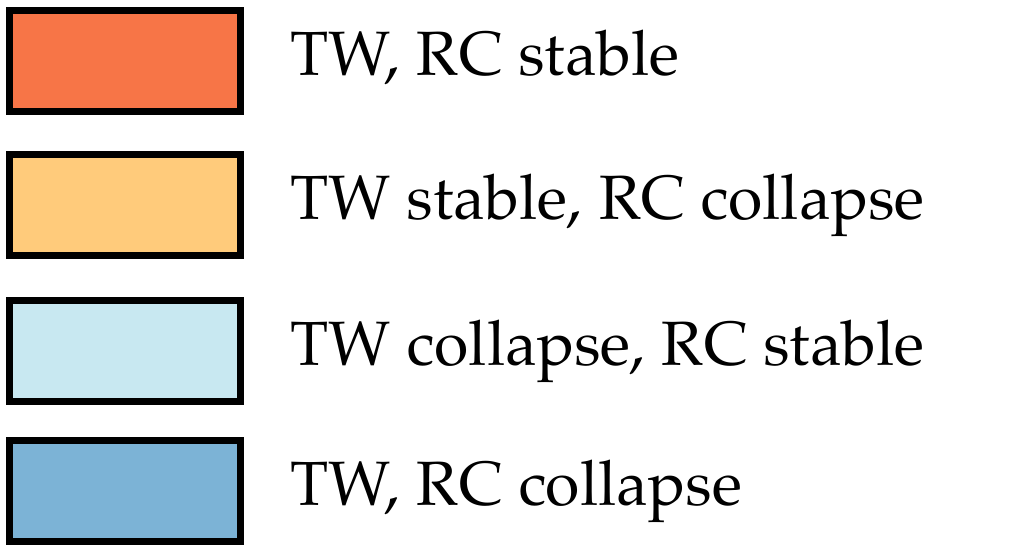}
     \hspace{2.5cm}
    \includegraphics[width=0.23\textwidth,trim = 0cm 13.5cm 15cm 0.0cm,clip]{auclair-desrotour_fig10g}
    \end{flushleft}
       \vspace{-0.2cm}
   \caption{Comparative stability diagrams as functions of logarithms of stellar flux (horizontal axis) and surface pressure (vertical axis) for various intensities of sensible heating, $\effconv =  10^{-4}, 10^{-2}, 10^{0}$ (from bottom to top). {\it Left:} Comparison between the general case ($\Lconv > 0$) and the purely radiative case characterised in \sect{ssec:radiative_case} ($\Lconv = 0$). {\it Right:} Comparison between this work and the model developed by \cite{Wordsworth2015} in the general case. The stellar flux is normalised by the modern Earth's value $\Fearth = 1366$\units{W~m^{-2}}, and the surface pressure is given in bar. The acronyms 'TW', 'RC', and 'W2015' are used for 'This work' (with sensible heating), 'Radiative case' (without sensible heating), and '\cite{Wordsworth2015}', respectively. Colours indicate the stability of the steady state predicted by the two models, similarly as in \fig{fig:map_stabilite_W2015}. These plots are obtained by solving \eq{eqsingle_noadv} ($\Ladv = + \infty$) and Eq.~(45) of \cite{Wordsworth2015} with parameters values given by \tab{tab:param_wordsworth2015}. Violet dots (top right panel) indicate the region of atmospheric stability computed by \cite{Wordsworth2015} from GCM simulations (Fig.~12), while small blue dots indicate the region of collapse.  }
       \label{fig:map_stabilite_effsen_W2015}%
\end{figure*}

\subsection{Dayside sensible heating increases stability}

As a second step, we investigate how the dayside sensible heating affects the atmospheric stability. To do so, we place ourselves in the asymptotic regime studied by \cite{Wordsworth2015}, where the atmosphere has the same temperature on the dayside and nightside hemispheres. This allows us to ignore the effects of advection for the moment ($\Ladv = + \infty$) and to use the simplified equation given by \eq{eqsingle_noadv} in the root-finding procedure instead of the general one (\eq{eqsingle_gen}). We solve this equation for various values of the efficiency parameter controlling the intensity of sensible heating ($\effconv$) and proceed similarly with \kc{Wordsworth's model \citep[][Eq.~45]{Wordsworth2015}.}

Figure~\ref{fig:map_stabilite_effsen_W2015} shows the stability diagrams derived from these calculations. The general case (i.e., with sensible heating) is compared with the purely radiative case treated in \sect{ssec:radiative_case} (\fig{fig:map_stabilite_effsen_W2015}, left column), and with the general case of W2015 (\fig{fig:map_stabilite_effsen_W2015}, right column). In addition, \kc{outcomes} computed from 3D GCM simulations for the $1~\Mearth$-planet in this early study \citep[][Fig.~12, top panel]{Wordsworth2015} are included (\fig{fig:map_stabilite_effsen_W2015}, top right panel). Violet dots indicate simulations where the atmosphere remained stable, and blue dots simulations where collapse occurred.

As discussed in \sect{sec:sensible_heating}, increasing the intensity of sensible heating acts to warm the atmosphere up. Consequently, the critical pressure decays as the efficiency of the dayside convective heat engine increases. However, this evolution is bounded. From the moment that $\Lconv \gg 1$, $\Taday \approx \Tday$ and $\pcrit$ reaches a minimum that is not sensitive to $\Lconv$, as observed for the nightside temperature in \figs{fig:temp_Lconv}{fig:temp_Ladv}. One may show analytically that the maximum amplitude of critical pressure variations is of 0.6 decades in the studied case. 

First, assuming that $\Taday = \Tday$ in the strongly convective asymptotic regime ($\Lconv \gg 1$), we obtain from \eq{Fnorm_Tnorm_thick2} that 

\begin{equation}
\Fnorm = \frac{\Alw}{\Ksw} \left[ 2 - \frac{\Alw}{\Klw}  + \frac{\Klw}{\Alw}   \right],
\end{equation}

\noindent which yields the nightside temperature in this regime,

\begin{equation}
\Tnight = \Teq \left( \frac{\Ksw}{\Klw} \right)^{1/4} \left[ 1 + \frac{1}{2} \left( \frac{\Klw}{\Alw} - \frac{\Alw}{\Klw} \right)  \right]^{-1/4}. 
\end{equation}

\noindent If the atmosphere is transparent in the shortwave and optically thin in the longwave, \rec{then} this expression simplifies to 

\begin{equation}
\Tnight \approx \Teq \left[ 2 \left( 1 - \Asurfsw \right) \taugrlw \right]^{1/4}. 
\label{Tnsen}
\end{equation}

The critical optical depth below which collapse occurs is determined by the equality $\Tnight \left( \taugrlw \right) = \Tcondcarbdiox$. This critical optical depth is denoted by $\taugrlwrad$ in the purely radiative regime and $\taugrlwconv$ in the regime dominated by sensible heating, and the corresponding critical pressures are denoted by $\pcritrad$ and $\pcritconv$, respectively. Thus, assuming that $\Tcondcarbdiox \approx \Tcondcarbdioxc$, and using both \eqs{Tsnrad}{Tnsen}, we get the ratio between these two boundaries 

\begin{equation}
\frac{\pcritrad}{\pcritconv} = \frac{\taugrlwrad}{\taugrlwconv} = 4, 
\end{equation}

\noindent  that is roughly $0.6$ decades. This corresponds to what is observed for $\effconv = 1$ in \fig{fig:map_stabilite_effsen_W2015} (top left panel).

 In W2015, no efficiency parameter such as $\effconv$ intervene to modulate the sensible heat flux except a geometrical coefficient ($\chi$), which is implicitly set to 1 here. This is why the critical pressure predicted by W2015 is the same in all panels of \fig{fig:map_stabilite_effsen_W2015}. Interestingly, setting $\effconv = 1.0$ reproduces the asymptotic behaviour of the critical pressure derived from W2015 at high stellar fluxes, although the two models use different approaches to scale the horizontal wind speed at planet's surface. Moreover, we note that the behaviour of the critical pressure predicted by \eq{eqsingle_gen} at low fluxes matches GCM simulations fairly well in this case. 

\begin{figure*}[htb]
   \centering
       \begin{flushleft}
   \hspace{0.30\textwidth} 
   $\effconv = 0$ \hspace{0.21\textwidth} $\effconv = 1.0$
 \end{flushleft}
  \vspace{-0.2cm}
  \hspace{0.5cm}
  \hspace{-0.23\textwidth}
  \raisebox{1.5cm}{ $\efftot  = + \infty$}
  \hspace{0.3cm}
    \includegraphics[width=0.018\textwidth,trim = 0.5cm 2.5cm 33.cm 1.9cm,clip]{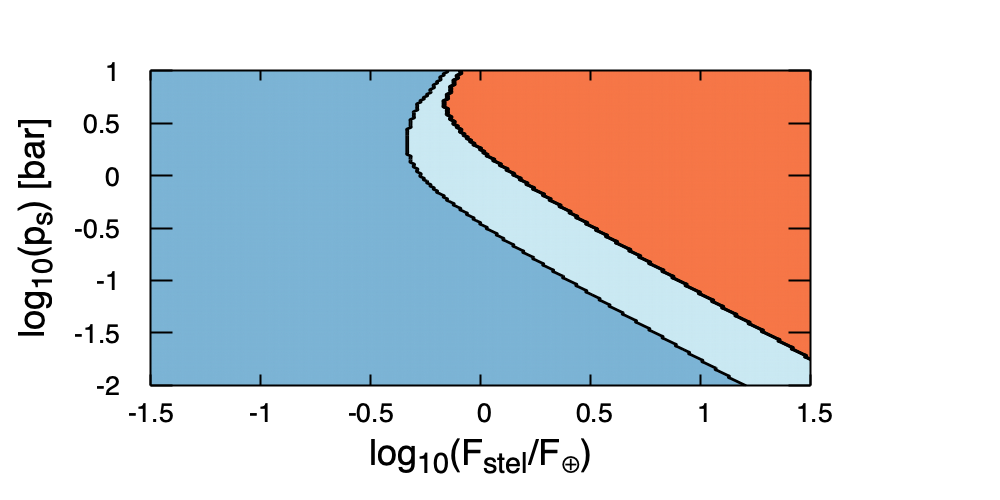} 
     \includegraphics[width=0.30\textwidth,trim = 2.5cm 2.5cm 5.9cm 1.9cm,clip]{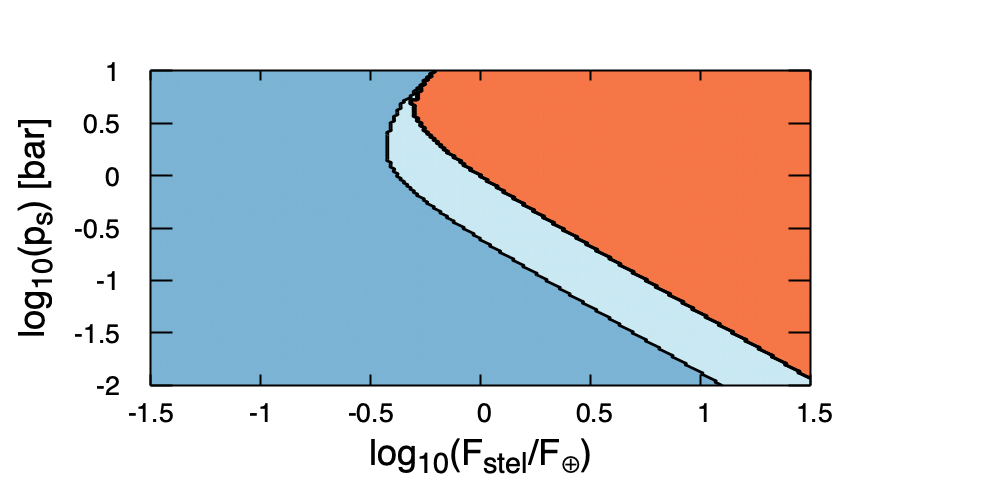} 
    \includegraphics[width=0.30\textwidth,trim = 2.5cm 2.5cm 5.9cm 1.9cm,clip]{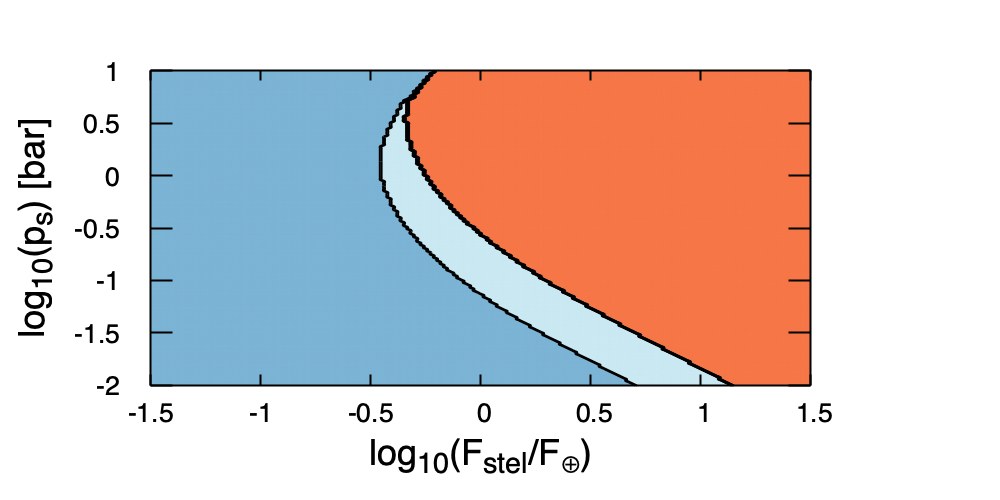} 
    \\
    \hspace{0.5cm}
      \hspace{-0.26\textwidth}
    \raisebox{1.5cm}{ $\efftot= \numprint{3e-3}$}
     \hspace{0.2cm}
    \includegraphics[width=0.018\textwidth,trim = 0.5cm 2.5cm 33.cm 1.9cm,clip]{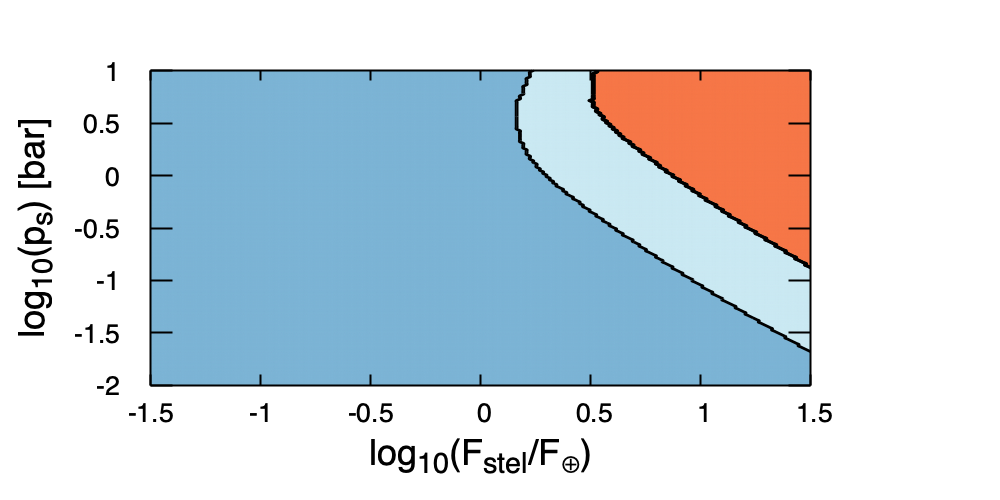} 
     \includegraphics[width=0.30\textwidth,trim = 2.5cm 2.5cm 5.9cm 1.9cm,clip]{auclair-desrotour_fig12a.png} 
    \includegraphics[width=0.30\textwidth,trim = 2.5cm 2.5cm 5.9cm 1.9cm,clip]{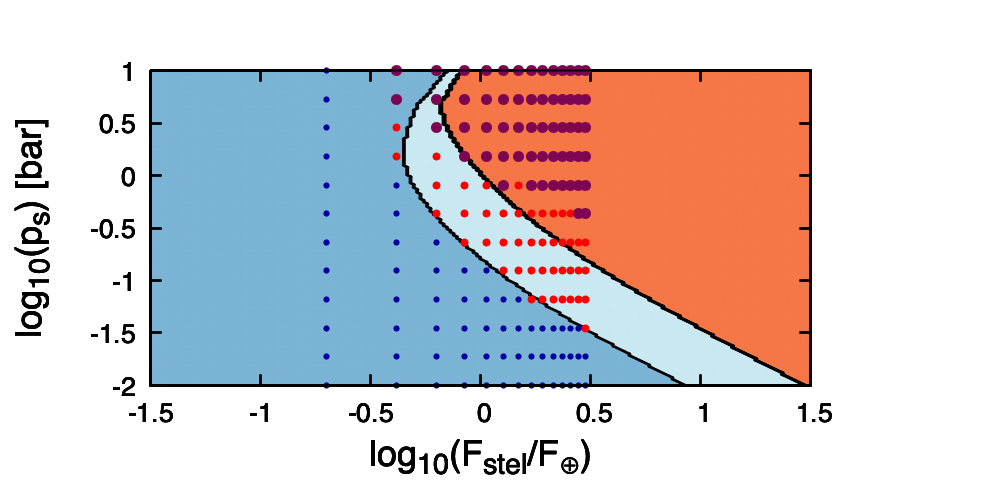} \\
    \hspace{0.5cm}
    \raisebox{1.5cm}{ $\efftot= 10^{-4}$}
     \hspace{0.2cm}
    \includegraphics[width=0.018\textwidth,trim = 0.5cm 2.5cm 33.cm 1.9cm,clip]{auclair-desrotour_fig12d.png} 
     \includegraphics[width=0.30\textwidth,trim = 2.5cm 2.5cm 5.9cm 1.9cm,clip]{auclair-desrotour_fig12d.png} 
    \includegraphics[width=0.30\textwidth,trim = 2.5cm 2.5cm 5.9cm 1.9cm,clip]{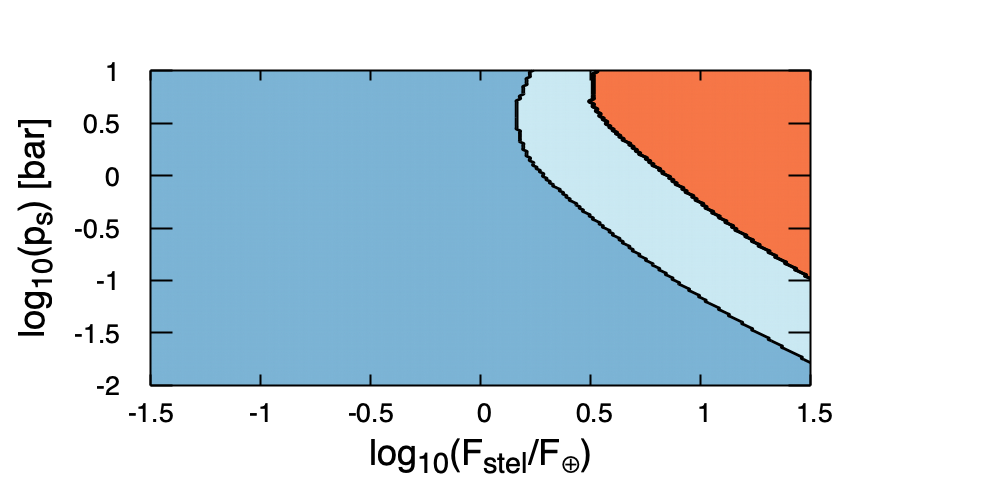}  \hspace{0.2cm}
    \includegraphics[width=0.20\textwidth,trim = 0cm 13.5cm 16.5cm 0.0cm,clip]{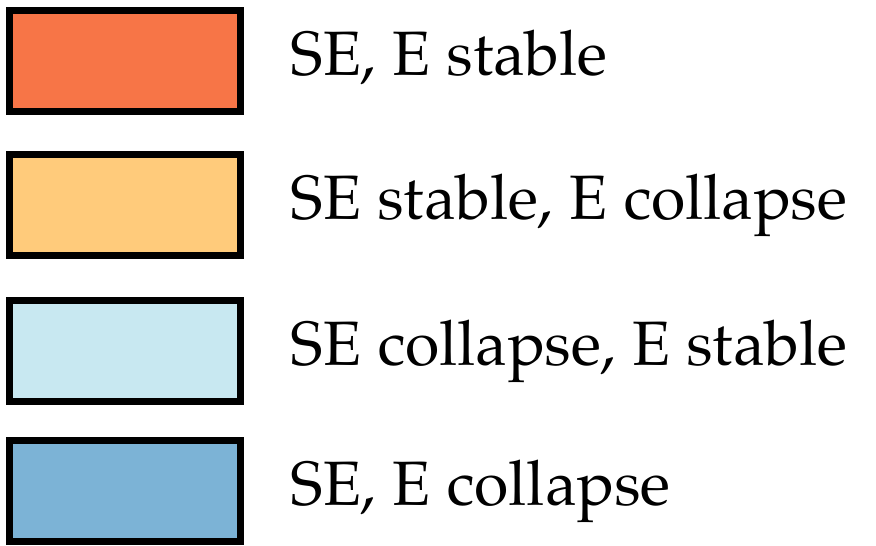}\\
    
     \vspace{-0.3cm}
    \begin{flushleft}
    \hspace{5.0cm}
    \small{$\logdix \left( \Fstar / \Fearth \right)$}  \hspace{3.6cm} \small{$\logdix \left( \Fstar / \Fearth \right)$} \\
    \end{flushleft}
       \vspace{-0.2cm}
   \caption{Comparative stability diagrams for $1~\Mearth$ and $10~\Mearth$ planets as functions of logarithms of stellar flux (horizontal axis) and surface pressure (vertical axis). The efficiency parameter controlling advective heat transport $\efftot$ take the values $\logdix \left( \efftot \right) =-4, -2.5, + \infty$ (from bottom to top). {\it Left column:} Case without sensible heating ($\effconv = 0$). {\it Right column:} Case with efficient sensible heating ($\effconv = 1.0$). The stellar flux is normalised by the modern Earth's value $\Fearth = 1366$\units{W~m^{-2}}, and the surface pressure is given in bar. The acronyms 'E' and 'SE' are employed for 'Earth' and 'Super-Earth', and designate the $1~\Mearth$ and $10~\Mearth$ planets, respectively. Colours indicate the stability of the steady state obtained in the two cases, similarly as in \fig{fig:map_stabilite_W2015}. These plots are obtained by solving \eq{eqsingle_noadv} ($\Ladv = + \infty$, top panels) and \eq{eqsingle_gen} (other panels) with $\klw = \numprint{8.5e-5}$\units{m^2~kg^{-1}} and parameters values given by \tab{tab:param_wordsworth2015} for the Earth-sized planet. In the case of the $10~\Mearth$ planet, the planet's radius and mass are set to $\Mpla = 10 \Mearth $ and $\Rpla = 1.88 \Rearth$, respectively. Data computed by \cite{Wordsworth2015} from GCM simulations (Fig.~12) are included in the middle right panel for comparison and indicate the region of atmospheric stability for the $1~\Mearth$ planet (red dots) and both planets (violet dots), and the region of atmospheric collapse (small blue dots).     }
       \label{fig:map_stabilite_effadv_W2015}%
\end{figure*}

\subsection{Increasing the planet's size decreases stability}
\label{ssec:planet_size}

In the text above, the atmosphere is assumed to be homogeneous in temperature ($\Tanight = \Taday$). We now assume that the day-night atmospheric temperature distribution is controlled by the mechanism of heat transport by atmospheric circulation, which depends on dayside and nightside temperatures in return. This allows us to study the link between the intensity of advective heat transport -- scaled by the efficiency parameter $\efftot$ -- and the atmospheric stability. Particularly, we aim to show how the 0-D theory, in spite of its limitations, may give us some insight to better understand the decrease of atmospheric stability that is observed as the planet size increases \citep[e.g.][Fig.~12]{Wordsworth2015}.

In order to benchmark the predictions of the model against \kc{outcomes} computed from GCM simulations, we reproduce the case treated by \cite{Wordsworth2015} and thus consider two planets: the Earth-sized planet characterised by \tab{tab:param_wordsworth2015} and a super-Earth of $10~\Mearth$ and $1.88~\Rearth$. As GCM simulations were performed using correlated-$k$ distributions for radiative transfers in this early work, there is no evident value for the equivalent gray-gas longwave opacity. We thus set the absorptivity to $\klw = \numprint{8.5e-5}$\units{m^2~kg^{-1}} for a pedagogical purpose, this value leading the analytic model to match GCM simulations for a simple value of $\efftot$, as shown in the following. The equation of the steady state in the general case (\eq{eqsingle_gen}) is then solved for various values of the efficiency parameters controlling sensible heating ($\effconv$) and heat transport by atmospheric circulation ($\efftot$). 

Diagrams comparing the atmospheric stability of the two planets are plotted in \fig{fig:map_stabilite_effadv_W2015} in the absence of sensible heating ($\effconv = 0$) and in the regime of strong dayside convection ($\effconv = 1.0$), for $\logdix \left( \efftot \right) = -4,-2.5, + \infty$ (this later case is treated by solving \eq{eqsingle_noadv} instead of \eq{eqsingle_gen}). Besides, we include the data computed from 3D GCM simulations \citep[][Fig.~12]{Wordsworth2015} in the figure (middle right panel). Violet dots designate simulations where the atmosphere remained stable in both cases, while red dots indicate those where the atmosphere of the $1~\Mearth$-planet only remained stable. The figure shows that, in all cases, the atmospheric stability decays as the planet size increases. 

We first consider the purely radiative regime (\fig{fig:map_stabilite_effadv_W2015}, top left panel).  In this case, the observed difference between the Earth-sized planet and the super-Earth comes from the dependence of the longwave atmospheric \kc{optical depth} on the planet mass and radius that is hidden in the surface gravity (\eq{taugr}). As the planet size increases, the atmospheric \kc{optical depth} decays, and so does the greenhouse effect. Consequently, the nightside surface temperature decays. The difference between the two planets does not change when sensible heating is included (top right panel) because this flux does not depend on the planet size at first order in the scaling given by \eq{Lconv} (there is a dependence hidden in $\Qconv$ but it is weakened by the $1/3$ exponent in the scaling of the horizontal wind speed). However, we observe that the gap of collapse pressure between the two planets predicted by GCM simulations is larger than that associated with the aforementioned dependence. Among the multiple reasons that may be invoked to explain this difference, we examine the hypothesis where it is related to a decrease of the efficiency of heat redistribution mechanisms through the example of large-scale advection. 

 Except in the asymptotic limit of strong horizontal mixing ($\Ladv \gg 1$), the intensity of the heat flux due to large-scale advection is modulated by the efficiency parameter $\efftot$, leading the atmospheric stability to decay with $\efftot$ (\fig{fig:map_stabilite_effadv_W2015}, from top to bottom). Furthermore, in the derived scaling of the heat transport by stellar and anti-stellar atmospheric circulation, the heat flux is quantified by $\Ladv \scale \ggravi^{-1/2} \Rpla^{-1} $ (\eq{Ladv}) and thus decays as the planet size increases. As a consequence, the intensity of the heat flux from the dayside to the nightside decays as one switches from the Earth-sized planet to the super-Earth, which tends to widen the gap between the two planets (\fig{fig:map_stabilite_effadv_W2015}, bottom panels). This effect is stronger in the transition regime ($\Ladv \sim 1$) than in asymptotic ones, as shown by the evolution of the nightside and dayside temperatures ratio plotted in \fig{fig:tempnorm_Ladv}.
 
The case $\logdix \left( \efftot \right) = -2.5$ (i.e. $\efftot \approx \numprint{3e-3}$, middle panels) highlights the divergence between the analytic model and GCM simulations \citep[][Fig.~12]{Wordsworth2015}. First, we note that the model tends to underestimate atmospheric stability at low stellar fluxes and high pressures, which is partly due to the simplified temperature profile used to derive radiative fluxes, as discussed in \sect{ssec:opacity_low_fluxes}. Second, the critical pressure predicted by the analytic model scales similarly with the stellar flux for both planets, while it is not the case in GCM results. This is obviously due to the numerous simplifications made in the present work, where the processes responsible for heat redistribution are described using rough scalings. Despite these limitations, the model approximately matches GCM simulations and captures the decrease of atmospheric stability caused by the increase of the planet size. It should be possible to improve the analytical theory with better scalings of the fluxes in future studies. 


\section{Role played by atmospheric structure}
\label{sec:RCSU_model}

In our zero-dimensional approach, the implications of the atmospheric structure on the nightside temperature are ignored, given that atmospheric temperature profiles are reduced to dayside and nightside bulk atmospheric temperatures. This approximation enabled us to derive radiative fluxes analytically, and to capture in a simplified way the non-linear dependence of the atmospheric opacity on the shortwave and longwave optical depths. In reality, the atmospheric structure is not the same on the dayside and on the nightside, and differs from the idealised isothermal temperature profile, as highlighted by analytical models \citep[e.g.][]{RC2012,KA2016} and GCM simulations \citep[e.g.][]{Leconte2013,Wordsworth2015}. Particularly, the evolution of the temperature profile with the stellar zenithal angle appears clearly in simulations performed using \texttt{THOR} (\fig{fig:W2015fig2_KAfig4_thor}, left panels). The substellar region is characterised by a quasi-adiabatic temperature gradient, which is due to the strong convection generated by the absorption of stellar heating near the ground. On the nightside, the structure inversion that may be observed indicates that the atmosphere is stable with respect to convection. 

As mentioned above, the role played by the atmospheric structure was already investigated in the framework of the analytical theory. For instance \cite{RC2012} proposed a one-dimensional radiative convective model to predict the atmospheric structure of a wide range of planets. \cite{KA2016} refined this approach in the case of tidally locked rocky planets by including in a simplified way the effect of subsidence on the nightside temperature profile and surface temperature, which led them to develop a two-column radiative-convective-subsiding (RCS) model. In this early work, the dayside temperature profile was supposed to be shaped by convection, and was thus specified in calculations. This assumption holds in the case of convective tropospheres, which corresponds to the regime of strongly irradiated atmospheres transparent in the shortwave. However, it may \kc{break down} if the absorption near the ground is not large enough to generate strong convection. In view of these limitations, the prospect of extending the theory to optically thicker atmospheres in the shortwave \kc{motivates us} to go past this assumption. 

Therefore, although the core of the present study is the zero-dimensional model, we make a first attempt to relax the specification of the dayside atmospheric structure in one-dimensional models for the purpose of this section. We aim to give here some insight about the transition from zero-dimensional to one-dimensional models where both the dayside and nightside temperature profiles are computed from atmospheric opacities. This leads us to develop a simplified two-column radiative-convective-subsiding-upwelling (RCSU) model by introducing dayside upwelling winds in the RCS model of \cite{KA2016}. The RCSU model is detailed as a first step, and used as a second step to compute dayside and nightside atmospheric temperature profiles and study atmospheric stability against collapse.

\subsection{A two-column RCSU model}

To treat the coupling between the atmospheric structure and mean flows, one should integrate self-consistently the momentum, mass conservation, thermodynamic and radiative transfer equations, which is achieved by three-dimensional GCM but cannot be envisaged in the simplified framework of one-dimensional models. We thus introduce vertical winds by using the thermodynamic equation solely, following the method by \cite{KA2016}. Written in pressure coordinates, this equation reads \citep[][]{Vallis2006}

\begin{equation}
\dd{\temp}{\time} + \Vvect \scalprod \gradp \temp + \pressvelocity \dd{\temp}{\pressure} = \frac{\Rspec \temp \pressvelocity}{\Cp \pressure} + \frac{\ggravi}{\Cp} \dd{\Fnet}{\pressure} + \frac{\ggravi}{\Cp} \dd{\Ddiff}{\pressure},
\label{thermo_eq}
\end{equation}

\noindent the notation $\gradp$ designating the horizontal gradient over an isobar, $\Vvect$ the horizontal velocity vector, $\pressvelocity \define \Dpart{\pressure}{\time}$ the pressure velocity (\kc{i.e. the vertical velocity in pressure coordinates;} $\Dpart{}{\time}$ stands for the material derivative), and $\Ddiff$ the vertical diffusive power flux. To relate pressure to optical depth consistently with the early work by \cite{KA2016}, we use the standard power law

\begin{equation}
\frac{\optdepthlw}{\taugrlw} = \frac{\optdepthsw}{\taugrsw} = \left( \frac{\pressure}{\psurf} \right)^{\nexptau},
\end{equation}

\noindent where the exponent $\nexptau$ specifies how optical depths increases with pressure. Introducing the dimensionless parameter $\betagas \define \Rspec / \left( \nexptau \Cp \right)$ and substituting the pressure coordinate by the optical depth in the longwave, \eq{thermo_eq} thus becomes

\begin{align}
 \frac{\psurf}{\nexptau \optdepthlw} \left( \frac{\optdepthlw}{\taugrlw} \right)^{\frac{1}{\nexptau}} \left( \dd{\temp}{\time} + \Vvect \scalprod \gradp \temp \right)  & \\ 
   + \pressvelocity \left( \dd{\temp}{\optdepthlw} - \frac{\betagas}{\optdepthlw} \temp \right) & = \frac{\ggravi}{\Cp} \left( \dd{\Fnet}{\optdepthlw} + \dd{\Ddiff}{\optdepthlw} \right).\nonumber
\end{align}

In steady state, $\dd{\temp}{\time} = 0$. Besides, we ignore the diffusion term $\dd{\Ddiff}{\optdepthlw}$ assuming that radiative transfers and advective heat transport predominates. In the general case, the horizontal advection term ($\Vvect \scalprod \gradp \temp$) is comparable to the vertical advection term ($ \pressvelocity \dd{\temp}{\optdepthlw}$), and should be taken into account. However, we ignore it for simplicity following \cite{KA2016}, since including horizontal advection would require to specify ad hoc horizontal wind speeds and temperature distributions, which goes beyond the scope of a simplified two-column approach. Nevertheless, one may think about clever ways to introduce this component in the modelling in future studies. The above approximations yield

\begin{equation}
\pressvelocity \left( \dd{\temp}{\optdepthlw} - \frac{\betagas}{\optdepthlw} \temp \right)  = \frac{\ggravi}{\Cp} \dd{\Fnet}{\optdepthlw},
\end{equation}

\noindent where the complex atmospheric circulation is now reduced to one single quantity, $\pressvelocity$, scaling the strength of vertical flows. We notice that $\pressvelocity >0$ corresponds to subsidence, and $\pressvelocity < 0$ to updraft, since pressure and optical depth decay with altitude. 

In this one-dimensional model, we consider a regime similar to that treated in \sect{sec:sensible_heating} in the framework of the zero-dimensional RC model. The heat transport from the dayside to the nightside is supposed to be efficient, and the stratosphere is consequently globally isothermal. However, the two cells introduced in the zero-dimensional model to characterise sensible heating and large-scale heat transport by atmospheric circulation separately are now encompassed into one single stellar and anti-stellar cell (see \fig{fig:convection_cell}), so that updraft takes place within the dayside substellar region and subsidence on the nightside. Moreover, we assume uniform vertical profiles for pressure velocity both in the updraft and subsidence regions. The corresponding amplitudes are denoted by $\pvup$ and $\pvsub$, respectively. The conservation of mass thus yields the relationship 


\begin{equation}
\Asub \pvsub = \Aup \pvup,
\label{pvup_pvsub}
\end{equation}

\noindent the notations $\Aup$ and $\Asub$ designating the updraft and subsidence areas, respectively. As shown by snapshots of vertical wind speeds obtained from GCM simulations in typical cases (\fig{fig:W2015fig2_KAfig4_thor}, right panels), there is an asymmetry between rising and sinking air motions. Particularly, air rises rapidly near the substellar point where heat absorption is the strongest, while it sinks slowly over a large area in cooler atmospheric regions. This asymmetry is quantified by the ratio $\Aup / \Asub$, which is a fixed parameter of the model. As it characterises the strength of convection, the pressure velocity of rising air in the substellar region is related to the typical horizontal wind speed $\Vconv$ introduced previously in the expression of the sensible heating flux (\eq{Fturb}) and quantified using the heat engine theory (\eq{Vconv}). A simple scale analysis based on the conservation of mass yields

\begin{equation}
\pvup \sim \frac{\psurf \Vconv}{\sqrt{\Aup}}.
\end{equation}

With a similar analysis, we derive the typical time for a fluid parcel to rise up and to subside,

\begin{equation}
\begin{array}{ll}
\tupd \define \dfrac{\psurf}{\pvup} , & \mbox{and} \  \tsub \define \dfrac{\psurf}{\pvsub}.  
\end{array}
\end{equation}

\noindent Because of the asymmetry between rising and sinking motions, $\tsub \gg \tupd$. \rec{For simplification, we assume in the following} that $\Asub = 2 \pi \Rpla^2$ and $\Asub / \Aup = \tsub / \tupd  \sim 10$ \rec{and ignore the subsidence happening on the dayside (see \fig{fig:W2015fig2_KAfig4_thor}, right panels)}. The steady state thermodynamic equation thus reads, for the dayside updraft region, 

\begin{equation}
\DD{\temp}{\optdepthlw} - \frac{\betagas}{\optdepthlw} \temp =  - \frac{\tupd}{\trad} \DD{\Fnet}{\optdepthlw} ,
\label{eq_thermo_day}
\end{equation}

\noindent and, for the dayside subsidence region,

\begin{equation}
\DD{\temp}{\optdepthlw} - \frac{\betagas}{\optdepthlw} \temp =   \frac{\tsub}{\trad} \DD{\Fnetlw}{\optdepthlw} ,
\label{eq_thermo_night}
\end{equation}

\noindent where $\trad$ is the timescale of radiative cooling defined by \eq{trad_tadv}. We note that the dayside net flux in the right-hand member of \eq{eq_thermo_day} is the sum of the longwave and shortwave components. For the later, we use the analytical solution given by \eq{Fnetsw_isoT}, which is not limited to the case of isothermal profiles since it does not depend on temperature. 

Combined with the day- and nightside net flux equations (\eq{rt_eq_order2}), \eqs{eq_thermo_day}{eq_thermo_night} form the system of equations governing the global heat redistribution and dayside and nightside temperature profiles. For both the dayside and nightside, two boundary conditions are required to integrate the radiative transfer equation, and one to integrate the thermodynamic equation, meaning that six boundary conditions have to be specified. As may be noticed, the simplified dynamics adopted to derive \eqs{eq_thermo_day}{eq_thermo_night} generates a singularity at $\optdepthlw  = 0$ in their left-hand members. For this reason, upper boundary conditions cannot be applied at $\optdepthlw = 0$, and we have to set arbitrarily the optical depth of the top of the atmosphere  $\tautoplw > 0$. We verify a posteriori that results do not vary with $\tautoplw$ from the moment that $\tautoplw \ll \taugrlw$. 

\kc{We remark} that \cite{KA2016} could \kc{preliminarily} calculate the level of the tropopause to which they applied nightside boundary conditions by assuming a convective dayside temperature profile. As the dayside temperature profile is not fixed in the present work, the pressure level of the nightside upper boundary cannot be determined in a similar way. Thus, upper boundary conditions have to be applied to the fixed level $\tautoplw$.


On the \recc{dayside substellar area of rising air}, we assume that there is no radiative source at infinity in the longwave, and \recc{subtract the amount of power available to drive atmospheric motion from the outgoing longwave radiation, assuming that dissipation occurs outside of the area.} \rec{In the scaling of the typical wind speed $\Vconv$ given by \eq{Vconv}, \recc{the power available to drive atmospheric motion} is quantified by $\effconv^3 \Qin $, \recc{where $\Qin$ is the theoretical power defined by \eq{Qconv}}, and $\effconv =1/2$ the efficiency coefficient evaluated by \cite{KA2016} for the wind speed using GCM simulations\footnote{\recc{We remark that including or not the term $\effconv^3 \Qin$ in the power budget has little impact on the obtained results given that the efficiency factor $\effconv^3$ is small.}}. \recc{We remind here that the day-night heat transport by atmospheric circulation is not taken into account in this section. The stratosphere is considered as globally isothermal, meaning that dayside and nightside atmospheric temperatures are equal at the upper boundary.}} The dayside surface temperature is defined by the blackbody radiation of the surface, \rec{meaning that $\Taday = \Tday $ at surface}. On the nightside, we assume the absence of downwelling longwave flux at infinity, as on the dayside. Besides, following \cite{KA2016}, we consider that the net flux in the longwave vanishes at the planet surface on the nightside -- which is a consequence of stable stratification (no sensible heat flux) -- and that the stratosphere is horizontally isothermal. Mathematically, these conditions read, for the dayside, 


\begin{align}
 \Fdownlw & = 0  				& & \mbox{at} \ \optdepthlw = \tautoplw,  \\
\label{bcday_2}
 \Fnetlw & = - \Fnetsw - \effconv^3 \Qin 		& & \mbox{at} \ \optdepthlw = \tautoplw, \\ 
\label{bcd_temp}
\Taday & = \Tday =  \left( \frac{\Fuplw}{\sigmaSB}  \right)^{1/4}  & & \mbox{at} \ \optdepthlw = \taugrlw, 
\end{align}

\noindent and, for the nightside,

\begin{align}
\Fdownlw & = 0 		& & \mbox{at} \ \optdepthlw = \tautoplw, \\ 
\Fnetlw & = 0 			& & \mbox{at} \ \optdepthlw = \taugrlw,  \\
\Tanight & = \Taday 		& & \mbox{at} \ \optdepthlw = \tautoplw.
\end{align}

\begin{figure*}[htb]
   \centering
        \begin{flushleft}
   \hspace{0.16\textwidth} 
   $ \psurf = 0.01$\units{bar} \hspace{0.15\textwidth} $\psurf = 0.1$\units{bar}  \hspace{0.15\textwidth}  $\psurf = 1$\units{bar}
 \end{flushleft}
  \vspace{-0.4cm}
    \includegraphics[width=0.023\textwidth,trim = 2.7cm 1.7cm 23.9cm 2cm,clip]{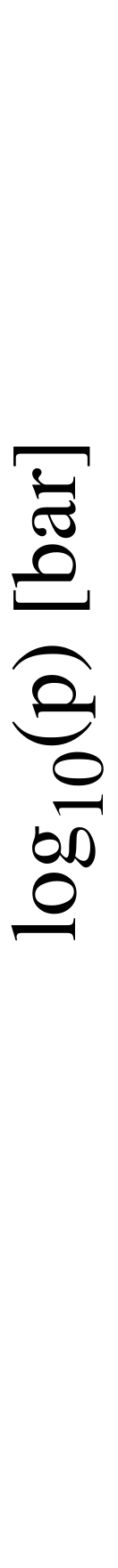} 
    \includegraphics[width=0.28\textwidth,trim = 4.3cm 1.7cm 6.7cm 2cm,clip]{auclair-desrotour_fig13a} 
   \includegraphics[width=0.28\textwidth,trim = 4.3cm 1.7cm 6.7cm 2cm,clip]{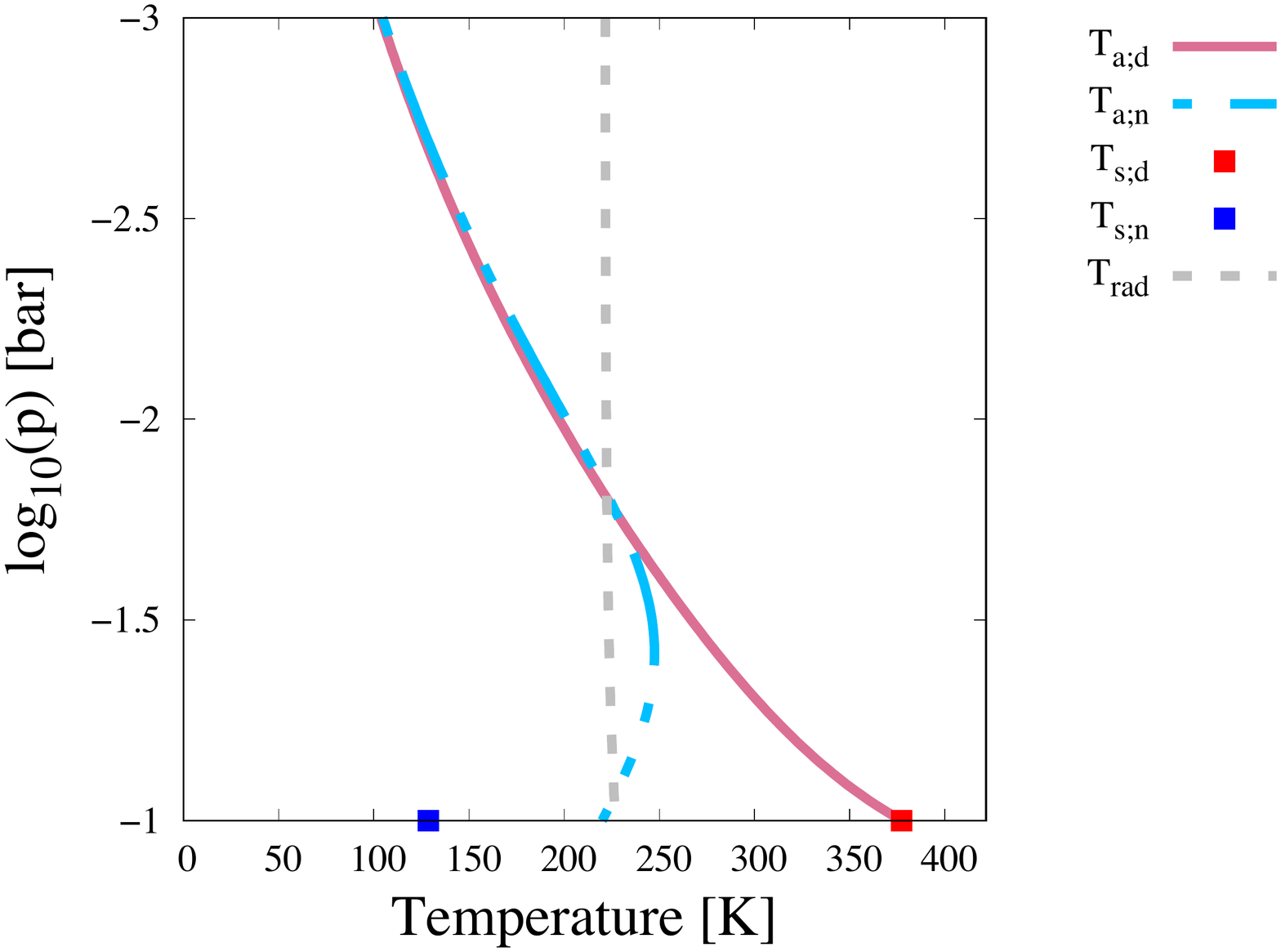} 
   \includegraphics[width=0.28\textwidth,trim = 4.3cm 1.7cm 6.7cm 2cm,clip]{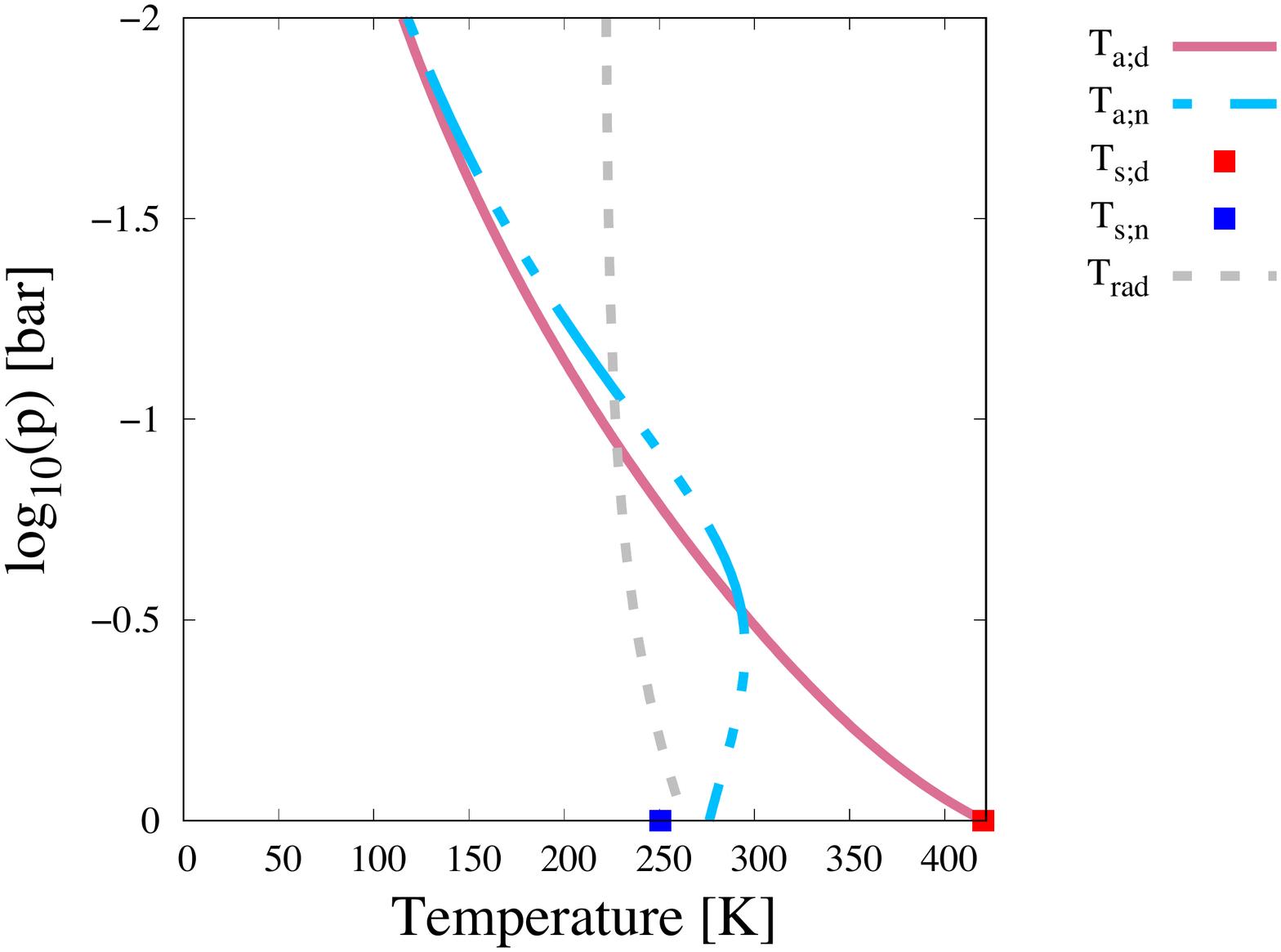}
    \includegraphics[width=0.07\textwidth,trim = 22.6cm 3.5cm 1.5cm 2cm,clip]{auclair-desrotour_fig13a}  \\

    \includegraphics[width=0.023\textwidth,trim = 2.7cm 2.5cm 23.9cm 2cm,clip]{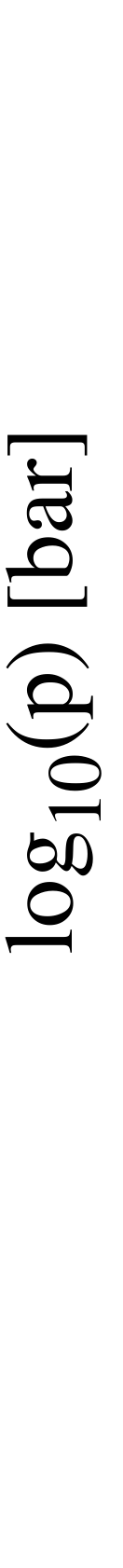} 
     \includegraphics[width=0.28\textwidth,trim = 4.3cm 2.5cm 7.5cm 2cm,clip]{auclair-desrotour_fig13d} 
   \includegraphics[width=0.28\textwidth,trim = 4.3cm 2.5cm 7.5cm 2cm,clip]{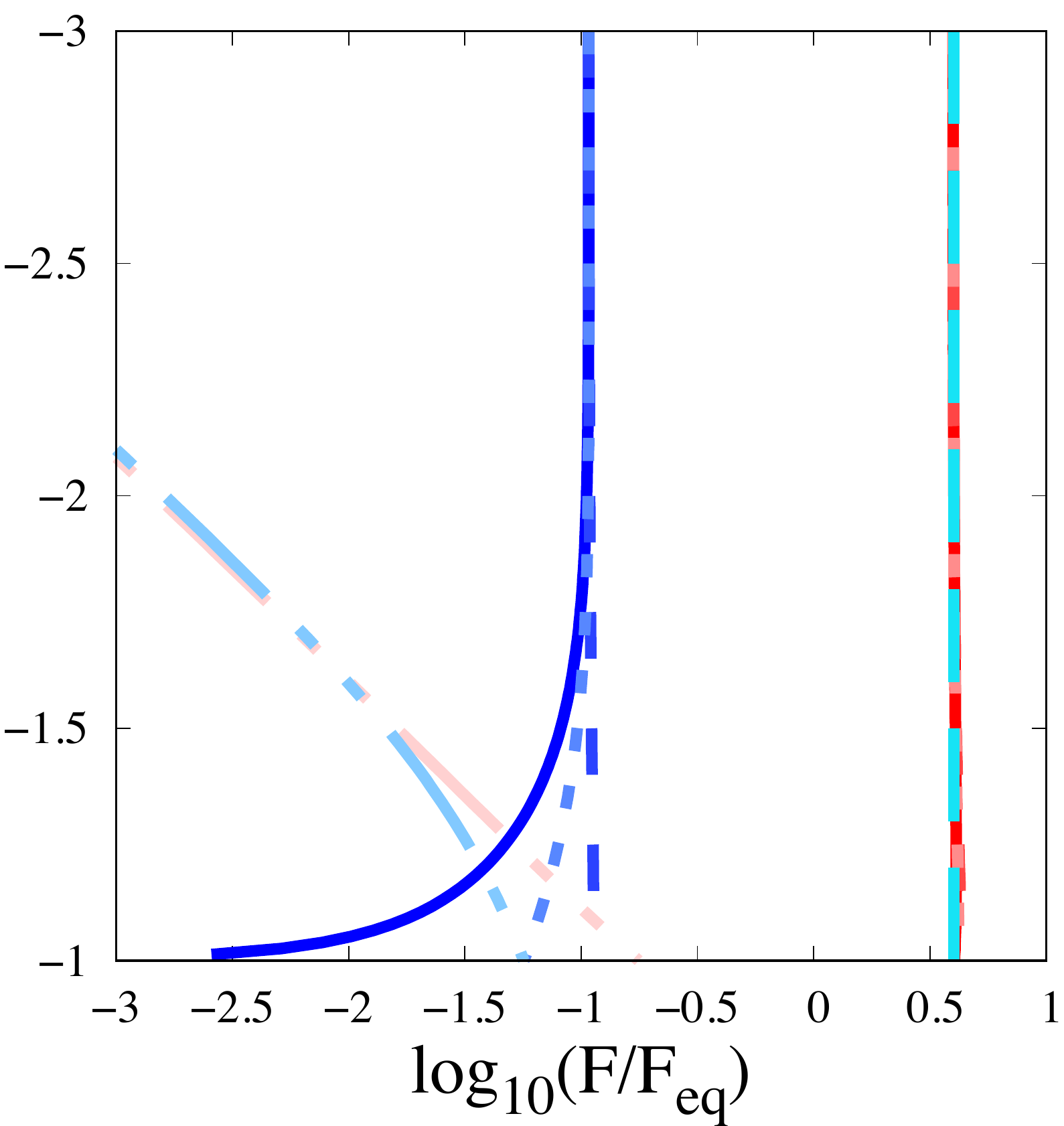} 
   \includegraphics[width=0.28\textwidth,trim = 4.3cm 2.5cm 7.5cm 2cm,clip]{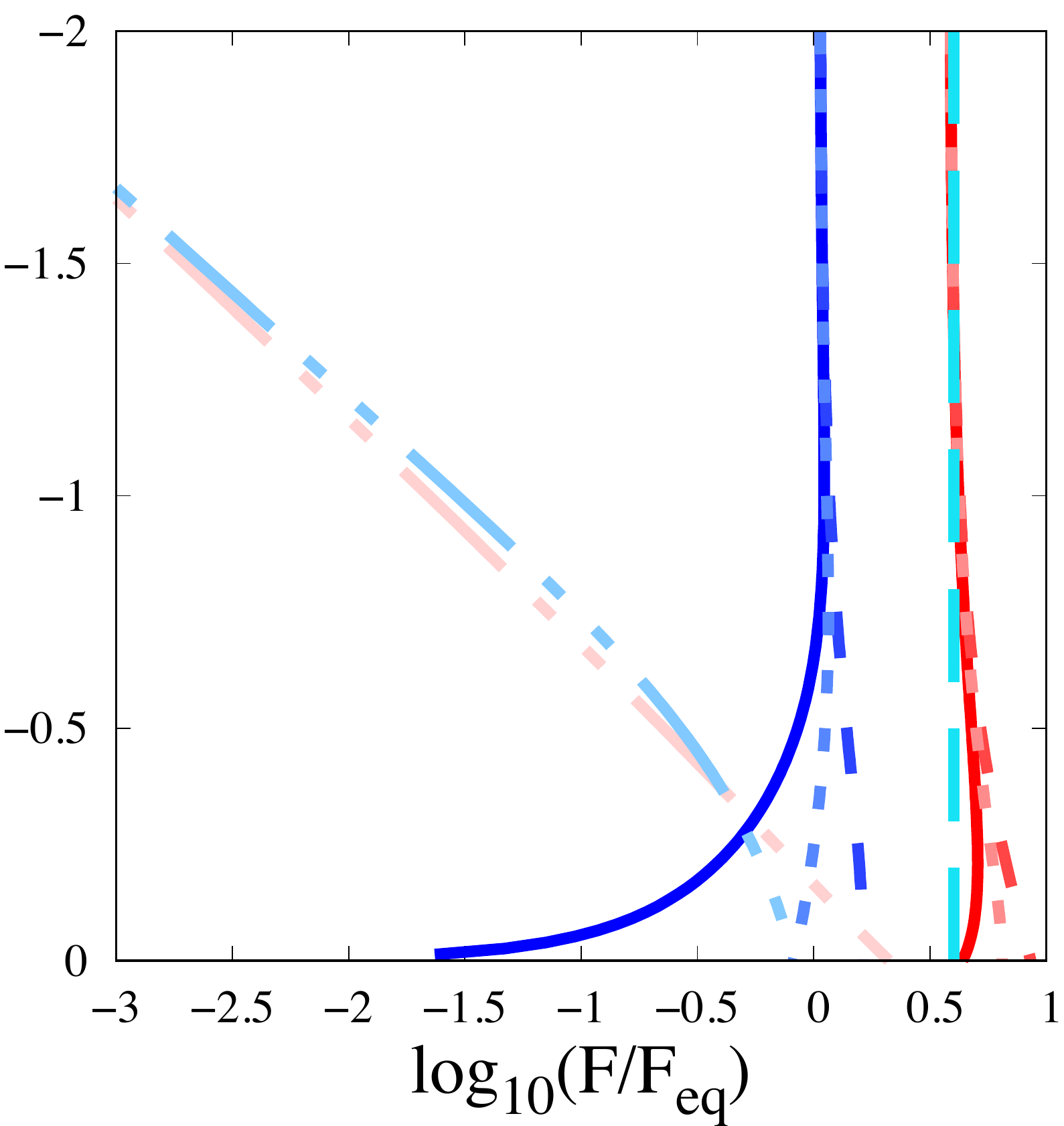}
    \includegraphics[width=0.07\textwidth,trim = 22.2cm 2.5cm 1.6cm 2cm,clip]{auclair-desrotour_fig13d}  \\
     \vspace{-0.0cm}
   \caption{Temperature structure and fluxes profiles calculated with radiative-convective-subsiding-upwelling (RCSU) model in reference case for various surface pressures as functions of pressure levels in logarithmic scale. {\it Left:} $\psurf = 0.01$~bar and $\taugrlw = 0.01$. {\it Middle:} $\psurf = 0.1$~bar and $\taugrlw = 0.1$. {\it Right:} $\psurf = 1$~bar and $\taugrlw = 1$. Dayside atmospheric (solid pink line), nightside atmospheric (dashed sky blue line), dayside surface (red square), nightside surface (blue square), and radiative equilibrium (dotted grey line) temperatures (K) are plotted in top panels. Dayside (shades of red) and nightside (shades of blue) radiative net, total, upwelling, and downwelling fluxes in the longwave, and net flux in the shortwave (dashed cyan line) are plotted in bottom panels.  Fluxes are normalised by the black body equilibrium flux $\Feq \define  \sigmaSB \Teq^4$, where $\Teq$ is the black body equilibrium temperature defined by \eq{Teq}. Calculations are performed using parameters values given by \tab{tab:param_wordsworth2015} with $\nexptau = 1$, which corresponds to Fig.~ 2 (right panel) of \cite{Wordsworth2015}. Complementary parameters of the RCSU model are set to $\Asub = 2 \pi \Rpla^2$ and $\Asub / \Aup = 10$. }
       \label{fig:RCSU_temp_profiles}%
\end{figure*}

The boundary value problem is solved over a fixed range of optical depths by means of a relaxation method. Starting from initial temperature profiles, the flux and temperature equations are integrated iteratively until convergence, an adaptive relaxation coefficient depending on the history of residuals being applied every iteration to stabilise the convergence process. At a given step, the dayside and nightside radiative transfer equations are integrated first with a finite difference scheme, by means of the shooting method \citep[][Sect.~18.1]{press2007numerical} and \kc{Thomas's} algorithm \citep[][Sect.~2.4]{press2007numerical}, respectively. Then, quantities parametrising the dynamics are calculated using auxiliary equations. Finally, the dayside and nightside thermodynamic equations are integrated, which yields new temperature profiles. The old temperature profiles are then incremented by the temperature difference between the old and new solutions weighted by the relaxation coefficient. This defines the temperature profiles for the next iteration, and so on.

\subsection{Temperatures and fluxes profiles}

We benchmark the two-column RCSU model against simulations performed by \cite{Wordsworth2015}, who used the three-dimensional \texttt{LMD} GCM \citep[][]{Hourdin2006} with grey gas opacities. These simulations correspond to the reference case described by parameters values of \tab{tab:param_wordsworth2015} with $\nexptau = 1$ and $\psurf = 0.01,0.1,1$~bar \citep[see][Fig.~2, right panel]{Wordsworth2015}. We remind \kc{ourselves} that the associated optical depths in the longwave are $\taugrlw = 0.01,0.1,1$, respectively, owing to the chosen effective longwave opacity ($\klw = 5 \times 10^{-5}$\units{m^{2}~kg^{-1}}). As may be noted, the substellar region is far smaller than the whole dayside hemisphere. Thus, the dayside atmospheric structure in this region is calculated by taking the value of the incident stellar flux at the substellar point, that is $\Fstar$. This differs both from \cite{Wordsworth2015} and \cite{KA2016}, who calculated dayside averaged profiles (with an averaged incident flux of $\Fstar/2$), and leads consequently to higher dayside temperatures and stronger fluxes. 

Temperatures and fluxes profiles computed using our RCSU model are plotted in \fig{fig:RCSU_temp_profiles}, which is the counterpart of Fig.~2 (right panel) in the study by \cite{Wordsworth2015}. For comparison, we indicate in plots the radiative equilibrium temperature $\Trad$, which corresponds to $\Fnetlw = 0$, and is expressed in the absence of scattering as \citep[e.g.][]{Pierrehumbert2011}

\begin{equation}
\Trad \define  \Tskin \left( 1 + \optdepthlw \right)^{1/4}.
\label{Trad}
\end{equation}

\noindent In this equation, $\Tskin $ designates the skin temperature, that is \kc{temperature the outer regions of the atmosphere would have in the absence of in situ heating by stellar absorption \citep[][Sect.~3.6, p.169]{Pierrehumbert2011}}. In the case where the atmosphere is transparent to incident stellar radiation, $\Tskin = 2^{-1/4} \left( 1 - \Asurfsw \right)^{1/4} \Teq$, which is typically the temperature of the stratosphere. 

We recover in \fig{fig:RCSU_temp_profiles} the behaviour described by the zero-dimensional RC model for surface temperatures, which are both increasing with the atmospheric optical thickness in the longwave because of greenhouse effect. Quantitatively, the predictions of the RCSU model for the nightside surface temperature match rather well the results obtained from GCM simulations, although the RCSU model tends to overestimate the nightside temperature with respect to the GCM in the case $\psurf = 1$~bar. The RCSU model also captures the variation of atmospheric structure between the dayside and the nightside. While convection on the dayside leads to a quasi-adiabatic temperature profile, the atmosphere is cooler and stably stratified on the nightside, which induces a smaller temperature gradient. Particularly, we retrieve here the nightside structure inversion observed in GCM simulations \citep[e.g.][]{Leconte2013,KA2016} and due to subsidence, as discussed by \cite{KA2016}. 

Because of the assumed vertically-uniform pressure velocity, both dayside and nightside atmospheric temperatures decay with pressure in order to compensate the increase of the singular term of the simplified thermodynamic equations (\eqs{eq_thermo_day}{eq_thermo_night}). In reality, vertical winds vanish above the troposphere, and there is no singularity. The tropopause approximately corresponds to the pressure level where $\Taday$ and $\Tanight$ join together. Above this pressure level, the temperature profile should approach the radiative equilibrium temperature profile $\Trad$ (\eq{Trad}) instead of decaying, as shown by GCM simulations \citep[e.g.][Fig.~2, right panel]{Wordsworth2015}. In spite of the unrealistic tendency observed in the stratosphere, the nightside temperature profile derived from the RCSU model in the troposphere is qualitatively similar to that obtained by \cite{KA2016} with their radiative-convective-subsiding model (Fig.~4 of the article). Thus we may now compare the predictions of the two models regarding atmospheric stability and test thereby the RCSU model.

\subsection{Atmospheric stability}

Considering the reference case of the study, we run grid calculations using the RCSU model to compute the nightside temperature of the planet, and determine how the stability of the steady state against collapse evolves with the incident stellar flux and surface pressure. Consistent with previous sections, the atmospheric stability is simply obtained by comparing the nightside surface temperature to the condensation temperature of $\carbondiox$, given by \eqs{TcondCO2_1}{TcondCO2_2}. We note that the boundary value problem becomes difficult to solve accurately in the optically thick limit because tiny changes of the upper boundary condition ($\optdepthlw = \tautoplw $) in Schwarzschild equation (\eq{rt_eq_order2}) substantially affect radiative fluxes at large $\optdepthlw$. As we start encountering convergence issues for $\taugrlw \gtrsim 10$, we set the upper bound of the surface pressure range to $\logdix \left( \psurf \right) = 0.5$ (bar).

\begin{figure}[htb]
   \centering
    \raisebox{1.5cm}{ \includegraphics[width=0.018\textwidth]{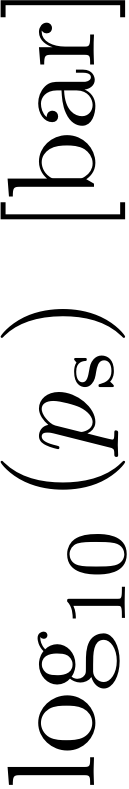}}
   \includegraphics[width=0.46\textwidth,trim =  2.5cm 2.5cm 5.9cm 1.9cm,clip]{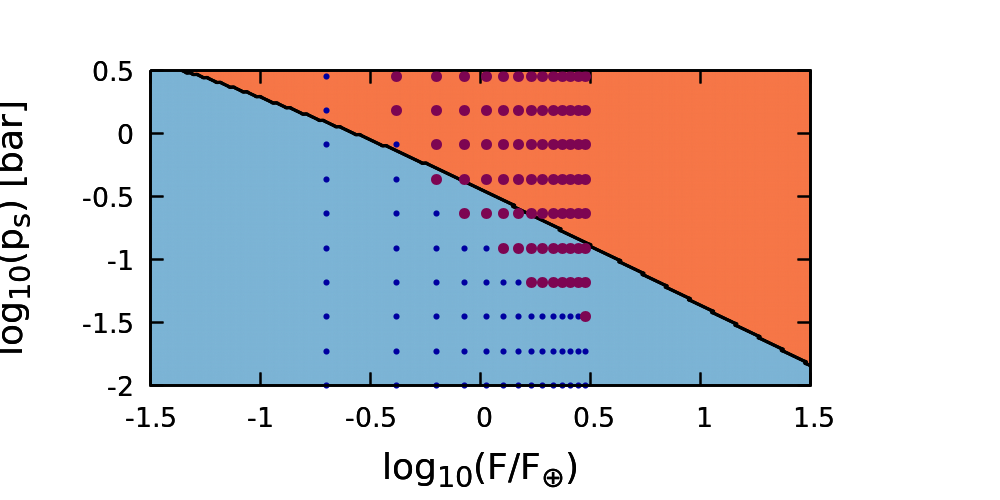} \\
   \small{$\logdix \left( \Fstar / \Fearth \right)$}
      \caption{Stability diagram computed with one-dimensional radiative-convective-subsiding-upwelling (RCSU) model in reference case. The orange area indicate a stable atmosphere, and the blue area collapse in the RCSU model. Data computed by \cite{Wordsworth2015} from GCM simulations (Fig.~12) are included for comparison. Violet dots indicate simulations where the atmosphere remained stable, while small blue dots indicate atmospheric collapse. Calculations are performed using parameters values given by \tab{tab:param_wordsworth2015} with $\nexptau = 1$, which corresponds to Fig.~ 2 (right panel) of \cite{Wordsworth2015}. Complementary parameters of the RCSU model are set to $\Asub = 2 \pi \Rpla^2$ and $\Asub / \Aup = 10$.}
       \label{fig:RCSU_stability}%
\end{figure}

Figure~\ref{fig:RCSU_stability} shows the obtained stability diagram, on which data computed by \cite{Wordsworth2015} from GCM simulations are also plotted. This diagram may be compared to those calculated using the zero-dimensional RC model (\fig{fig:map_stabilite_effsen_W2015}) and to the results obtained by \cite{KA2016} with their RCS model \citep[][Fig.~12, left panel]{KA2016}. Although it includes the day-night difference of atmospheric structure, the RCSU model does not match GCM simulations, which were well approached by the RCS model of \cite{KA2016}. Particularly, the model does not capture the behaviour caused by the non-linear dependence of temperature on the atmospheric optical thickness, whereas this behaviour was captured by the 0-D approach, as shown by \fig{fig:map_stabilite_effsen_W2015} (top right panel). The scaling of the collapse pressure given by the RCSU model actually corresponds to the purely radiative case of the 0-D model, that is $\pcrit \scale \Fstar^{-1}$ (\eq{pcrit_scaling}). 

These limitations of the theory are apparently due to the oversimplified dynamics assumed to derive the differential equations of dayside and nightside temperatures (\eqs{eq_thermo_day}{eq_thermo_night}). Particularly, the pressure velocity \kc{(i.e. the velocity in pressure coordinates)} is supposed uniform over the whole air column while it should tend to zero above the tropopause. As a consequence, the model does not describe properly the stratosphere. The inclusion of shortwave opacities in the two-column RCS approach thus still remain an open question for future studies. Moreover, temperature and winds profiles strongly vary with the stellar zenithal angle on the dayside, which cannot be properly modelled in the framework of a 1D theory. This encourages the development of 2D intermediate models treating the coupling between the large-scale atmospheric circulation and radiative transfers in a self-consistent way in the regime of slowly rotating planets.

\section{Discussion and conclusions}
\label{sec:conclusions}

In order to better understand the mechanism of atmospheric collapse, which is of fundamental importance to characterise surface conditions of tidally locked rocky planets, we developed a \kc{hierarchy of models} building on pioneering works of the analytical theory \kc{in the spirit of \cite{Held2005}}. We aimed thereby to enrich and to consolidate the theory by means of cross-comparisons between multiple approaches ranging from 0-D scalings to 3D GCM simulations. Models of increasing complexity were thus developed following a step-by-step method to treat the case of dry atmospheres hosted by slowly rotating rocky planets by including incrementally the couplings between \kc{radiative transfer}, convection and large-scale circulation. In these models, \kc{radiative transfer is} described using the two-stream, dual-band and grey gas approximations, while the heat transport by atmospheric circulation is treated in the framework \kc{of heat engine theory} following \cite{KA2016}. 

We first revisited the early study by \cite{Wordsworth2015}, which is based on a 0-D approach. Assuming an isothermal atmosphere, we derived analytically the short- and longwave net flux profiles in presence of absorption and scattering, as well as the atmospheric, dayside and nightside surface temperatures characterising the steady state in the purely radiative case. This enabled us to quantify at first order how scattering and shortwave absorption modify the equilibrium temperatures and atmospheric stability against collapse by acting on greenhouse effect. Typically, longwave and shortwave absorption both increase greenhouse effect, and thus the atmospheric stability, which is not the case of scattering. While scattering in the shortwave induces anti-greenhouse cooling and favours collapse, scattering in the longwave tends to stabilise the atmosphere by generating scattering greenhouse effect. The model captures these dependences in a simplified way. 

It also captures the particular behaviour of the collapse pressure (or critical pressure) at low stellar fluxes due to the non-linear dependence of the atmospheric thickness on the longwave opacity. Although highlighted by \cite{Wordsworth2015} using GCM simulations, this behaviour was absent of the theory proposed by the author since the regime of optically thin atmospheres solely was considered in this early study. It results from the fact that the absorbed power ceases to grow linearly with the atmospheric optical depth at the planet's surface as one leaves the optically thin regime. The present study shows that, in spite of being a rough approximation of the atmospheric structure, the isothermal atmosphere assumption is sufficient to capture the induced scaling of collapse pressure with stellar flux that may be observed in GCM simulations. 

As a second step, we included in the model the turbulent exchanges due to convection within the dayside boundary layer in the asymptotic regime where the atmosphere is globally isothermal (efficient large-scale heat transport). The sensible heat flux between the atmosphere and surface was scaled considering the atmosphere as a heat engine controlled by the atmospheric and dayside surface temperatures. In this framework, we derived the steady-state equation that governs the global heat redistribution, and solved it numerically to recover the early results obtained by \cite{Wordsworth2015} and \cite{KA2016} in the case of $\carbondiox$-dominated atmospheres. Besides, we derived analytically a lower bound for the collapse pressure. This lower bound corresponds to the regime of strong convection, which maximises greenhouse effect. 

As a third step, we ceased to treat the atmosphere as a globally isothermal layer and examined the case where the dominating heat transport mechanism is coupled with dayside and nightside temperatures. By way of an example, we supposed the heat to be advected from the dayside to the nightside by a stellar and anti-stellar atmospheric circulation, which is the typical dynamical regime of slow rotators. The associated heat flux was thus quantified from a scale analysis by treating the stellar and anti-stellar cell as a heat engine, as done for dayside convection. This simplified modelling accounts for the fact that the efficiency of heat transport by large-scale advection decays as the planet size increases. It allowed us to derive a generalised version of the steady state equation including the coupled effects of \kc{radiative transfer}, sensible heating, and large-scale atmospheric circulation. 

In GCM simulations, the atmospheric stability decays as the planet-size increases. Our generalised 0-D model approximately captures this behaviour and relate it to two conjugated effects: (i) for given gas opacity and surface pressure, the optical depth at the planet's surface decays as the surface gravity increases, which weakens greenhouse effect; and (ii) the ratio of the advective and radiative timescales decay, which decreases the hemisphere-averaged amount of heat transported from the dayside to the nightside. 

The 0-D approach may certainly be improved in future studies by introducing refined scalings for heat fluxes and mean flows, which would better capture the dependence of the collapse pressure on the physical properties of the star-planet system. However, the model already allows us to characterise asymptotic regimes using a small number of dimensionless control parameters, such as -- for instance -- $\Lconv$ and $\Ladv$, which compare dayside sensible heating and large-scale heat transport with radiative cooling, respectively. Furthermore, the fact that the root-finding procedure solving the steady state equation can be massively parallelised makes it possible to widely explore the parameter space. This clearly encourages the development of analytic 0-D models as a complement of studies based on GCM simulations. 

As a last step, we investigated the transition between 0-D and 1D models by studying how the atmospheric circulation affects the dayside and nightside atmospheric structure. Starting from the two-column radiative-convective-subsiding (RCS) model proposed by \cite{KA2016}, we developed a radiative-convective-subsiding-upwelling (RCSU) model. In this approach, the dayside and nightside temperature profiles are modified by vertical winds. While upwelling winds due to convection in the substellar region drive the temperature gradient towards the adiabatic profile, subsidence on the nightside causes a structure inversion leading to stable stratification. In order to extend the theory to the case of non-transparent atmospheres in the shortwave, we let the dayside temperature profile be integrated simultaneously with the nightside temperature profile, instead of specifying its shape, which was done by \cite{KA2016}. 

From a quantitative point of view, the RCSU model does not match well the behaviour of the collapse pressure obtained from GCM simulations, and rather predict a scaling similar to that given by the 0-D model in the purely radiative case. This apparently results from the simplifications made in the dynamics to include the effect of vertical winds. Particularly, in both subsidence and updraft areas, the vertical wind speed was supposed to be uniform over the whole air column, although the nature of the dynamics is not the same in tropospheric and stratospheric layers. Moreover, horizontal advection was ignored notwithstanding the fact that it is comparable to vertical advection in the general case. Future studies will have to remedy to both of these rough approximations to derive self-consistently realistic temperature profiles and collapse pressure at the same time with the 1D two-column approach, which still remains an open question. Nevertheless, in spite of its limitations, the RCSU model is a promising first attempt that allowed us to recover the features identified by early works, and particularly the nightside structure inversion previously captured by the RCS model \citep[][]{KA2016}. 

\kc{For future works, it} is crucial to improve the treatment of the dynamics in the analytic theory since it is a major aspect of the problem. This suggests to develop an intermediate class of 2D atmospheric models describing the complex coupling between mean flows and the thermodynamics in the asymptotic regime of slowly rotating tidally locked planets, the circulation being symmetric with respect to the star-planet axis in this regime. These simplified 2D GCMs would also be useful to quantify timescales associated with the atmospheric collapses, which cannot be achieved with steady states models, and requires a substantial computational effort when 3D GCMs are used.  

In addition to the dynamics, \rec{one should evolve} towards improved modellings of radiative transfers including wavelength-dependent opacities and accounting for the relationship between atmospheric absorption and stellar spectra, which is a blind spot of the dual-band approximation. This aspect is of fundamental importance when the stellar and planetary radiative fluxes overlap, that is typically in the case of temperate exoplanets orbiting cool dwarf stars. TRAPPIST-1 planets are a representative example of such a configuration. This question could be addressed for instance by introducing simplified dynamics in 1D radiative transfer codes. 

Similarly as their predecessors, the simplified 0-D and 1D analytical models detailed in the present work are limited to dry thermodynamics and slow rotation, which corresponds to the simplest physics. The presence of water is however a feature of uppermost interest in the study of Earth-sized exoplanets located in the habitable zone of their host star because of its implications on climate and prebiotic chemistry. Thus, this feature should be introduced in the analytic theory in order to consolidate the conclusions of works essentially based on GCM simulations \citep[e.g.][]{MS2010,Leconte2013}. 

Particularly, the presence of a slab ocean induces an additional heat transport both by latent heat flux in the atmosphere and by heat diffusion in the ocean itself \citep[][]{Edson2011}, and tends thereby to stabilise the atmosphere against collapse. Additionally, water vapour leads to the formation of clouds, which affects the atmospheric albedo, thermal structure, and general circulation in a complex way. To complete the theory, similar inclusions may be done for the heat transport by propagating gravity waves, and super-rotation in the regime of rapid rotators. Most of these aspects were discussed in early works \citep[see e.g.][]{Leconte2013}.

Finally, we shall emphasise that steady states do not necessarily exist, and if so, are not necessarily stable. For example, \cite{Edson2011} showed evidence of the bistability characterising the dependence of the wind speeds on the planet spin by highlighting an abrupt transition between two regimes. This transition leads to a significant change of velocity. Regarding this type of questions, the analytic theory has a strong asset with respect to GCMs since it offers the opportunity to characterise formally the existence and stability of steady states.

\begin{acknowledgements}
\rec{The authors thank the reviewer, D.~D.~B. Koll, for helpful comments that contributed to improve the manuscript. They} acknowledge financial support from the European Research Council via the Consolidator grant \texttt{EXOKLEIN} (grant number 771620). \rec{K.~Heng also acknowledges partial financial support from the National Swiss Foundation, the PlanetS National Center of Competence in Research, the Center for Space \& Habitability and the MERAC Foundation.} This research has made use of NASA's Astrophysics Data System.
\end{acknowledgements}


\bibliographystyle{aa}  
\bibliography{references} 

\appendix

\section{Turbulent heat exchanges}
\label{app:turbulent_heat_exchanges}

At the planet surface, the heat flux resulting from turbulent friction is written as 

\begin{equation}
\Dconv \define \rhoatm \Cp \left( \mean{\deltatemp \deltaVr} \right)_{\isurf},
\label{Fturb1}
\end{equation}

\noindent where $\rhoatm = \psurf / \left( \Rspec \Tatm \right)$ is the density of the atmosphere at the planet's surface, $\deltatemp$ and $\deltaVr$ are fluctuations of temperature and radial velocity, respectively. 

Fluctuations of the radial flow are induced by turbulence due to the interaction between the dominating horizontal flow and the planet surface. Thus, we can write the energy transport in \eq{Fturb} as $\left( \mean{\deltatemp \deltaVr} \right)_{\isurf} = \Vfric \tempfric $, where $\Vfric$ and $\tempfric$ correspond to frictional velocity and temperature, respectively. Using the flux gradient theory with the mixing length hypothesis, these quantities are defined as 

\begin{equation}
\begin{array}{ll}
 \Vfric \define \mixingL \abs{\dd{\meanV}{\zz}} , & \mbox{and} \ \tempfric \define \mixingL \abs{  \dd{\meantemp}{\zz} },
\end{array}
\label{frictional_vel_temp}
\end{equation}

\noindent where $\zz$ is the altitude with respect to the surface, $\mixingL$ the turbulence mixing length, and $\meanV$ and $\meantemp$ the average horizontal velocity and temperature, respectively. 

The turbulence mixing length is calculated from the empirical scaling law of \cite{Blackadar1962},

\begin{equation}
\mixingL = \frac{\mixingLmax \karman \zz}{\karman \zz + \mixingLmax},
\label{ML_blackadar}
\end{equation}

\noindent the parameter $\karman \approx 0.4$ being the von K\'arm\'an constant and $\mixingLmax$ the maximum  attainable mixing length in the boundary layer. By substituting \eq{ML_blackadar} into \eq{frictional_vel_temp}, and integrating the equations, we obtain 

\begin{align}
\Vfric = & \ \frac{\karman}{\karman \mixingLmax^{-1} \left( \zz - \roughheight \right) + \ln \left( \zz / \roughheight \right) } \meanV \left( \zz \right)  , \\
\tempfric = & \ \frac{\karman}{\karman \mixingLmax^{-1} \left( \zz - \roughheight \right) + \ln \left( \zz / \roughheight \right) } \left[  \meantemp  \left( \zz \right) - \meantemp \left( \roughheight \right)   \right],
\end{align}

\noindent where $\roughheight$ is the roughness height. Thus, considering that $\zz$ is the thickness of the boundary layer and introducing the bulk drag coefficient 

\begin{equation}
\Cd = \left[ \frac{\karman}{\karman \mixingLmax^{-1} \left( \zz - \roughheight \right) + \ln \left( \zz / \roughheight \right) } \right]^2,
\label{Cd_drag}
\end{equation}

\noindent  we can write the vertical turbulent heat flux (from the surface to the atmosphere) as 

\begin{equation}
\Dconv = \Cd \Cp \rhoatm \absV \left( \Tsurf -  \Tatm \right).
\label{Fturb2}
\end{equation}

By assuming $\mixingLmax = + \infty$ (no fixed maximum for the turbulence mixing length), we recover the scaling law of the mixing length initially proposed by von K\'arm\'an on the basis of laboratory experiments, $\mixingL = \karman \zz$ \citep[][]{vonKarman1931}, and the classical expression for the bulk drag coefficient given by \eq{bulk_coeff}. The parameters $\zz$ and $\roughheight$ are free parameters in the general case. They are set to $\zz = 10$~m and $\roughheight = 1$~cm in \cite{Wordsworth2015}. 

\section{Solving the steady state equation in the general case}
\label{app:solving_equation}

Equation~(\ref{eqsingle_gen}) defines a root-finding problem. This problem is formulated as 

\begin{equation}
\fTanorm \left( \Tanorm \right) = 0,
\label{zero_finding}
\end{equation}

\noindent where $\fTanorm$ is the function defined by the left-hand member of \eq{eqsingle_gen}, that is

\begin{align}
\fTanorm \left( \Tanorm \right) \define & \left[ \frac{1 - \Agrsw}{\Alw} \left( \frac{2 - \frac{\Alw}{\Klw} }{\Ladv} \right)^{\frac{4}{3}}  \!  \!  \frac{  \Tanorm^{16/3}  }{\left( 1 - \Tanorm \right)^2} + 1 \right] \Tnorm^4 \left( \Tanorm \right) - \frac{\Klw}{\Alw}  \nonumber \\
    &  - \Lconv   \!  \left( \frac{2 - \frac{\Alw}{\Klw} }{\Ladv} \right)^{\frac{11}{9}}  \!  \!  \! \left(  \frac{\Tanorm^{\frac{8}{3}}}{1 - \Tanorm} \right)^{\frac{11}{6}}  \! \!  \!   \left[ 1 - \Tnorm \left( \Tanorm \right) \right]^{\frac{4}{3}} \Tnorm^3 \left( \Tanorm \right)   \!  .  \nonumber
\end{align}

To solve \eq{zero_finding}, we first have to determine the range of $\Tanorm$ where $\Tnorm$ is defined, which corresponds to the range where the denominator of the expression given by \eq{Tnorm_Tanorm} is strictly positive. This is a preliminary zero-finding problem, which is treated using the standard secant method \citep[e.g.][]{press2007numerical}. We thus obtain the lower bound of the range, namely $\Tanorminf$. Equation~(\ref{zero_finding}) is then solved by using a combination of the dichotomy and secant method within the interval $\Tanorminf < \Tanorm \leq 1$.

We remark that $\fTanorm$ is subject to very sharp variations in the vicinity of $\Tanorminf$ while it smoothly varies with $\Tanorm$ in the rest of the interval. As a consequence, the coordinate $\Tanorm$ is not well appropriate to the root-finding procedure. We rather use the logarithm of the normalised distance, 

\begin{equation}
\DTanorm \define \logdix \left( \frac{\Tanorm - \Tanorminf}{1 - \Tanorminf} \right),
\end{equation}

\noindent the notation $\logdix$ designating the decimal logarithm. The problem thus becomes $ \fTanormlog \left(  \DTanorm \right) = 0$, where $\fTanorm \left( \DTanorm \right) = \fTanorm \left( \Tanorm \right)$. Finally, to deal with the possible issues resulting from the variations of $\fTanorm$ over several orders of magnitude, we apply the root finding method to the problem

\begin{equation}
\left( \fsmooth \rond \fTanormlog \right) \left( \DTanorm \right) = 0,
\end{equation}

\noindent where $\rond$ is the function composition operator and $\fsmooth$ the smoothing function defined as a function of the variable $\XX$ by

\begin{equation}
\fsmooth \left( \XX \right) \define \sign \left( \XX \right) \logdix \left( 1 + \abs{\XX} \right),
\end{equation}

\noindent the notation $\sign$ referring to the sign function ($\sign \left( \XX \right) = -1,1$ if $\XX<0$ or $\XX>0$, respectively, and $\sign \left( 0 \right) = 0$). 

\begin{figure}[htb]
   \centering
   \includegraphics[width=0.45\textwidth,trim = 1.5cm 2.5cm 1.5cm 1.0cm,clip]{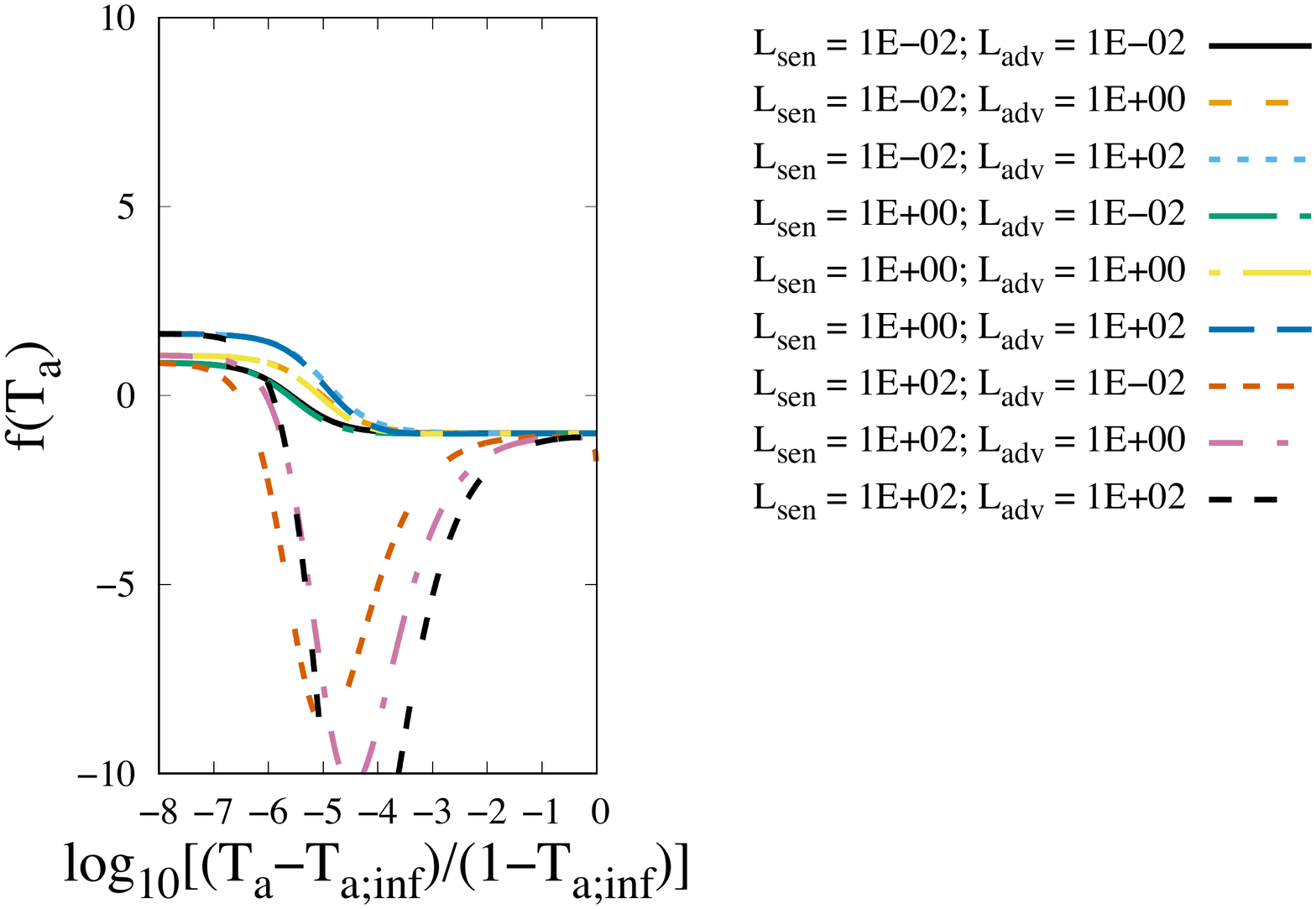}
      \caption{Left-hand member of \eq{eqsingle_gen} as function of normalised temperature difference $\Tanorm - \Tanorminf$ in logarithm scale for $ \Lconv, \Ladv = 0.01,1,100 $. The function is denoted by $f$. The parameters $\Lconv$ and $\Ladv$ are the non-dimensional control parameters of sensible heating and advected heat transport defined by \eqs{Lconv}{Ladv}, respectively. Parameters values: $\betalw = \betasw = 1$, $\taugrlw = 10$; $\taugrsw = 0.1$; $\Fstar = 1366$\units{W~m^{-2}}; $\Asurfsw = 0.2$. }
       \label{fig:function_Tanorm}%
\end{figure}

For an illustrative purpose, $\fTanorm$ is plotted in \fig{fig:function_Tanorm} as a function of $\DTanorm$ for different values of the nondimensional parameters controlling atmospheric circulation, $\Lconv$ and $\Ladv$. In these calculations, the atmosphere is optically thick in the longwave ($\taugrlw = 10$) and thin in the shortwave ($\taugrsw = 0.1$). The figure shows that there is one unique solution to the steady state equation in each of the treated cases. Besides, roots are found to be very close to $\Tanorm$ with values ranging within the interval $10^{-7} \lesssim \Tanorm - \Tanorminf \lesssim 10^{-4}$. This interval depends on the atmospheric opacity, and increases when the optical depth in the longwave decays.

\end{document}